\documentclass[12pt,a4paper]{report}
\usepackage[algochapter]{algorithm2e}
\usepackage{scienceThesis}
\usepackage{verbatim}
\usepackage{amsfonts}
\usepackage{amsmath}
\usepackage{amssymb}
\usepackage{amsthm}
\usepackage{pdfpages}
\usepackage{lipsum}
\usepackage[acronym,toc,nonumberlist]{glossaries}
\usepackage{breqn}

\theoremstyle{definition}

\theoremstyle{definition}

\theoremstyle{definition}

\theoremstyle{definition}

\theoremstyle{definition}

\theoremstyle{definition}

\usepackage{fancyhdr}
\makeatletter
\renewcommand{\chaptermark}[1]{\markboth{\small\textsc{\@chapapp}\ \thechapter:\ \sc{#1}}{}}
\makeatother
\usepackage{sectsty}
\chapterfont{\large\sc\centering}
\chaptertitlefont{\sc\centering}
\subsubsectionfont{\centering}

\usepackage{hyperref}
\hypersetup{
	bookmarksnumbered,
	pdfstartview={FitH},
	citecolor={black},
	linkcolor={black},
	urlcolor={black},
	pdfpagemode={UseOutlines}
}
\usepackage{lmodern}
\usepackage[T2A,T1]{fontenc}
\usepackage[english]{babel}
\usepackage{amsmath}
\usepackage{amsfonts}
\usepackage{amssymb}
\usepackage{graphicx}
\usepackage{tabularx}
\usepackage{listings}   
\newcommand{\shah}{III}
\newcommand{\etal}{\textit{et al.}}
\newcommand{\vis}{\varphi}
\usepackage{booktabs}
\usepackage{multirow}
\usepackage{float}
\usepackage{placeins}

\SetKwFor{ForAllP}{for all}{do in parallel}{end}
\usepackage[none]{hyphenat}
\usepackage[numbers,sort&compress]{natbib}
\usepackage{IEEEtrantools}

\newcolumntype{L}{>{\raggedright\arraybackslash}X}
\newcolumntype{P}[1]{>{\raggedright\arraybackslash}p{#1}}
\begin{document}

\title{High-Performance Image Synthesis for Radio Interferometry}
\author{Daniel Muscat}
\date{August 2013}
\supervisor{Dr Kristian Zarb Adami}

\department{Physics}
\degree{M.Sc}
\acknowledgements{As in most ambitious and successful projects, success is only possible with the help of others. \\ \\
I wish to thank my supervisor, Dr Kristian Zarb Adami, for his 24x7 support, dedication and patience.  Special thanks to Dr Oleg Smirnov and Dr Filippe Abdalla, who provided valuable insights and advice. Thanks also to Mr Alessio Magro who gave advice and help on GPU programming.  A big thanks goes to Dr John W. Romein for supplying me with the test code and data set related to his work, used in this thesis. Thanks also to Dr Anna Scaife and Dr Cyril Tasse for their help. \\ \\ 
My family also gave their contribution. A big thank you  goes to my parents Joseph and Maria Dolores Muscat for their support. Their support was given during a period when they also had to face a number of health problems. My father also  helped to creating some of the graphics included. A last but not least heartfelt thanks goes to my girlfriend Ms Nicola Attard, who, notwithstanding the challenging environment to our relationship, stood by me and gave me her full support.
}

\abstract{A radio interferometer indirectly measures the intensity distribution of the sky over the celestial sphere. Since measurements are made over an irregularly sampled Fourier plane, synthesising an intensity image from interferometric measurements requires substantial processing. Furthermore there are distortions that have to be corrected.\\ \\
In this thesis, a new high-performance image synthesis tool (imaging tool) for radio interferometry is developed. Implemented in C++ and CUDA, the imaging tool achieves unprecedented performance by means of Graphics Processing Units (GPUs). The imaging tool is divided into several components, and the back-end handling numerical calculations is generalised in a new framework. A new feature termed \textit{compression} arbitrarily increases the performance of an already highly efficient GPU-based implementation of the \textit{w-projection} algorithm. \textit{Compression} takes advantage of the behaviour of oversampled convolution functions and the baseline trajectories. A CPU-based component prepares data for the GPU which is multi-threaded to ensure maximum use of modern multi-core CPUs. Best performance can only be achieved if all hardware components in a system do work in parallel. The imaging tool is designed such that disk I/O and work on CPU and GPUs is done concurrently.\\ \\
Test cases show that the imaging tool performs nearly 100$\times$ faster than another general CPU-based imaging tool. Unfortunately, the tool is limited in use since \textit{deconvolution} and \textit{A-projection} are not yet supported. It is also limited by GPU memory. Future work will implement \textit{deconvolution} and \textit{A-projection}, whilst finding ways of overcoming the memory limitation.}

\include{}

\chapter{Introduction to Radio Interferometry and Image Synthesis}

The first interferometer used in astronomy dates back to 1921 when Albert Abraham Michelson together with Francis G. Pease made the first diameter measurements of the star Betelgeuse \cite{Michelson1921}. The setting used, known as the \textit{Michelson Interferometer}, is today the basis of modern interferometers. Michelson had been discussing the use of interferometry in astronomy for at least 30 years before the experiment was done \cite{CI_Michelson_stellar,Michelson1920}. Throughout the century,  the technology of radio interferometers has made enormous advancement and up to this day extraordinarily ambitious projects have been commissioned, and more are on the pipeline. An example of a recently commissioned radio interferometer is the \textit{LOw-Frequency ARray} (LOFAR) \cite{Haarlem2013}. The \textit{Square Kilometre Array} (SKA) \cite{2013,Dewdney2013,P.E.Dewdney2011,P.E.Dewdney2010,R.T.Schilizzi2007} is the most ambitious project currently under-way.

Radio interferometers can be used for various purposes, even for applications outside the scope of astronomy, such as geophysics \cite{Carter1988,Robertson1987}. The scope of this thesis is limited to the use of the interferometer as a measuring device for the intensity distribution of the sky over the celestial sphere. The measured quantity of the interferometer is known as \textit{visibility}, and the thesis' main focus is on how to recover the said intensity distribution from such measured visibility data.

An interferometer is made up of an array of N$\geq$ 2 antennas, and differently from a single dish antenna it achieves sub-arcsecond resolutions with high accuracy. The maximum angular resolution  of a single dish $\theta_m$ is limited by its diffraction limit. The limit is inversely proportional to the diameter $D$ of the dish, that is $\theta_m\propto 1/D$. On the other hand, the angular resolution of the interferometer is limited by the distance between the furthest two antennas in the array $B_{max}$ in the same inverse proportional way, that is $\theta_m\propto 1/B_{max}$. Achieving sub-arcsecond resolution in single-dish antennas requires large diameters that are prohibitory. For interferometers, achieving sub-arcsecond resolution is just a matter of having antennas as far away as possible from each other. If necessary, part of the array can be orbiting in space such as in the case of the \textit{Very Long Baseline Interferometry} (VLBI) \textit{Space Observatory Programme} (VSOP) \cite{Hirabayashi1998}.

The basic measurement device (based on the \textit{Michelson interferometer}) is composed of just two elements. Each possible two element combination of N $\geq$ 2 antennas forms a two element independent measuring device. During an observation, the antenna array tracks a position on the celestial sphere normally within the field of view of the observation. Each basic element makes a measurement of the intensity distribution of the sky in the form of visibility values. $N(N-1)/2$ visibility readings are done simultaneously by the whole $N$-element antenna array and each reading differs in the geographical set-up of the basic device. The geographical set-up is described by the \textit{baseline} which is the vector covering the distance between the two antennas. Earth rotation changes the directions of the baseline (assuming a frame of reference stationary with the celestial sphere), and this is taken advantage of by taking subsequent visibility readings \cite{Ryle1962}. As it will be shown later on in this chapter, a Fourier relationship exists between visibility as a function of baseline and the intensity distribution, provided that certain conditions are met. To take advantage of such a relationship is not an easy computational task and is today an active area of research which this thesis is part of.

This introduction aims to give a brief on the theory of interferometry, defines the measurement equation of the interferometer and discusses an image synthesis pipeline commonly used. The brief serves as a preamble for the discussion on motivations, aims and objectives of this thesis, which is done in the penultimate section. The chapter is concluded by giving an outline of the thesis. 

The theory presented in this chapter is based on Thompson \etal\ \cite{thompson2008interferometry} and some online sources, notably presentations given in a workshop \cite{Perley2010} organised by the \textit{National Radio Astronomy Observatory}(NRAO), and a course \cite{Condon} given by the same organisation.

\section{Analysis of a two-element interferometer}
\begin{figure}
   \centerline{\includegraphics[scale=0.65]{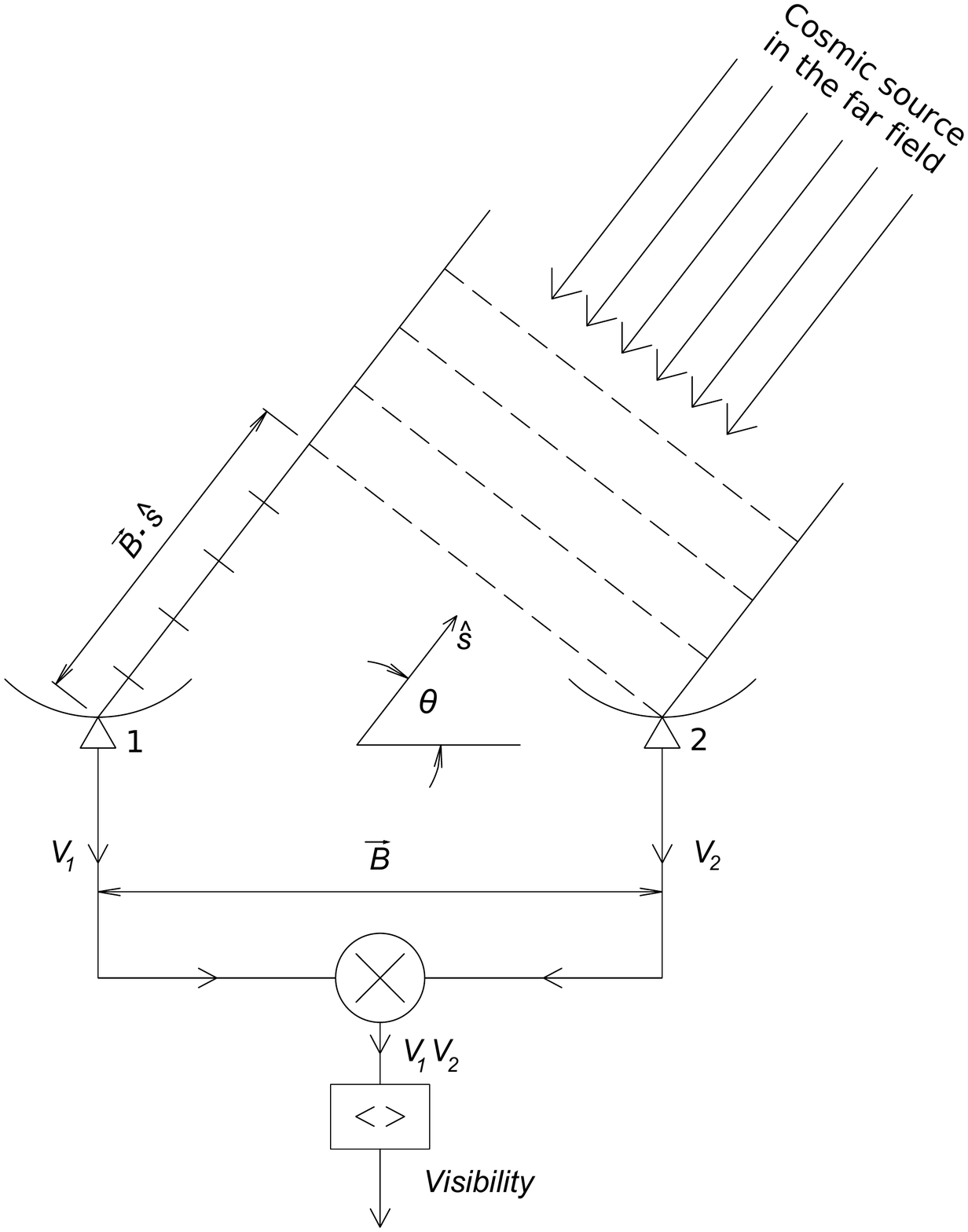}}
   \caption[A basic 2-element interferometer]{A basic two-element interferometer, with baseline $\vec{B}$ observing an arbitrary cosmic source in the direction of the unit vector $\hat{s}$.}
\label{fig:simpleinterferometer}
\end{figure}

A basic two-element interferometer is depicted in Figure \ref{fig:simpleinterferometer}. It is observing an arbitrary cosmic source in the sky, in the direction of the unit vector $\hat{s}$. The baseline is defined with the vector $\vec{B}$. The interferometer is assumed to have a quasi-monochromatic response at a central frequency $\nu_c=\omega_c(2\pi)^{-1}$. Cosmic sources are in the far field of the interferometer, and so the incoming waves at the interferometer can be considered planar over the baseline. Clearly the output of antenna 1 $(V_1)$ is the same as that of antenna 2 $(V_2)$, except that $V_1$ is lagging behind by say $\tau_g$ seconds. $\tau_g$ is called the \textit{geometric delay}. The wave needs to travel an extra distance $\vec{B}\cdot\hat{s}$ to reach antenna 1 thus, letting $c$ represent the speed of light, $\tau_g$ equates to:

\begin{equation}
\tau_g=\frac{\vec{B}\cdot\hat{s}}{c} =\frac{|\vec{B}|sin\theta}{c}
\label{local:a}
\end{equation}
Since the interferometer is quasi-monochromatic then, the voltage output of antenna 1 and antenna 2 denoted by $V_1$ and $V_2$ respectively can be written as follows:

\begin{subequations}
\begin{equation}
V_1=V\cos(\omega_c t)
\end{equation}    
\begin{equation}
V_2=V\cos(\omega_c[t+\tau_g])
\end{equation}    
\end{subequations}
where $t$ is time and $V$ is proportional to the intensity of the source. 

\subsection{The correlator}

To produce a visibility reading, the two voltages get correlated by first multiplying and then averaging over a period of time known as the \textit{integration time}. As it will be shown in this section, such a correlation is not enough since some intensity data is lost. A \textit{complex correlator} is used to make a second correlation between the output of the antennas with one of the outputs phase delayed by 90 degrees.  The two outputs of the complex correlator are presented as one complex quantity which is the visibility. Visibility is defined formally in equation \ref{equ:visibility}.

Multiplication of $V_1$ with $V_2$ results in:

\begin{equation} 
V_1V_2=V^2 \cos(\omega_c t)\cos(\omega_c[t+\tau_g])=\left(\frac{V^2}{2}\right)[\cos(2\omega_c t+\omega_c\tau_g) + \cos(\omega_c\tau_g)]
\end{equation}

$V_1V_2$ has a constant term and a sinusoidal term with respect to $t$. Provided that averaging is done over a long enough time interval, the sinusoidal term averages out to 0 implying that the average $\left<V_1V_2\right>$ equates to:

\begin{equation}
\left<V_1V_2\right> =\left(\frac{V^2}{2}\right)\cos(\omega_c\tau_g)=\cos \left(\frac{\omega_c|\vec{B}|}{c} sin \theta \right)
\label{equ:fringe}
\end{equation}
\begin{figure}
   \centerline{\includegraphics[scale=0.5]{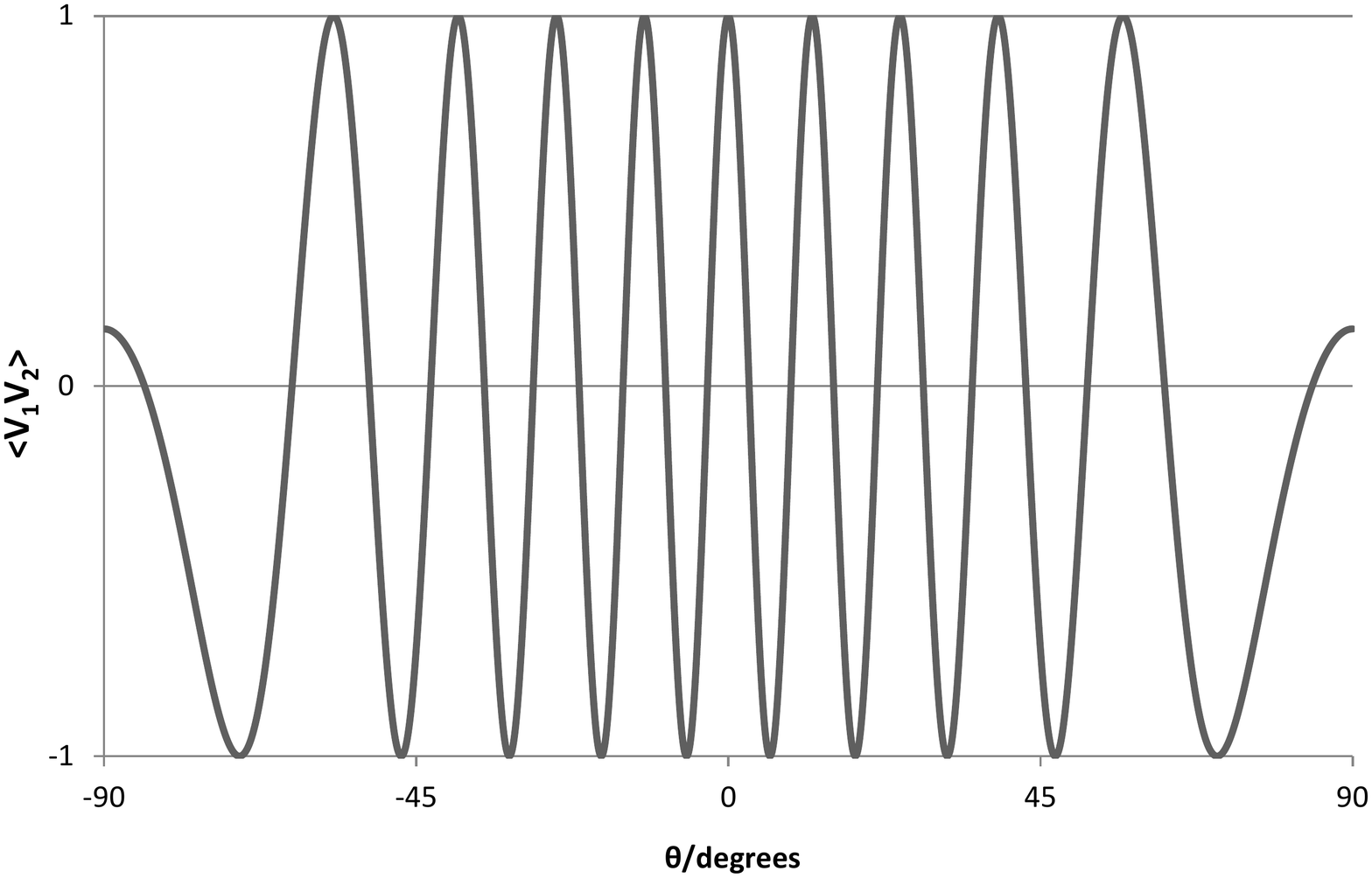}}
   \caption[Plot of a fringe]{Plot of fringe equation \ref{equ:fringe} with $V=1$ and $\omega_c|\vec{B}|/c=30$}
\label{fig:fringe}
\end{figure}

The correlator response is dependent on the direction $\theta$ of the source  with a sinusoidal behaviour known as a \textit{fringe}. The $\omega_c\tau_g$ term is known as the \textit{fringe phase}.

Figure \ref{fig:fringe} shows a plot of a fringe. It clearly indicates that, for some values of $\theta$, the correlator gives no output. This implies loss of data. If an intensity distribution $I(\theta)$ is considered, then it is easy to show that the even part of the distribution is lost.

Let $I_e(\theta)$ and $I_o(\theta)$ be the even and odd part of $I(\theta)$. Using equations \ref{equ:fringe} and integrating over $\theta$, $\left<V_1V_2\right>$ results to be proportional to:
\begin{subequations}
\begin{equation}
\left<V_1V_2\right>\ \propto \int_{2\pi} I(\theta)\cos(\omega\tau_g)d\theta=\int_{2\pi} I_o(\theta)\cos(\omega\tau_g)d\theta
\end{equation}
since
\begin{equation}
\int_{2\pi} I_e(\theta)\cos(\omega\tau_g)d\theta=0
\end{equation}
\end{subequations}

The second correlation made over the two antenna outputs, with one of the output phase delayed by 90 degrees, generates an even fringe and will respond to the even intensity distribution. This implies that the \textit{complex correlator} responds to all the intensity distribution.

\subsection{Effects of channel bandwidth and the \textit{delay centre}}
\label{sec:channelbadwidth}
If it is assumed that the interferometer is responsive over a bandwidth $\Delta\nu$, whereby intensity is constant throughout, then, based on equation \ref{equ:fringe}, the correlator output is:

\begin{subequations}
\begin{equation}
\left<V_1V_2\right>=(\Delta\nu)^{-1}\int_{\nu_c-\Delta\nu/2}^{\nu_c+\Delta\nu/2} \left(\frac{V^2}{2}\right)\cos\left(2\pi\nu \tau_g \right)d\nu
\end{equation}
\begin{equation}
\left<V_1V_2\right>=\left(\frac{V^2}{2}\right)\cos(\omega_c\tau_g)\text{sinc}(\pi\tau_g\Delta\nu)
\end{equation}
where
\begin{equation}
\text{sinc}(\pi\tau_g\Delta\nu)=\frac{\sin(\pi\tau_g\Delta\nu)}{\pi\tau_g\Delta\nu}
\end{equation}
\end{subequations}

The term $\text{sinc}(\pi\tau_g\Delta\nu)$ is known as the \textit{fringe washing function} and degrades the response of the interferometer. The degradation is dependent on the frequency bandwidth $\Delta\nu$ and should be kept as narrow as possible, especially for wide-field imaging. An interferometer splits up a wide bandwidth into many frequency channels each of narrow bandwidth, and treats each channel independently.  

The \textit{fringe washing function} has no effect over the interferometer response for $\tau_g=0$. This is because  signals of different frequencies will reach the correlator in phase. The \textit{geometric delay} ($\tau_g$) can be controlled by inserting an extra delay $\tau_0$ in the circuit between the antenna receiving the leading signal and the complex correlator. For $\tau_0=\tau_g$, the \textit{fringe washing function} is nullified.  Since $\tau_g$ is a function of $\hat{s}$, one can choose a direction for which the \textit{fringe washing function} is nullified. This direction is termed the \textit{delay centre}.

\section{The \textit{UVW} co-ordinate system}
\label{sec:UVW}

Many variables in interferometry are expressed against a reference direction known as the \textit{phase reference centre} or \textit{phase centre}. The \textit{phase centre} is fixed to the sky and is common practice that it is set to point to the same direction of the \textit{delay centre}.

\begin{figure}
\centerline{\includegraphics[scale=0.5]{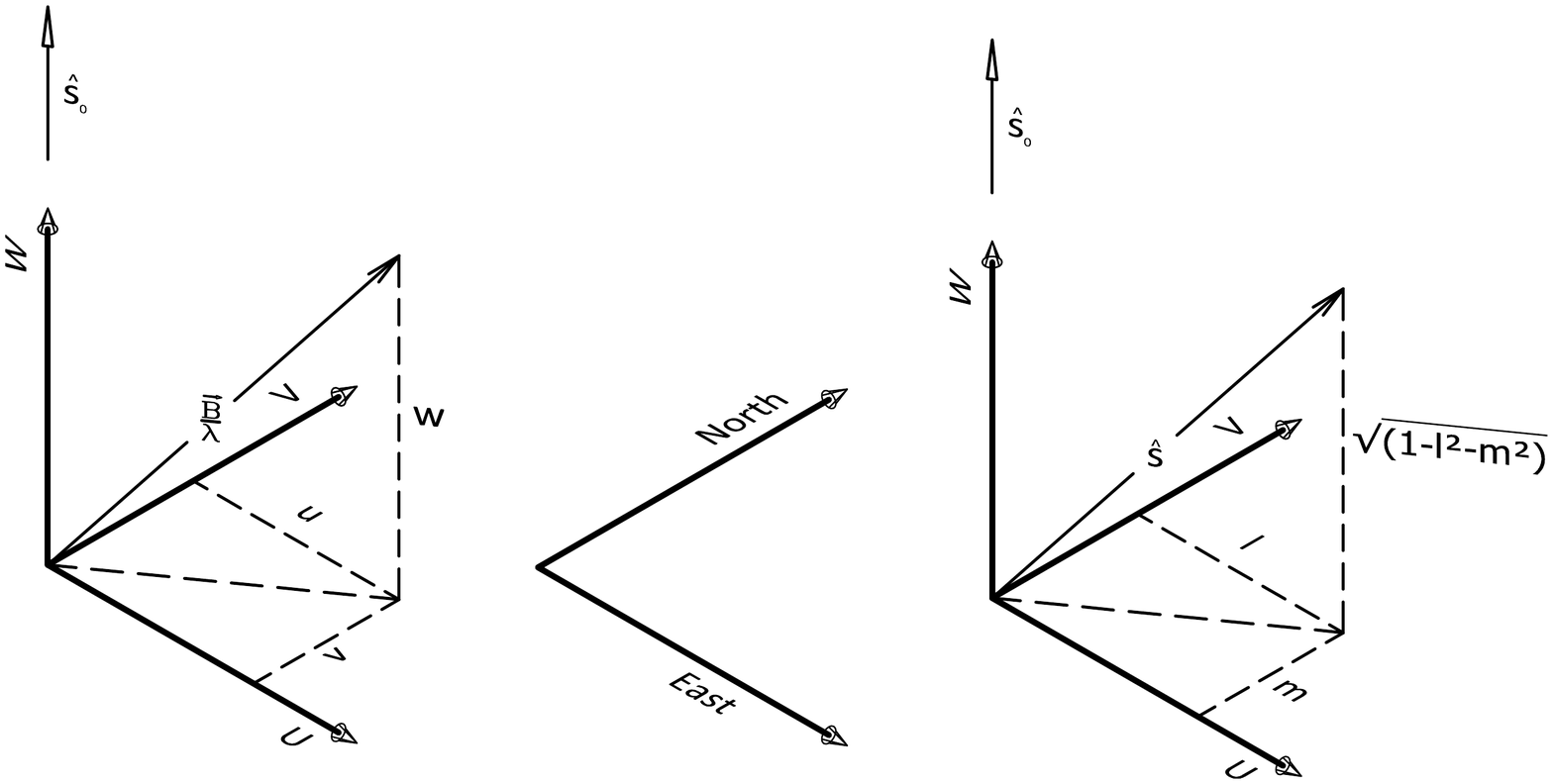}}
\caption[The UVW co-ordinate system]{The UVW co-ordinate system. The diagram on the left shows how a baseline is expressed in its $(u,v,w)$ components while the diagram on the right shows how the direction cosines $l$ and $m$ express a position on the celestial sky pointed to by the unit vector $\hat{s}$. The unit vector $\hat{s}_0$, represents the \textit{phase centre}.}
\label{fig:uvw}
\end{figure}

The general co-ordinate system used in interferometry is the $UVW$ co-ordinate system and is depicted in Figure \ref{fig:uvw}. It is a right handed Cartesian system where axes $U$ and $V$ are on a plane normal to the \textit{phase centre} and the  $W$-axis in the direction of the \textit{phase centre}. The $U$-axis is in the East-West direction while the $V$-axis is in the North-South direction. 

One of the main uses of the co-ordinate system is to measure the baseline against the \textit{phase centre}. The baseline components expressed over the $UVW$-axes defined by $(u,v,w)$ are depicted in Figure \ref{fig:uvw}. The components are normally given in units of number of wavelengths such that:

\begin{equation}
(u,v,w)=\vec{B}_\lambda=\frac{\vec{B}}{\lambda}
\label{equ:compbaseline}       
\end{equation}

where $\lambda$ is the channel wavelength, and $\vec{B}_\lambda$ is the baseline expressed in number of wavelengths. 

\begin{figure}[h]
   \centerline{\includegraphics[scale=0.6]{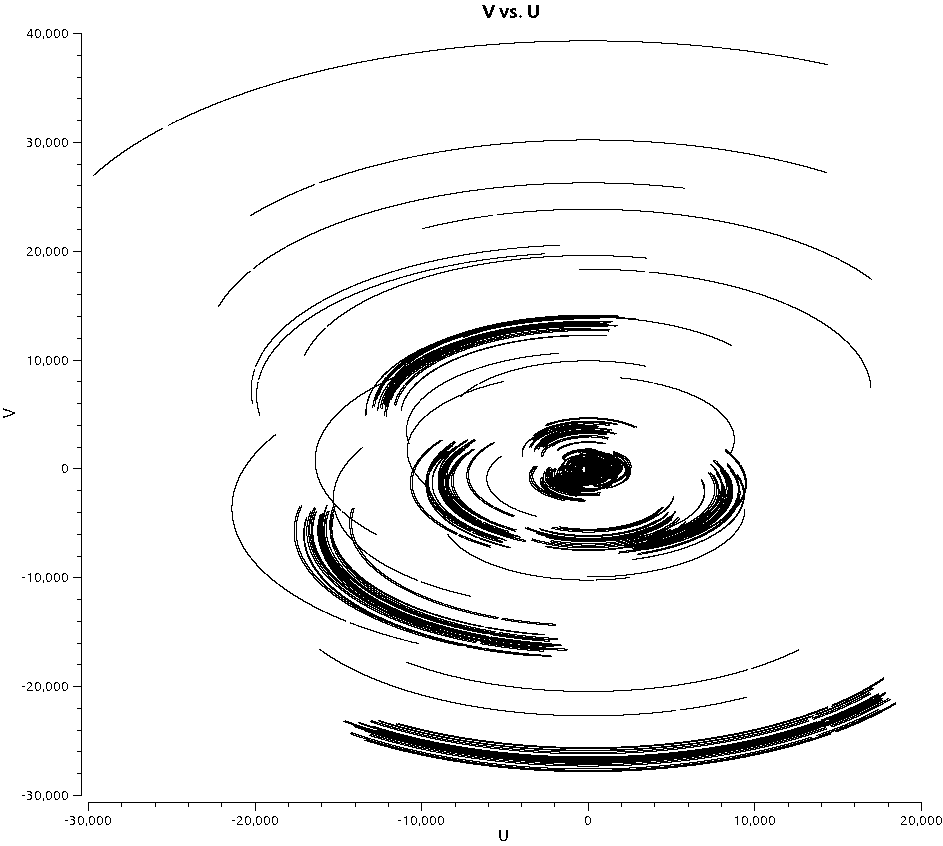}}
   \caption{UV-coverage of a true LOFAR observation}
\label{fig:uvcoverage}
\end{figure}

As visibility measurements are taken consecutively in time, the baseline vector rotates in the UVW-space since the interferometer resides on the Earth surface. The UV-coverage of an observation is defined as the set of $(u,v)$ values for which the interferometer makes a visibility measurement. Figure \ref{fig:uvcoverage} depicts the UV-coverage of a true LOFAR observation. Each baseline will form an elliptical trajectory, while the interferometer samples (measures) visibility in time. A short baseline tends to form a trajectory near the centre, while a longer one tends to form a trajectory further out.  

A clarification on the units of the $(u,v,w)$ components is now proper. Expressing $(u,v,w)$ in units of number of wavelengths is mathematically convenient and is the general approach found in literature \cite{thompson2008interferometry}. Nevertheless it is not always practical to use, since values are dependent on channel frequency. The MeasurementSet \cite{Kemball2000} is a common format used to store visibility measurements. It specifies $(u,v,w)$ data to be stored in meters rendering it valid over all frequency channels. In this thesis the $(u,v,w)$ components are expressed in number of wavelengths unless stated otherwise.

When describing a position on the celestial sphere the direction cosines $l$ and $m$ are used as shown in Figure \ref{fig:uvw}. They are defined with respect to the U and V axes such that the direction vector $\hat{s}$ is described as: 

\begin{equation}
\hat{s}=(l,m,\sqrt{1-l^2-m^2})
\label{equ:cosines}
\end{equation}

\section{The measurement equation}
\label{sec:measurment}
\begin{figure}[h]
   \centerline{\includegraphics[scale=0.5, trim=0cm 0cm 0cm 0cm, clip=true]{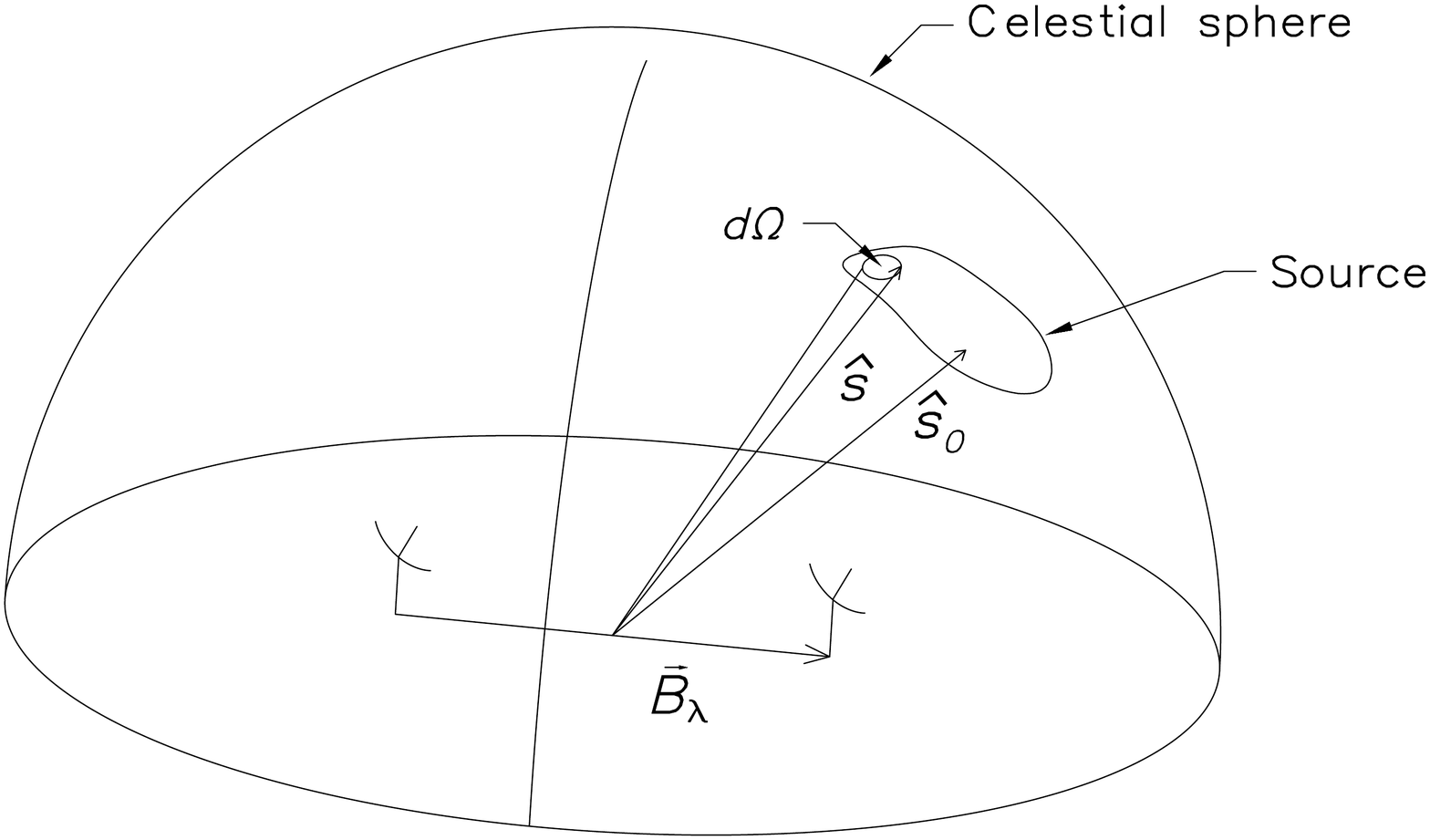}}
   \caption{Interferometer diagram used in section \ref{sec:measurment}}
\label{fig:interferometer}
\end{figure}

Consider Figure \ref{fig:interferometer}. $\hat{s}_0$ represents the \textit{phase reference position}. The unit vector $\hat{s}$ retains its definition, that is, a pointer towards an arbitrary position in the celestial sphere. $d\Omega$ is a solid angle, and $\vec{B}_\lambda$ is the baseline expressed in number of wavelengths.

Not considered in the previous section, is the non-uniform polarisation dependent reception behaviour of the antennas known as the \textit{A-term}. Let $A_N(\hat{s})$ represent a normalised version of the \textit{A-term} such that $A_N(\hat{s}_0)=1$. The complex visibility $\mathcal{V}(\vec{B}_\lambda)$ is defined as:
\begin{equation}
\mathcal{V}(\vec{B}_\lambda)=\int_{4\pi} A_N(\hat{s})I(\hat{s})e^{-j2\pi{\vec{B}_\lambda\cdot(\hat{s}-{\hat{s}}_0)}} d\Omega
\label{equ:visibility}
\end{equation}

This definition is in-line with that given in \cite{thompson2008interferometry} that follows the sign convention of the exponent used in \cite{born1999principles,Bracewell1958}. $I(\hat{s})$ is the intensity distribution.

The complex visibility, as defined in equation \ref{equ:visibility}, has dimensions of flux density ($Wm^{-2}Hz^{-1}$). The dimensions of the intensity distribution are $Wm^{-2}Hz^{-1}sr^{-1}$. In astronomy,  the \textit{Jansky}(Jy) is a commonly used unit defining the flux density where 1 Jansky=$10^{-26}Wm^{-2}Hz^{-1}$.

As described in \cite{thompson2008interferometry} $d\Omega$ equates to:

\begin{equation}
d\Omega=\frac{dldm}{\sqrt{1-l^2-m^2}}
\end{equation}

Using equations \ref{equ:compbaseline} and \ref{equ:cosines} it results that:

\begin{subequations}
\begin{equation}  
\vec{B}_\lambda\cdot{\hat{s}}=ul+vm+w\sqrt{1-l^2-m^2}
\end{equation}
\begin{equation}  
\vec{B}_\lambda\cdot{\hat{s}_0}=w
\end{equation}
\end{subequations}

Substituting in equation \ref{equ:visibility} the following results: 
\begin{equation}
\mathcal{V}(u,v,w)=\int_{-\infty}^\infty \int_{-\infty}^\infty {\frac{A_N(l,m)I(l,m)e^{-j2\pi(ul+vm+w(\sqrt{1-l^2-m^2} -1)}}{\sqrt{1-l^2-m^2}}} dldm
\label{equ:measurment}
\end{equation}

Equation \ref{equ:measurment} is the \textit{\textbf{measurement equation}} of the interferometer. It defines the relationship between the interferometer measured quantities $\mathcal{V}(u,v,w)$ and the intensity distribution of the sky.

\section{Image Synthesis}
\label{sec:imagesynthesis}
This thesis hardly gives importance to the effects of $A_N(l,m)$ and the quotient term $\sqrt{1-l^2-m^2}$ in the measurement equation \ref{equ:measurment}. For convenience,  these two terms are subsumed in the intensity distribution which will be referred to, as the \textit{measured intensity distribution} $I_{meas}(l,m)$. 

\begin{equation}
I_{meas}(l,m)=\frac{I(l,m)A_N(l,m)}{\sqrt{1-l^2-m^2}}
\end{equation}

The measurement equation \ref{equ:measurment} is re-written as:

\begin{equation}
\mathcal{V}(u,v,w)=\int_{-\infty}^\infty \int_{-\infty}^\infty {I_{meas}(l,m)e^{-j2\pi(ul+vm+w(\sqrt{1-l^2-m^2} -1)}} dldm
\label{equ:measurmentrewritten}
\end{equation}

This section reviews how $I_{meas}(l,m)$ can be recovered from visibility measurements made by the interferometer. The term $2\pi w(\sqrt{1-l^2-m^2}-1)$, is referred to as the \textit{w-term}. If it equates to 0 then a Fourier relationship between visibility expressed in terms of $u$ and $v$ ($w$ ignored) and the \textit{measured intensity distribution} is obtained, that is:
\begin{equation}
\mathcal{V}(u,v)=\mathcal{F} I_{meas}(l,m)  \text{\ \ if\ \ } w(\sqrt{1-l^2-m^2}-1)=0
\label{equ:fourier-w-term}
\end{equation}

where $\mathcal{F}$ is the Fourier operator.

\begin{figure}
   \centerline{\includegraphics[scale=1, trim=1cm 1.0cm 2.2cm 1.0cm, clip=true]{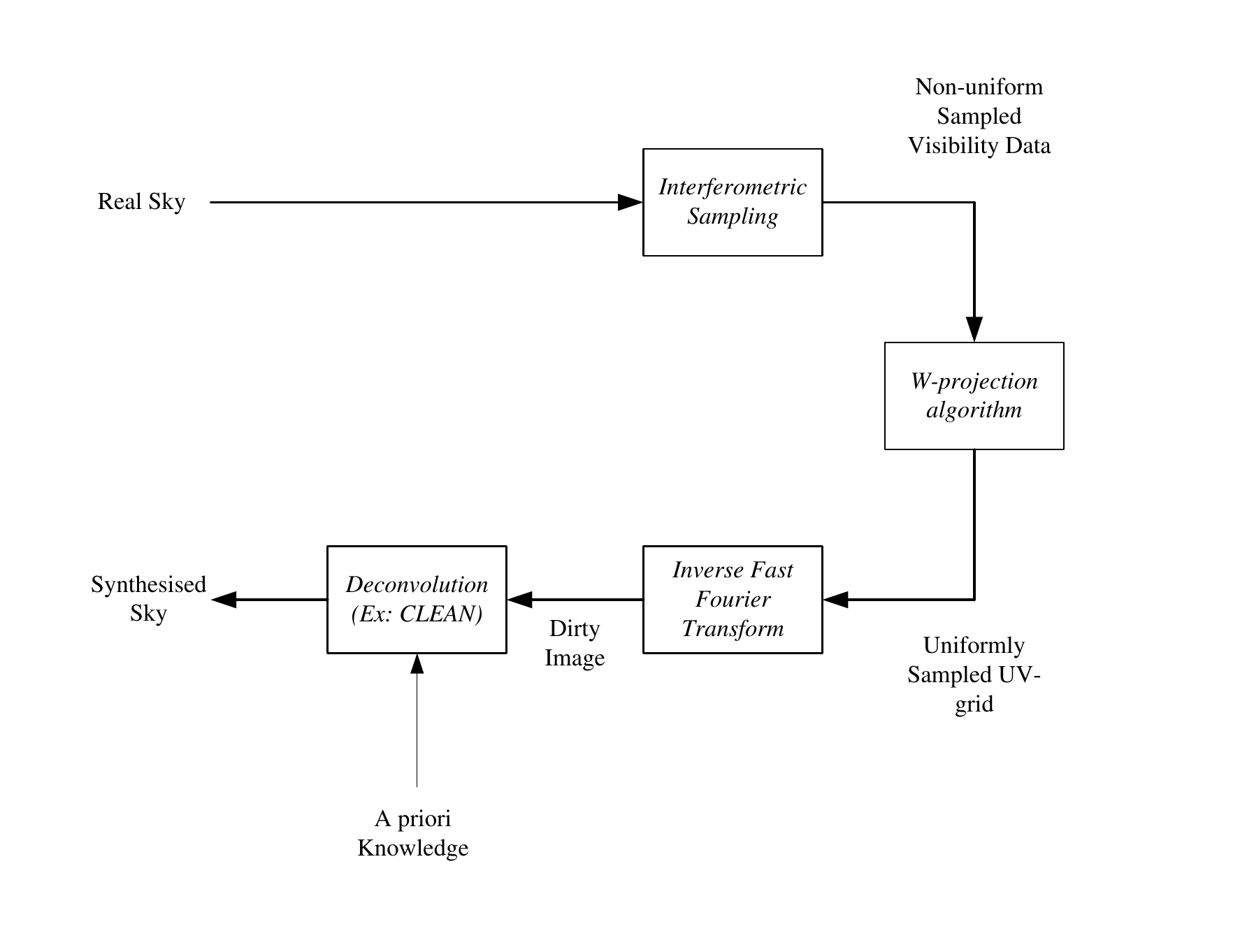}}
   \caption{A simple imaging pipeline for interferometric data}
\label{fig:imagepipeline}
\end{figure}

Most of the image synthesis techniques, exploit this Fourier relationship in order to use the computationally efficient \textit{Inverse Fast Fourier Transforms (IFFT)} algorithms. However, there are various issues that need to be circumvented as to apply such algorithms, which mandates some pre- and post-processing of data. The next subsections gives a brief on the main issues and how they are commonly handled. Based on this brief, Figure \ref{fig:imagepipeline} depicts an imaging synthesis pipeline that is commonly used.

\subsection{Non-uniform sampling of the visibility plane}
\label{sec:non-uniformsampling}

Fast Fourier Transform algorithms require that the visibility UV-plane (also referred to as the UV-grid) is uniformly sampled. As already discussed in section \ref{sec:UVW} the UV-coverage of the interferometer is not uniform, implying that some processing is required to generate a uniformly sampled UV-grid.  

\textit{Convolutional gridding} is the technique used in interferometry to transform the non-uniform UV-coverage into a uniform one. Details on this technique are given in section \ref{sec:convgridding}. Measured data is convolved with an appropriately chosen function and the result is sampled as to generate a uniform UV-grid. After an IFFT takes place, the convolution is reversed by a simple element-by-element division on the intensity image. One notes that \textit{convolutional gridding} is a topic actively researched upon, especially in its application in medical sciences \cite{OSullivan1985,Jackson1991,Eklund2013}.

\subsection{Incompleteness}
\label{sec:incompletness}
The non-uniform UV-coverage implies that the measured visibility data is incomplete. The intensity distribution can never be recovered in full by relying solely on the measured visibility samples. The output intensity image that is generated by the \textit{convolution gridding} algorithm is called a \textit{dirty image} in view that it contains artefacts caused by the non-uniform incomplete UV-coverage. 

Incompleteness is handled by iterative methods commonly known as \textit{deconvolution}. Based on a priori knowledge on the intensity distribution, \textit{deconvolution} is applied over the dirty image in order to recover the intensity distribution. A classic example of a \textit{deconvolution} algorithm is CLEAN \cite{Hogbom1974,Clark1980} which assumes that the intensity is made up of point sources. Starck \etal\ \cite{Starck2002} give a review of other \textit{deconvolution} methods.

\subsection{The \textit{w-term}}
\label{sec:wterm}
As pointed out by Cornwell \etal\ \cite{Cornwell2008}, if the \textit{w-term} is much less than unity it can be neglected and a two dimensional Fourier relationship results. When synthesising a narrow field of view, the \textit{w-term} can in many cases, be neglected. Neglecting the \textit{w-term} when synthesising a wide-field of view, causes substantial distortions.

An advanced method that corrects the \textit{w-term}, is the \textit{w-projection} \cite{Cornwell2008} algorithm. A visibility sample is projected on the $w=0$ plane by means of a convolution. The \textit{w-term} for the \textit{w=0} plane is 0, so effects of the \textit{w-term} get nullified by the projection. The algorithm is integrated with the \textit{convolutional gridding} algorithm and is explained in detail in section \ref{sec:w-projection}.

Other alternatives to \textit{w-projection} exploit other characteristics of the measurement equation. For example,  it is possible to express the measurement equation as a 3-dimensional Fourier transform \cite{Clark1973,Cornwell1992}. \textit{Snapshots} \cite{Cornwell2008,Cornwell2012,Cornwell1992,Ord2010} consider a coplanar array whereby at any given time $w$ is a linear combination of $u$ and $v$ for all the measurements made by the coplanar array \cite{brouw1971data,Cornwell1992}. Thus, for short periods of observation time, the \textit{w-term} causes only a distortion in the \textit{lm-plane} co-ordinates which is corrected by a linear transform. \textit{Facets} consider the wide-field intensity image as a sum of smaller images (\textit{facets}) over which the \textit{w-term} can be neglected. There are two types of \textit{facets}: the \textit{image-plane facets} and the \textit{uvw-space facets}. These are reviewed in \cite{Cornwell2008}.

Hybrid algorithms using \textit{w-projection} with any of these alternative algorithms are also possible and are discussed in section \ref{sec:hybrid}.

\section{Motivation, aims and objectives of the thesis}

As new powerful radio telescopes are being built up, the computational demand of the imaging pipeline described in Figure \ref{fig:imagepipeline}, is increasing to exuberant levels. It is estimated that Phase 2 of the \textit{Square Kilometre Array} (SKA) will need as much as 4 Exaflops of computational power \cite{Cornwell2009a}, with 90\% of computational resources taken by the \textit{gridding} algorithm \cite{Cornwell2009}. Phase 2 is expected to be commissioned by around 2020 and is predicted that though technology would have advanced by that time, a super-computer delivering computational rates in Exaflops  would be at the top of the TOP 500 \cite{Top500} list \cite{Cornwell2009a,R.NewmanMay2011}.

Motivated by such computational requirement, this thesis aims in giving a contribution to the application of high performance computing in radio interferometry. The main objective is to develop a new high-performance imaging tool. It is being called the \textit{malta-imager} or \textit{mt-imager} for short. It relies on CUDA\footnote{CUDA stands for \textit{Compute Unified Device Architecture} and is a parallel programming framework for GPUs developed by NVIDIA\textregistered} \cite{Nickolls2008} compatible NVIDIA\textregistered Graphical Processing Units (GPUs) to synthesise images by means of \textit{w-projection}. \textit{Deconvolution} is outside the scope of this work. 

The infrastructure handling all numerical calculations is generalised in a framework, which is being called the \textit{General Array Framework} (GAFW). It is designed to handle different hardware such as GPUs and \textit{Field Programmable Gate Arrays} (FPGAs). It is being implemented for CUDA compatible GPUs only so as to serve the main objective. The \textit{mt-imager} is built over this framework to perform all GPU-based computation.

\section{Outline of thesis}

This introductory chapter discussed the main concept of image synthesis in radio interferometry. The next chapter, that is Chapter \ref{chap:litreview}, gives a literature review. The focal point of the thesis is the implementation of \textit{w-projection} and \textit{convolutional gridding}. A mathematical treatment and a review of reported implementations of these algorithms are given. The chapter is concluded by giving a detailed description of a gridding algorithm over GPUs proposed by Romein \cite{Romein2012}. It is a pivotal algorithm to this thesis and will be often referred to as Romein's algorithm.

The \textit{General Array Framework} is the subject of Chapter \ref{chap:gafw}. The framework is discussed in detail and terms are defined. 

Chapter \ref{chap:mtimager} is the main chapter whereby the \textit{mt-imager} is discussed in detail. Note that the chapter uses the terms and definitions given in previous chapters. The imaging tool is divided into components that are independent of each other. The high-level design is first discussed, and then details of the tool are given through a discussion on each component. 

Chapter \ref{chap:results} reports on the performance obtained by the tool and shows that the main objective of the thesis has been achieved with success. The chapter also reports some detailed experimental analyses on the performance of Romein's algorithm as implemented in the imaging tool. These analyses are meant to enrich the knowledge on the algorithm. The chapter is concluded by proposing future work.

The thesis is concluded in Chapter \ref{chap:conclusion}.

\chapter{Gridding and W-Projection}
\label{chap:litreview}

This chapter reviews \textit{convolutional gridding} and the \textit{w-projection} algorithm. Some mathematical treatment is given together with other background knowledge. A literature review on performance and implementation of these algorithms is included together with some notes on GPU programming.

\section{Convolutional gridding}
\label{sec:convgridding}

In interferometry, \textit{convolutional gridding} is the method used to transform the non-uniform sampled \textit{UV-plane} (visibility plane) to a uniform sampled one such that \textit{Inverse Fast Fourier Transform} (IFFT) algorithms can be applied (refer to subsection \ref{sec:non-uniformsampling}). Visibility measurements are convolved with an appropriately chosen convolution function $C(u,v)$. The resultant \textit{UV-plane} is then sampled uniformly, over which an IFFT is applied. The result is an image in the \textit{lm-plane} (intensity distribution plane), which is corrected from convolution by a simple element-by-element division. The output is a \textit{dirty image} of the intensity distribution, which is the true intensity aliased by the non-uniform incomplete UV-coverage. 

In this section, \textit{convolutional gridding} is discussed via a mathematical treatment. The concept of weighting is introduced, and some topics on the convolution function $C(u,v)$ are reviewed.

\subsection{Mathematical review}
\label{sec:convgriddingmath}
  
The mathematical treatment given in this section is based on, and adapted from, Jackson \etal\ \cite{Jackson1991}. 

The \textit{measured intensity distribution} $I_{meas}(l,m)$ considered here, has been defined in section \ref{sec:imagesynthesis} as follows:

\begin{equation}
I_{meas}(l,m)=\frac{I(l,m)A_N(l,m)}{\sqrt{1-l^2-m^2}}
\end{equation}   

For this treatment,  the \textit{w-term} is ignored such that Fourier relationship expressed in equation \ref{equ:fourier-w-term} is true, that is:

\begin{equation}
\mathcal{V}(u,v)=\mathcal{F} I_{meas}(l,m)
\end{equation}

Let $n$ be the number of visibility measurements that the interferometer samples, and let $i$ be a zero-base index such that $u_i$ and $v_i$ are the $u$ and $v$ values of a measured visibility point. The sampling function $P(u,v)$ of the interferometer is defined as:

\begin{equation}
P(u,v)=\sum_{i=0}^{n-1}\delta(u-u_i,v-v_i)
\end{equation}

where $\delta(u,v)$ is the two dimensional Dirac function. 

As discussed further on, it is desirable to control the shape of the sampling function. This is done by introducing a weight $\varepsilon_i$ for each sampled visibility, such that the weighted sampling function $P_\varepsilon(u,v)$ equates to:

\begin{equation}
P_\varepsilon(u,v)=\sum_{i=0}^{n-1}\delta(u-u_i,v-v_i)\varepsilon_i
\label{equ:weight}
\end{equation}

$\mathcal{F}^{-1}P_\varepsilon(u,v)$ is known as the \textit{synthesised beam}, \textit{dirty beam} or \textit{point spread function} (PSF). It is here denoted as $psf(l,m)$.

\begin{equation}
psf(l,m)=\mathcal{F}^{-1}P_\varepsilon(u,v)
\label{equ:convpsf}
\end{equation}

With interferometric sampling and use of weights, the visibility plane $\mathcal{V}(u,v)$ is transformed into a non-uniform sampled plane $P_\varepsilon(u,v)\cdot\mathcal{V}(u,v)$. The desire is to transform the non-uniform sampled plane to a uniform UV-grid $\mathcal{V}_{grid}(u,v)$, sampled at regular intervals $\Delta u$ and $\Delta v$ on the U and V axes respectively. This is achieved by convolving $P_\varepsilon(u,v)\cdot\mathcal{V}(u,v)$ with an appropriately chosen convolution function $C(u,v)$ and then uniformly sampling the output.

\begin{equation}
\mathcal{V}_{grid}(u,v)=\{[P_\varepsilon(u,v)\cdot \mathcal{V}(u,v)]*C(u,v)\}\cdot\shah\left(\frac{u}{\Delta u},\frac{v}{\Delta v}\right)
\end{equation}
where the operator $*$ is a two-dimensional convolution, and  $\shah(u/\Delta u,v/\Delta v)$ is the Shah function defined as:

\begin{equation}
\shah\left(\frac{u}{\Delta u},\frac{v}{\Delta v}\right)=\sum_{i=-\infty}^{\infty}\sum_{k=-\infty}^{\infty}\delta(u-i\Delta u,v-k\Delta v) 
\end{equation}

An inverse Fourier transform is applied on $\mathcal{V}_{grid}(u,v)$ which equates to: 
\begin{equation}
\mathcal{F}^{-1}\mathcal{V}_{grid}=\{[{psf}(l,m)*I_{meas}(l,m)]\cdot c(l,m)\}*\shah(l\Delta u,m\Delta v)
\end{equation}
where $c(l,m)=\mathcal{F}^{-1}C(u,v)$ is referred to as the tapering function.

Convolution with the Shah function causes a replication over $\mathcal{F}^{-1}V_{grid}(u,v)$ at intervals $(\Delta u)^{-1}$ in $l$ and $(\Delta v)^{-1}$ in $m$. The synthesised field of view is thus dependent on the choice of $\Delta u$ and $\Delta v$. Only one replica is truly calculated which is mathematically equivalent to multiplying $\mathcal{F}^{-1}V_{grid}(u,v)$ with a rectangular function. To complete the process, the image is divided by $c(l,m)$ so as to reverse the convolution and produce the non-normalised dirty image $I_{dirty}(l,m)$. 
\begin{equation}
I_{dirty}(l,m)=\frac{(\{[psf(l,m)*I_{meas}(l,m)]\cdot c(l,m)\}*\shah(l\Delta u,m\Delta v))\cdot\text{rect}(l\Delta u,m\Delta v)}{c(l,m)}
\label{equ:convgrid}
\end{equation}
where
\begin{equation}
\text{rect}(l,m)= \begin{cases}1 & \text{if\ } |l|<0.5\ \text{and}\ |m|<0.5\\0 & \text{otherwise} \end{cases}
\end{equation}

Equation \ref{equ:convgrid} describes the output of the \textit{convolutional gridding} algorithm.

\subsection{Deconvolution and weighting schemes}
\label{sec:weighting}
The PSF aliases  $I_{meas}(l,m)$ in such a way that $I_{meas}(l,m)$ cannot be recovered through direct methods. This is the effect of data incompleteness. Iterative methods known as \textit{deconvolution} are applied to try to recover $I_{meas}(l,m)$ to the best possible accuracy (refer to section \ref{sec:incompletness}). The effectiveness is dependent on the form of the PSF, which is desired to be as compact as possible. If too few visibility measurements samples are considered, then, the PSF might be too wide for a proper recovery. In the extreme case, where only one measurement is considered, the PSF is infinitely wide. 

The weight introduced in equation \ref{equ:weight} plays a pivotal role in controlling the form of the PSF. As visible in Figure \ref{fig:uvcoverage}, there is a higher density of visibility measures in regions covered by shorter baselines. This tends to overemphasise the long spatial wavelength of the PSF causing wide skirts. The \textit{uniform weighting} scheme caters for this issue by weighting each measurement with a value inversely proportional to the neighbourhood density \cite{Thompson1974}. In another scheme, called \textit{natural weighting}, the density issue is given second priority, and  visibility data is weighted inversely proportional to its variance so as to obtain the best signal to noise ratio \cite{thompson2008interferometry}. Short baseline data will still be overemphasised, but an in-between scheme, known as \textit{robust weighting} \cite{Briggs1995}, tries to compromise between natural and uniform weighting schemes.

\subsection{The tapering function}
\label{sec:tapering}
The form of the tapering function $c(l,m)$ is crucial, since the replication caused by the convolution with the Shah function, forces $c(l,m)$ to alias the image. Aliasing can be fully suppressed if the tapering function covers the whole field of view defined by $\text{rect}(l\Delta u,m\Delta v)$ and then goes to 0 outside the region. The infinite sinc function (in the \textit{UV-plane}) is the best choice for $C(u,v)$ \cite{OSullivan1985} but it is computational unattainable. Various functions have been studied \cite{Jackson1991} and the general choice in interferometry is the spheroidal prolate function \cite{SchwabOct1980,Slepian1961,Landau1961}.

The convolution function $C(u,v)$  has to be finite, or in other words it has to have a finite \textit{support}. The \textit{support} of the convolution function is defined as the full width of the function in grid pixels. Note that in literature, support might be defined differently as explained by Humphreys and Cornwell \cite{HumphreysB.}.
   
It is a common approach that the convolution function is not calculated during gridding but is pre-generated prior the gridding process \cite{Romein2012,HumphreysB.} and stored in memory. The function has to be sampled at a higher rate than $\mathcal{V}_{grid}(u,v)$. Such sampling is referred to as \textit{oversampling}. The \textit{oversampling factor} $\beta$ is defined as: 

\begin{equation}
\beta=\frac{\Delta u}{\Delta u_{conv}}=\frac{\Delta v}{\Delta v_{conv}}
\label{equ:local:o}
\end{equation}

where $\Delta u_{conv}$ and $\Delta v_{conv}$ represent sampling intervals at which the convolution function is being sampled. For convenience, it is assumed that the \textit{oversampling factor} is invariant over the U and V axes.

In general an implementation of a gridder does not perform any interpolation on the numerical data of the convolution function. It merely chooses the nearest numerical values while gridding a record\footnote{In this thesis, the term \textit{record} is used to describe a single- or  multi-polarised visibility measurement.}. This changes the form of the convolution function which can be modelled by the following equation:

\begin{equation}
\tilde{C}(u,v)=\left[C(u,v)\cdot \shah\left(\frac{u}{\Delta u/\beta},\frac{v}{\Delta v/\beta}\right)\right]* \text{rect}\left(\frac{u}{\Delta u/\beta},\frac{v}{\Delta v/\beta}\right)
\end{equation}

where $\tilde{C}(u,v)$ is the oversampled version of $C(u,v)$.

Applying the Inverse Fourier transform, the following is obtained:

\begin{equation}
\mathcal{F}^{-1}\tilde{C}(u,v)=\left[c(l,m)*\shah\left(\frac{l}{\beta/\Delta u},\frac{m}{\beta/\Delta v}\right)\right]\cdot\text{sinc}\left(\frac{\pi l\Delta u}{\beta}\right)\text{sinc}\left(\frac{\pi m\Delta v}{\beta}\right)
\label{equ:coversampled}
\end{equation}

The sinc functions change the form of the convolution function. Their affects are corrected for, by dividing the image with the sinc functions after the inverse Fourier transform takes place. 

\section{W-projection}
\label{sec:w-projection}

In subsection \ref{sec:convgriddingmath}, the \textit{w-term} was ignored while describing the \textit{convolutional gridding} algorithm. When the term is much less than unity, such an assumption is acceptable \cite{Cornwell2008}. Otherwise ignoring the \textit{w-term} leads to substantial distortions. This section discusses a recent algorithm known as \textit{w-projection} \cite{Cornwell2008} that handles the \textit{w-term} through convolution. A visibility record can be projected to the $w=0$ visibility plane, by convolving with a $w$-dependent function. \textit{W-projection} builds over the \textit{convolutional gridding} algorithm by applying the stated fact. A mathematical treatment of \textit{w-projection} are the subject of the next subsection. It is followed by other \textit{w-projection} related topics which are performance, and hybrid algorithms (with \textit{w-projection}).      

\subsection{Mathematical treatment}
\label{sec:wprojmathtreatmnet}
Frater and Docherty \cite{Frater1980} show that a relationship exists between any constant $w$ plane and the $w=0$ plane. Following is a mathematical proof based on \cite{Frater1980}: 

Let $\mathcal{V}_w(u,v)$ represent visibility over a plane with constant $w$ and let $g_w(l,m)$ represent the \textit{w-term} in its exponential form.
\begin{equation}
g_w(l,m)=e^{-j2\pi  w(\sqrt{1-l^2-m^2}-1)}
\end{equation}

Using the measurement equation \ref{equ:measurmentrewritten} it can be shown that:

\begin{equation}
\mathcal{V}_w(u,v)=\int_{-\infty}^\infty \int_{-\infty}^\infty {I_{meas}(l,m)g_w(l,m)e^{-j2\pi(ul+vm)}} dldm
\end{equation}

which implies the following Fourier relationship:

\begin{equation}
\mathcal{V}_w(u,v)=\mathcal{F}[I_{meas}(l,m)g_w(l,m)]
\label{equ:fourierequ}
\end{equation}

Clearly $g_0(l,m)=1$ and thus

\begin{equation}
\mathcal{V}_0(u,v)=\mathcal{F}I_{meas}(l,m)
\end{equation}

Substituting in equation \ref{equ:fourierequ} and applying the convolution theorem it results that:

\begin{equation}
\mathcal{V}_w(u,v)=\mathcal{V}_0(u,v)*G_w(u,v)
\label{equ:localconv}
\end{equation}

where 

\begin{equation}
G_w(u,v)=\mathcal{F}g_w(l,m)
\end{equation}

Any visibility plane $V_w(u,v)$ with constant $w$ is related to the $w=0$ plane by a convolution with a known function $G_w(u,v)$.

Clearly $\overline{g}_w(l,m)=[g_w(l,m)]^{-1}$ exists implying that $\overline{G}_w(u,v)=\mathcal{F}\overline{g}_w(l,m)$ also exists and thus the convolution in equation \ref{equ:localconv} can be inverted to put $\mathcal{V}_0(u,v)$ subject of the formula.

\begin{equation}
\mathcal{V}_0(u,v)=\mathcal{V}_w(u,v)*\overline{G}_w(u,v)
\label{equ:localconv2}
\end{equation}

Any visibility plane with constant $w$ can be projected to the $w=0$ plane via a convolution with the w-dependent function  $\bar{G}_w(u,v)$. 

Cornwell \etal\ \cite{Cornwell2008} argue that as a consequence of equation \ref{equ:localconv2} any visibility point can be re-projected on the $w=0$ plane. This forms the basis of the  \textit{w-projection} algorithm which builds over \textit{convolutional gridding} by convolving each visibility record with a $w$-dependent function $C_w(u,v)$. It is defined as:

\begin{equation}
C_w(u,v)=\overline{G}_w(u,v)*C(u,v)
\end{equation}                

The Point Spread Function (defined in section \ref{sec:convgriddingmath}) is calculated using \textit{w-projection} as follows: 

\begin{equation}
psf(l,m)=\mathcal{F}^{-1}\sum_{i=0}^{n-1}\delta(u-u_i,v-v_i)*\overline{G}_{w_i}(u,v)\varepsilon_i
\end{equation}
\begin{equation}
psf(l,m)=\mathcal{F}^{-1}\sum_{i=0}^{n-1} \overline{G}_{w_i}(u-u_i,v-v_i)\varepsilon_i
\label{equ:wprojpsf}
\end{equation}
 
Equation \ref{equ:convgrid}, which is the output of the \textit{convolutional gridding} algorithm is valid for \textit{w-projection} provided that equation \ref{equ:wprojpsf} is used for $psf(l,m)$. 

$\overline{G}_w(u,v)$ is not directly solvable and thus the convolution functions\footnote{The plural is used in view that there is a different convolution function for each unique value of $w$.} $C_w(u,v)$ have to be solved numerically \cite{Cornwell2008}. They need to be pre-generated and stored in an oversampled format. The gridding process will convolve with  $\tilde{C}_w(u,v)$ which is the oversampled version of $C_w(u,v)$.      

\begin{equation}
\tilde{C}_w(u,v)=\left\{[\overline{G}_w(u,v)*C(u,v)]\cdot \shah\left(\frac{u}{\Delta u/\beta},\frac{v}{\Delta v/\beta}\right)\right\}* \text{rect}\left(\frac{u}{\Delta u/\beta},\frac{v}{\Delta v/\beta}\right)
\label{equ:C_W}
\end{equation}  

The aliasing and correction arguments given in section \ref{sec:tapering} still hold. Sampling in $w$ is also required, and this will also generate some aliasing. Cornwell \etal\ \cite{Cornwell2008} argues that aliasing can be reduced to tolerable values by scaling in $\sqrt{w}$ rather than linearly. Support of $\tilde{C}_w(u,v)$ is dependent on $w$ and increases with increasing $w$ \cite{Cornwell2012}. This can lead to prohibitive memory requirements for storage of $\tilde{C}_w(u,v)$ \cite{Bannister2013,Cornwell2012}. It is a known issue in \textit{w-projection} and there is research going on in order to handle the problem, such as the work presented by Bannister and Cornwell \cite{Bannister2013}.

It is to be noted that based on the same principles described here, it is possible to correct the \textit{w-term} in the intensity plane. Applying an inverse Fourier transform on equation \ref{equ:fourierequ} it results that:

\begin{equation}
I_{meas}(l,m)=\frac{\mathcal{F}^{-1} \mathcal{V}_w(u,v)}{g_w(l,m)}
\label{equ:stacking}
\end{equation}

The \textit{w-stacking} algorithm \cite{Cornwell2012} is based on equation \ref{equ:stacking}. Visibility data is partitioned in $w$ and gridded on separate constant $w$ visibility planes. Correction is applied after a Fourier transformation of the plane. At the end,  all planes are added together.

\subsection{Hybrid algorithms}
\label{sec:hybrid}

In section \ref{sec:wterm} alternatives to \textit{w-projection} were discussed. These alternatives are not mutually exclusive with \textit{w-projection}. Hybrid solutions using \textit{w-projection} are possible and have been successfully implemented. For example, \textit{facets} and \textit{w-projection} have integrated together in CASA \cite{McMullin2007} where \textit{w-projection} is applied over wide \textit{facets}. The recently proposed \textit{w-snapshots} algorithm \cite{Cornwell2012} uses \textit{w-projection} to project visibilities on a plane over which the \textit{snapshots} algorithm can be applied. The plane is chosen by least square fit and changed whenever the visibility reading being projected over the plane deviates too much. 

\subsection{Performance}

The basic idea of \textit{w-projection} is the one currently giving the best performance. Bhatnagar \cite{Bhatnagar2010} claims that \textit{w-projection} is faster than \textit{facets} by a factor of 10. Bannister and Cornwell \cite{Bannister2013} claim that \textit{w-projection} is the most computationally efficient algorithm. 

Variants or hybrids based on \textit{w-projection} can give better performance. For example, Kogan\cite{KoganOctober82012} shows that \textit{w-stacking} might be computationally faster than \textit{w-projection}. Cornwell \etal\ \cite{Cornwell2012} claim that \textit{w-snapshots} algorithm is faster and more scalable than \textit{w-projection}.

\section{Notes on GPU programming using CUDA}
This section summarises key aspects of NVIDIA GPU programming using CUDA. For details, reference should be made to documents \cite{2011a,n2012} supplied by NVIDIA. 

\subsection{Thread configuration}

GPUs are parallel devices that can achieve impressive computational power by handling thousands of execution threads concurrently. Mandated by hardware design, threads are grouped in blocks, and each block is executed by one multi-processor. Each multi-processor concurrently handles a few of these blocks and blocks are scheduled to run only after other blocks finish execution. Threads within a block are guaranteed to be all running when a given block is scheduled on a multi-processor. For modern GPUs,  the maximum threads per block is 1024, implying that the modern GPU can handle tens of thousand of threads concurrently.

An execution code over a GPU is referred to as a \textit{kernel}, and in this thesis this term is used only for this purpose. The term \textit{thread configuration} is used to describe the set-up of threads and blocks for the \textit{kernel}. Many of the \textit{thread configurations} mentioned in this thesis are in such a way that a thread is set to process an element without the need to interact with other threads. In such thread configurations it is to be assumed that blocks are set with the maximum number of threads, and enough number of blocks are requested to execute at full occupancy. It is also to be assumed that there can be thread re-use, in the sense that a given thread will process  a set of elements one after each other. It has the same effect of a thread being destroyed after an element is processed and re-created to handle a new element when a new block is scheduled.

\subsection{Memory}

Many GPU algorithms tend to be \textit{memory bound} meaning that the main limiting factor is access to memory. In the design of such algorithms, the strive is often the optimisation memory usage.

The GPU provides different types of memory. Table \ref{tab:GPUmemory} lists the relevant ones for this thesis together with some salient features.

\begin{table}[h]
  \centering
    \begin{tabularx}{\textwidth}{ P{3cm}P{2.6cm}P{1.8cm}LP{3cm} }
    \hline
    \textit{Memory Type} & \textit{Location (on/off chip)}& \textit{Access} & \textit{Scope} & \textit{Lifetime} \\
    \hline
    Register & On & R/W & 1 thread & Thread \\
	Shared  & On & R/W  & All threads in block & Block \\
	Global & Off & R/W & All threads and host & Host allocation \\
	Texture & Off & R  & All threads and host & Host allocation \\
    \hline
    \end{tabularx}%

\caption[An incomplete list of different memory types available on an NVIDIA GPU]{An incomplete list of different memory types available on an NVIDIA GPU, together with some salient features. Memory types, which are not relevant to this thesis, are not listed.} 
\label{tab:GPUmemory}
\end{table}

Global memory is the only Read/Write memory that is persistent beyond the lifetime of a kernel and it is also the largest in size (some few Gigabytes). It also suffers from high latency. Shared memory is much faster than global memory but limited in scope and size (a few tens of kilobytes). Registers are even faster as  they can be accessed at zero extra clock cycles per instruction in most execution scenarios. They are much more limited in scope than shared memory and are a scarce resource.

Shared memory and registers can be used to store temporary data, that need to be accessed quickly. For example, if a set of data is required to be read by all threads in a block, then it is general practice to load the data set in shared memory.

Textures are read-only memory structures, and in this thesis, they are used for their on-chip cache that enables efficient access to read-only data from global memory. They provide further functionalities, not used in this thesis, and not reviewed here.

\section{Related work on high-performance gridding} 

Literature on high-performance implementation of \textit{w-projection} and \textit{convolution gridding} for radio interferometry is scarce. This contrasts with medical science research whereby GPUs are a common topic of research for use in medical equipment \cite{Pratx2011,Eklund2013,Schiwietz2006}.

Edgar \etal\ \cite{Edgar2010} report work on a CUDA gridder for the \textit{Murchison Wide Field Array} (MWA) \cite{Bowman2012}. A thread is associated with a unique point on the grid. Each thread scans through a list of visibility records, calculating the contribution (if any) that the record gives to the grid point under the thread responsibility.  Most of the records do not give any contribution to a grid point, and there is substantial waste of time in the scan. This waste is reduced by grouping records by their position on the grid, such that each gridder thread scans only some of the groups.

Humphreys and Cornwell \cite{HumphreysB.} discuss a gridder used for benchmarking. A thread is assigned for each point of the convolution function. To avoid race conditions, a single run of the kernel only grids few records which do not overlap when convolved.

Another benchmark study is presented by Amesfoort \etal\ \cite{Amesfoort2009}. Race conditions are evaded by allocating a private grid for each thread block. This study makes the gridder unusable for radio interferometry in view that memory requirement for such a configuration are much higher than the global memory available in modern GPUs.

Romein \cite{Romein2012} has recently proposed as algorithm which this thesis regards as a breakthrough on the subject. It is used in the development of the thesis' imaging tool. From here on this algorithm will be referred to, as Romein's algorithm. It exploits the trajectory behaviour of the baseline that was explained in section \ref{sec:UVW}. Details on the algorithm are given in section \ref{sec:romein}.

Yatawatta \cite{Yatawatta2013} at the \textit{Netherlands Institute for Radio Astronomy} (ASTRON)\cite{ASTRONhomepage} developed a new imaging tool called \textit{excon}. It grids visibility measurement over GPUs using \textit{w-projection} and \textit{w-snapshots}.

GPUs are not the only device considered for the high performance implementation of \textit{w-projection} and \textit{convolutional gridding}. For example, Verbanescu \etal\ \cite{Varbanescu2008}, Verbanescu\cite{Varbanescu2010} and Amesfoort \etal\ \cite{Amesfoort2009} consider an implementation over the Cell B.E processor while Kestur \etal\  \cite{Kestur2011} reports on a framework for gridding over FPGAs.

\section{Romein's algorithm}
\label{sec:romein}
As already stated, Romein's algorithm refers to the algorithm presented in \cite{Romein2012}. This thesis makes use of this algorithm and hence it is reviewed in some detail here. It should be noted that chapter \ref{chap:results} presents new analysis of this algorithm that are not published in \cite{Romein2012}.

Records measured by a given baseline are sorted by time and presented consecutively to the algorithm. In this way, while records are being gridded one after the other, the region in the UV-grid that is updated by a convolved record moves with the baseline trajectory.

\begin{figure}
 \centerline{\includegraphics[scale=0.4, trim=0cm 0cm 0cm 0cm, clip=true]{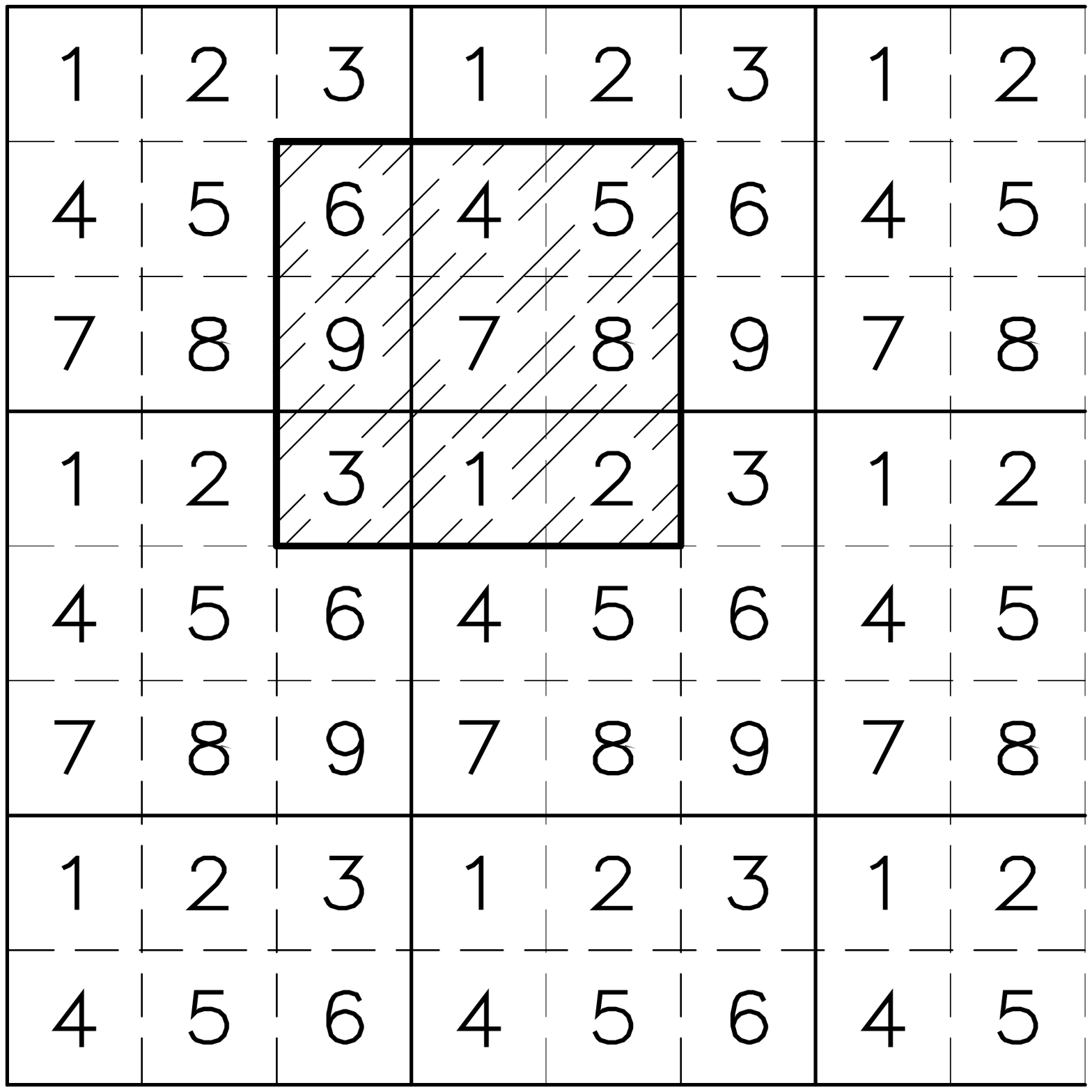}}
   \caption[Thread configuration of Romein's algorithm.]{Thread configuration of Romein's algorithm for a hypothetical  9 $\times$ 9 UV-grid while convolving with a 3 $\times$ 3 convolution function. The shaded area represents the convolution function being gridded. The number in each box represents a thread. Since sub-grids are equal in size to the convolution function size (or less if sub-grid is at the edge of the grid) then a thread updates one and only one grid point when convolving a visibility record.}
\label{fig:threadromain}
\end{figure}

The grid is split up in sub-grids of sizes equal to the size of the convolution function that will be used to grid. Threads are assigned to take care of one grid point in each sub-grid as shown in Figure \ref{fig:threadromain}. Thanks to this configuration a thread updates one and only one grid point in convolving a visibility record. By virtue of the baseline trajectory behaviour, it is quite likely that, subsequent records will need to update the same grid point. Updates are accumulated in the registry, until the grid point moves out of the region being updated. At this point,  the thread commits the accumulated update to global memory and commences a new accumulation for a new grid point in a different sub-grid.

Explanation of some implementation details used in the test case presented by Romein \cite{Romein2012} for this algorithm, now follows. Only the main details that are reused in this thesis are pointed out. 

All grid point updates related to a record are handled by one CUDA block. This means that the block requires an amount of threads equal to the convolution function size. GPUs impose a maximum on the number of threads per block (maximum of 1024 threads per block for the latest architectures) which is in many cases smaller than the size of a convolution function. A given thread is thus allowed to cater for more than one grid point in each sub-grid. Ideally all the grid points under the responsibility of a thread should be handled concurrently but due to registry limitations in modern GPUs, this is not always possible. Instead, a group of records is scanned several times by the thread so as to cater for all the grid points entrusted to it. All the threads in a block will need to read the same record data for several times, and by GPU best practices, data is pre-loaded in stored memory so as to have a fast access.

Different CUDA blocks grid different groups of records. This mandates the use of atomic additions to commit the accumulated grid point updates to the grid. Single precision atomic additions are intrinsically supported by GPUs, but double precision atomic additions are not supported. Despite the fact that a software implementation of double precision atomic additions is possible, such an implementation is inefficient and heavily impairs the algorithm's performance. In conclusion, the algorithm works efficiently only for single-precision.

The implementation makes use of textures for retrieving convolution function data stored in global memory.

Romein \cite{Romein2012} reports a maximum  of 93 Giga Grid point updates per second\footnote{The metric is explained in section \ref{sec:updatemetric}.} for the test implementation. Gridding was done over a quad-polarised grid using a GeForce GTX680 GPU. This result is revised in chapter \ref{chap:results} in view that enhancements made in this thesis, give more knowledge about the algorithm.

\chapter{The General Array FrameWork}
\label{chap:gafw}
The \textit{General Array FrameWork} (GAFW) provides an infrastructure for numerical calculations. The framework has features meant to facilitate  high-performance and has a simple user\footnote{In this context the "user" is the developer of an application built over the \textit{General Array Framework}.} interface. It adapts to the underlying hardware (CPUs, GPUs, FPGAs etc) since all control logic is handled by the framework through an engine.

The framework's design promotes collaboration between scientists with basic knowledge in programming, and developers specialised in high performance computing. A layered approach allows for two distinct development areas, each suited for the respective collaborators mentioned above. The concept of this framework is based on the mathematical concepts of \textit{arrays} and \textit{operators} with minimal framework-specific jargon. This makes it easily comprehensible and manageable by scientists.

In this thesis, a C++ \cite{ISO2012} implementation has been developed that supports multi-GPU systems and forms the basis of the imaging tool that will be discussed in the next chapter. The use of the framework in the imaging tool simplified the development of the tool.

\section{Main concepts}

The framework mimics the way humans tend to make a calculation. In most cases, an equation is first defined, and afterwards numerical data is entered to obtain a result. In other words, humans tend to understand mathematical relationships before doing any calculation. A similar computational scenario is presented in this framework. The application defines how a calculation is done and then makes a separate request for a calculation over a set of input data. Since calculation definitions are separate from the actual calculation, they can be reused over and over again for different input data.

A calculation is defined using a \textit{calculation tree}. The \textit{calculation tree} is a graph of \textit{operators} that get connected together via \textit{arrays}. \textit{Operators} define logic that generates numerical data, while each  \textit{array} represents an N-dimensional ordered set of numerical data of a specific type (ex integers, complex numbers, etc). \textit{Arrays} are set as input and/or output to \textit{operators}.   

\begin{figure}[h]
    \centerline{\includegraphics[scale=1, clip=yes, trim=1cm 0.3cm 1cm 0.3cm]{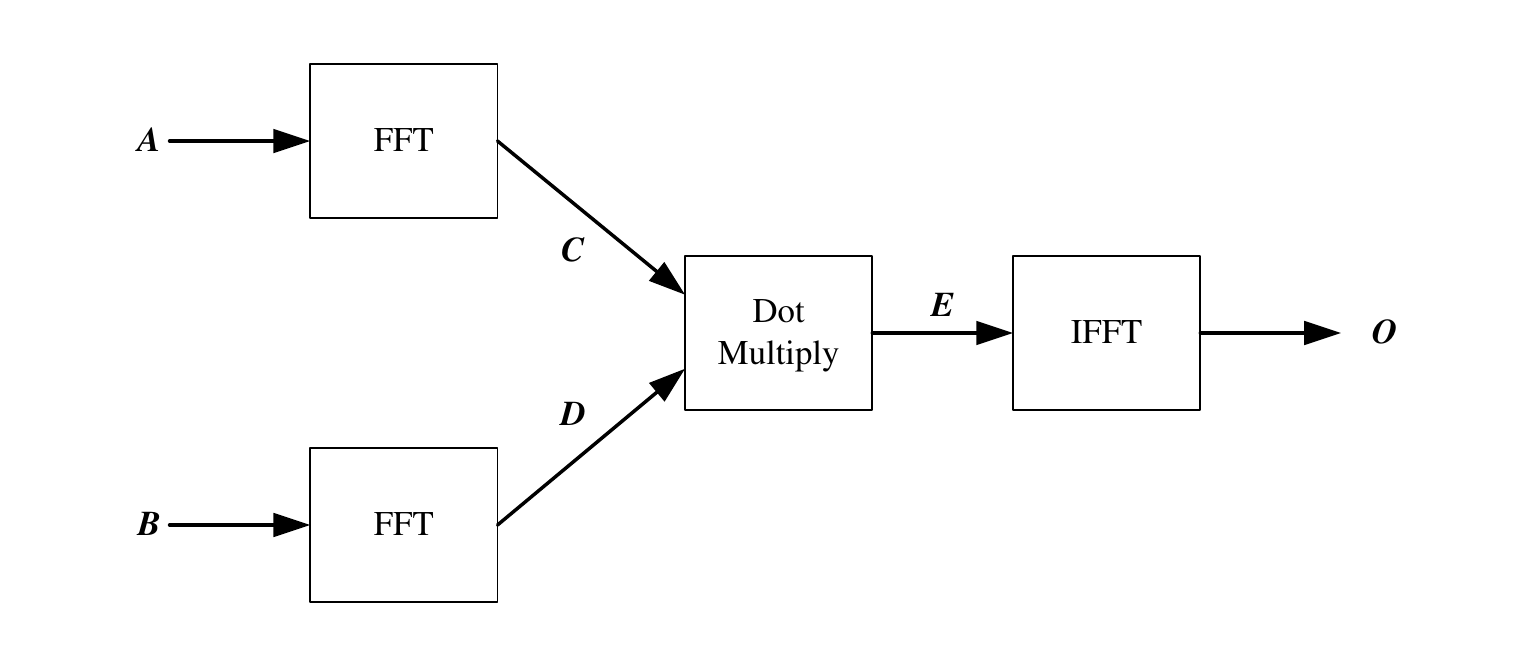}}
  \caption[An example of a \textit{General Array Framework} calculation tree]{An example of a calculation tree. This tree defines convolution by multiplying elements in the Fourier domain. Boxes represent \textit{operators} while \textit{arrays} are represented by arrows.}
\label{fig:convtree}

\end{figure}

Figure \ref{fig:convtree} gives an example of a simple tree. It defines convolution using FFTs. \textit{Arrays} are represented by arrows.

For convenience, the entry point for the generated numerical data for an \textit{array} is provided by a third object called the \textit{result}. Each array has associated to it one and only one \textit{result} object.

\textit{Result} objects provide also the entry point for calculation requests. Calculation requests are \textit{result-centric} and not \textit{tree-centric} in the sense that the request is to calculate the result. Based on the defined \textit{calculation trees}, the framework decides which operations (that is, execution logic described by \textit{operators}) are required to calculate and obtain the requested result.
 
\textit{Result} objects provide mechanisms to inform the framework on what to do with the given data. An example is whether the application intends to retrieve the data or not. This has a significant impact on performance since a memory copy from GPU to host is required.

Calculations are done asynchronously with respect to the application thread. Such behaviour is crucial to achieve high performance and maximise the usage of available resources. Once the application requests a calculation, the framework validates and proceeds in the background. The application thread can load more data and request further calculations while GPUs are executing code in the background. 

Two final points on the general concept of the framework are the \textit{factory} and object identification.  A \textit{factory} is used to make things as easy as possible to the user and ensure full control of the framework over its own objects. The \textit{factory} maintains all framework objects, including their creation and destruction. Framework objects are identified by a two-string structure which is referred to as an \textit{identity}. A registry service provided by the framework enables the application to use the objects' \textit{identity} instead of C++ pointers or references, when communicating with the framework.

\section{A layered approach}

\begin{figure}[h]
  \centerline{\includegraphics[scale=1,trim=0.9cm 0.9cm 0.9cm 0.9cm,clip=yes]{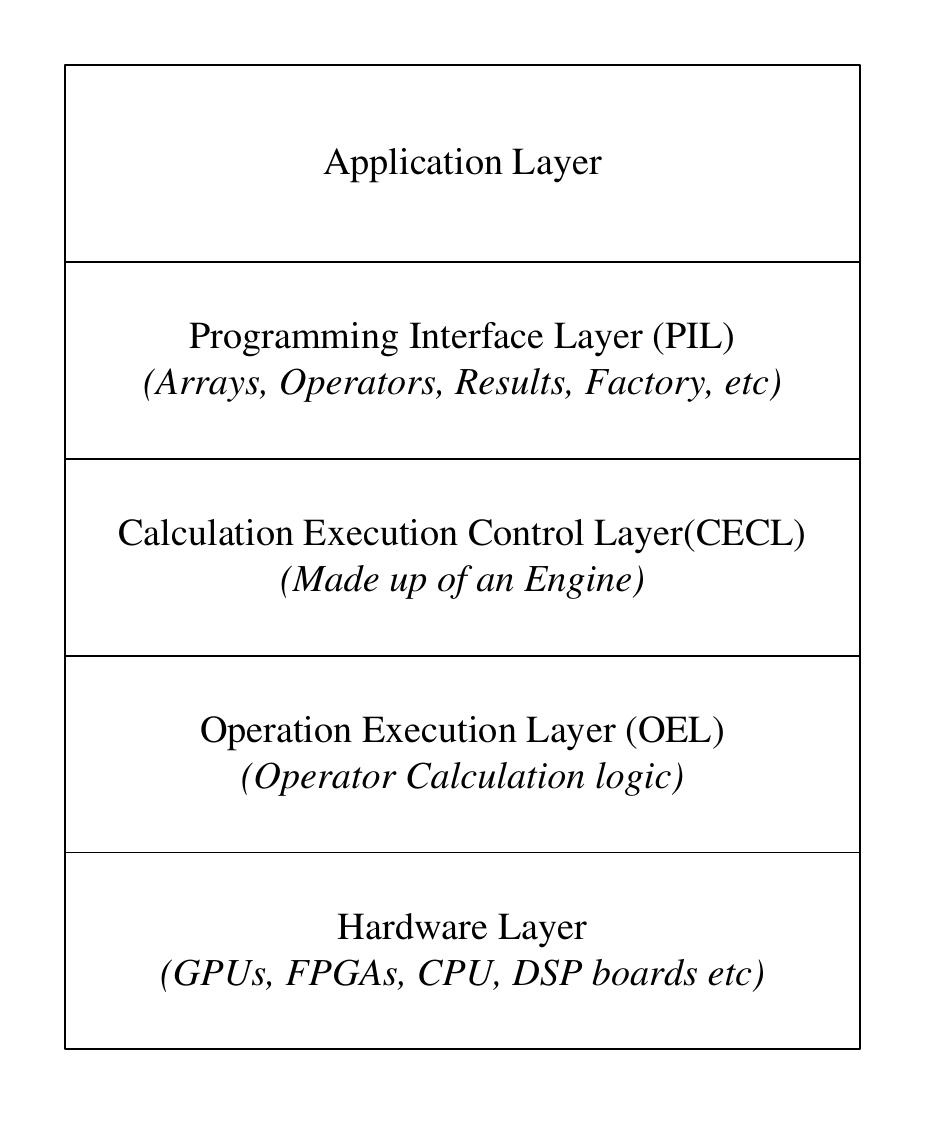}}
  \caption[Layers of the \textit{General Array Framework}]{Layers of the \textit{General Array Framework}. Each layer is aware only of the layer directly below it and gives service to the layer directly above it. The only exception to the rule is the CECL which has direct access to the \textit{Hardware Layer}.}
\label{fig:gafwlayer}
\end{figure}

The framework can be modelled by five stacked layers as shown in Figure \ref{fig:gafwlayer}. Each layer is described hereunder:

\begin{description}

\item{\textbf{\textit{The Application Layer:}}} As its name implies, this layer represents the application built over the GAFW. The application constructs \textit{calculation trees}, inputs numerical data to the system, requests calculations and retrieves results. Note that the application layer is unaware of the underlying hardware as it is up to the framework to make the necessary adaptations to execute calculations on the underlying hardware. This keeps the development of the application layer simple.

\item{\textbf{\textit{The Programming Interface Layer (PIL)}}} defines all functionality that the framework provides to the application layer. Its design is intended to offer a remarkably simple interface to the application layer.  Section \ref{sec:pil} discusses the \textit{programming interface layer} in greater detail. 

\item{\textbf{\textit{The Calculation Execution Control Layer (CECL)}}} contains all the logic to perform a calculation. It is well aware of the underlying hardware and adapts to it. As already pointed out, only GPUs are currently supported, and it adapts calculations to the number of GPUs available. The layer is made up of a multi-threaded engine, providing all GPU control logic to execute a calculation. It requires direct access to the underlying hardware.

\item{\textbf{\textit{The Operation Execution Layer(OEL)}}} is responsible to execute operator code on the hardware.

\item{\textbf{\textit{The Hardware Layer}}} is the hardware itself. As pointed out many times, CUDA compatible GPUs are the only supported hardware. It is aimed that in the future, there will be support for other hardware such as FPGAs, DSP boards, CPUs and clusters.    

\end{description}

\section{The programming interface layer}
\label{sec:pil}

This section discusses more concepts and details of the framework as seen from the point of view of the application layer. Each subsection discusses a concept, object or service given by the framework.  

\subsection{The factory}

Since a priority in the design of the framework is simplicity, the handling of objects is mandated to the framework and stripped off from the application. The framework delivers such service through a \textit{factory}. Only one instance of the \textit{factory} can exist at any moment in the application life cycle, and is expected to exist throughout the whole lifetime of the application. Whenever the application requires the creation and destruction of framework objects, it does so by making a request to the framework's \textit{factory}.

The factory simplifies the management of \textit{operators} for the application layer, since \textit{operator} initialisation and destruction logic can vary from one \textit{operator} to an other.  Further details are discussed in section \ref{sec:operators}.

The use of the \textit{factory} is beneficial for forward compatibility with future releases of the framework. Since all the initialisation/destruction logic is in the absolute control of the framework, any future enhancements of the framework that changes such logic can be implemented with no change to the application layer code. The application will still be able to compile against new versions of the framework.

The \textit{factory} represents the framework, and its instantiation is analogous to the instantiation of the framework. Once the factory is set up, there are no other procedures to follow to initialise the framework. The application can immediately build \textit{calculation trees} and subsequently request calculations. 

\subsection{Identification of objects and the registry service}
\label{sec:identity}
Framework objects are identified using a two-string structure called an \textit{identity}. User-defined objects can also have an \textit{identity}, and the framework is well structured to support such user-defined objects. 

An \textit{identity} is defined by two strings: a \textit{name} and an \textit{object-name}. The \textit{name} identifies the type of object such as \textit{"array"} or \textit{"result"} while the \textit{object-name} identifies the object itself. In Figure \ref{fig:convtree} the \textit{object-names} \textit{"A"} and \textit{"B"} are used to represent two \textit{arrays} in the calculation tree. Every object is expected to have  a unique \textit{object-name}, which is given by the application layer. \textit{Names} are given by the framework and are useful to distinguish between different operators (such as \textit{"FFT"} or \textit{"Dot Multiply"}).

\textit{Object-names} have a hierarchical naming scheme\footnote{\textit{Names} do not have a naming scheme.}. A dot (.) is used to separate levels in the hierarchy. There are no rules regarding how the hierarchy is set up, but, if the object is to be registered, its parent needs to be registered beforehand. The hierarchy is essential to avoid conflicts in \textit{object-names}. For example, if two unrelated trees are developed, there is the possibility that, by mistake, the same \textit{object-name} is chosen for objects in the two trees. If each tree uses its own namespace, by defining its own unique branch in the hierarchy, then the issue is avoided. Note that the framework provides  mechanisms to support application defined objects to work in their own namespace.  

The use of \textit{identities} gives the possibility of communicating with the framework on objects using \textit{object-names} or names instead of C++ pointers or references.  A registry service is available in the framework that keeps a mapping between \textit{object-names} and respective objects. Framework objects are automatically registered on creation. Other application defined objects can be registered manually. This strategy alleviates the application from maintaining C++ memory pointers to objects that it needs.

\subsection{Arrays}

An \textit{array} represents the movement of data within a tree. It has an \textit{identity} and is automatically registered by the \textit{factory} on creation. An \textit{array} is defined by its dimensions and the type of numbers it holds (ex integers, complex numbers, etc). 

An \textit{array} within a tree can be in three modes: input, output or bound. The modes are mutually exclusive, and \textit{arrays} can only be in one mode.

An \textit{array} is in input mode if numerical data is manually pushed in by the application layer. The \textit{array} provides the entry point to input such data.

An \textit{array} can be bound with a \textit{result} object. Such a bind instructs the framework to use the last generated data represented by the \textit{result} object as input to a subsequent calculation in which the bound \textit{array} is involved.

If an \textit{array} is neither bound nor in input mode then the array is in output mode and is expected to be set as an output of an operator. The application is not normally expected to define the dimensions and type of data for such arrays, since \textit{operators} are expected to have logic to determine such properties automatically during validation (refer to section \ref{sec:validation}).

\subsection{Operators}
\label{sec:operators}

A GAFW \textit{operator} describes logic that generates numerical data. For example, an operator might describe the logic of a Fast Fourier Transform. 

\textit{Operators} have an \textit{identity} (discussed in section \ref{sec:identity}). The \textit{identity's name} of an \textit{operator} is an essential property and is one of the main reasons why an \textit{identity} includes a name.  Different operators are regarded as different object types and must have a different \textit{name}. The application communicates its request for a new \textit{operator} object using the \textit{name} of the \textit{operator}. Good meaningful \textit{names} are those that describe the execution code represented by the operator such as \textit{"FFT"}. 

In the general case, different \textit{operators} are implemented using a different class. Thanks to \textit{identities} and object-oriented polymorphism, this complexity is hidden from the application which is aware only of the base class.    

An \textit{operator} has three forms of input data: input \textit{arrays}, pre-validator \textit{arrays} and \textit{parameters}. Input-less operators are legitimate.

\begin{description}
\item{\textbf{\textit{Input arrays}}} are an ordered list of \textit{arrays} that describe the data on which the \textit{operator} operates. The numerical data that the \textit{array} represents is made available on the GPU memory during the operation execution.

\item{\textbf{\textit{Pre-validator arrays}}} are another ordered list of \textit{arrays}. Different from the normal input \textit{arrays}, the data is not presented to the operator during execution. They are only applicable during the validation phase (refer to section \ref{sec:validation}). The framework guarantees that they get validated before the \textit{operator}. It is expected that the \textit{operator} validation logic will need to obtain some information from the \textit{array} (such as dimensions or data type) in order to execute its own validation logic.     

\item{\textbf{\textit{Parameters}}} are a set of variables that are set up prior to calculations. They can be regarded as configuration data for the \textit{operator}. For example,  an \textit{operator} doing zero padding would need to know how much zero padding is needed. This information can be given to the \textit{operator} through parameters.  
\end{description}

An \textit{operator} has only one form of output, which is an ordered list of \textit{arrays} in output mode. 

An \textit{operator} can request some additional allocation of memory on the GPU for its own private use while executing. This memory is referred to as a buffer.  

\textit{Operators} exist in the PIL and OEL layers (refer to Figure \ref{fig:gafwlayer}). As mentioned earlier, the base class is presented to the application layer. The PIL, therefore, only defines the basic interactions of the \textit{operator} with the application layer. As for the OEL, the operator defines all validation and execution logic. 

The procedure to develop an \textit{operator} is simple. The \textit{operator} is defined by a new C++ class that inherits the \textit{operator} basic class. Two methods need to be overloaded, one providing validation logic and another one providing the kernel submission logic. In most cases,  a new CUDA kernel needs to be coded. This is the only hard part in the whole procedure since it requires skilled expertise in CUDA programming. The final step is to register it in the framework by its \textit{name}. The registration process requires coding for class instantiation and \textit{operator} object destruction.

\subsection{Results}
\label{sec:GAFW:result}

\textit{Result} objects provide the entry points for controlling and retrieving calculated data and requesting calculations. Each \textit{array} has associated to it one and only one \textit{result} object and contains all the generated numerical data related to the \textit{array}.

Since calculation logic is automated within the framework, the application needs to give instructions on how to handle the data. These instructions are listed and explained below: 

\begin{description}
\item{\textbf{Application requires results:}}
The application has to inform the framework about its intention of retrieving the calculated data. This is particularly essential for performance because extra memory copies from GPU to host memory are required to allow the application to access the result. It also removes any unnecessary  memory allocation on the host. 
\item{\textbf{Data re-usability:}}
This defines the intention of the application to request subsequent calculations that will re-use data generated. On such instruction,  the framework keeps a copy of the result, possibly on GPU memory, for re-use. 
\item{\textbf{Overwrite:}}
Setting a result for overwrite instructs the framework that prior to executing the corresponding operation on the GPU, it should initialise output data to the last calculated result. 
\end{description}

These instructions can be given to any \textit{result} object participating in a calculation and not only to the  \textit{result} object through which a calculation is requested. For example, referring to the \textit{calculation tree} in Figure \ref{fig:convtree}, if \textit{O} is requested to be calculated and \textit{C} is set to be a required result, then the framework will copy the result of \textit{C} to the host and make it available to the application.     

\subsection{Proxy results}
As their name implies, \textit{proxy results} are meant to serve as proxies to \textit{result} objects. \textit{Proxy result} objects behave as if they are genuine \textit{result} objects (the \textit{result} class is inherited). The proper \textit{result} object which is proxied, is configurable at any time, and can be changed at will. The main use of these objects is in situations where a fixed \textit{result} object needs to be presented, but needs to be changed every now and then behind the scenes.

\subsection{Modules}
\label{sec:gafw-modules}
GAFW modules are intended to contain a defined calculation in one object. They are application defined and need to inherit the module's base class. Inputs and outputs of modules are in the form of \textit{result} objects. The module must provide all the logic to request calculations to the framework.

\subsection{Statistics}
\label{sec:gafw-statistics}
It is advantageous to have a statistics system in a high-performance application. For example, it is helpful to have execution timings of operations executed on the GPU. The framework generates its own statistics that are pushed to the application for further processing.

A single statistic is contained in a unique object whose class inherits a general base class. An interface is defined in such a way that  statistic objects can be posted through it. The interface needs to be implemented by the application. It is also expected that posted statistics are processed in a background thread. This design allows the application to use the infrastructure for its own generated statistics.

\section{The engine}

\begin{figure}[h]
    \centerline{\includegraphics[scale=1, trim=1cm 1cm 1cm 0.9cm,clip=yes]{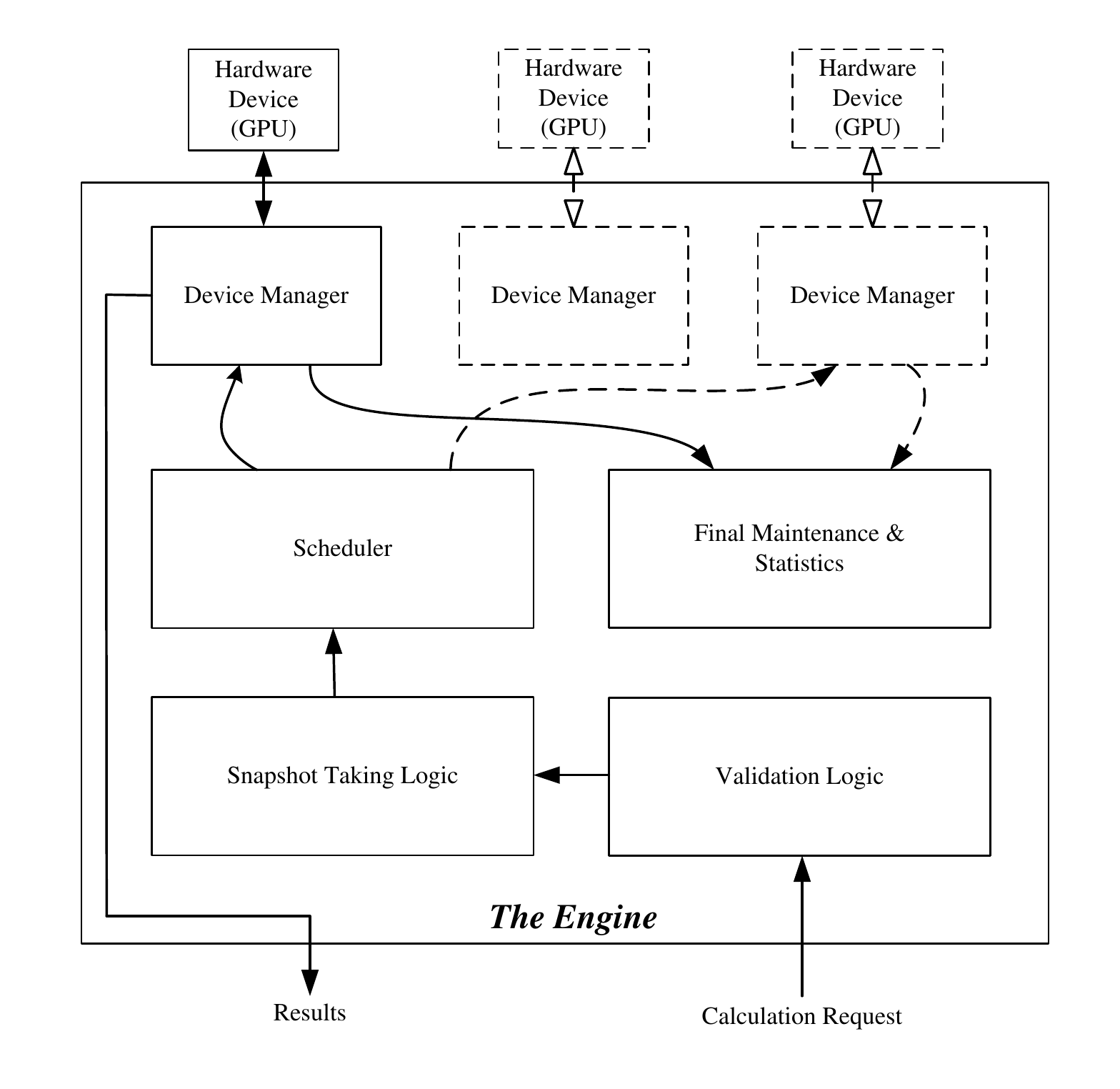}}
  \caption{High-level design of the engine}

\label{fig:engine}

\end{figure}

The \textit{Calculation Execution Control Layer} is made up of an engine whose function is to execute a calculation request. This section discusses the engine and how it executes a calculation over GPUs. Figure \ref{fig:engine} portrays a high-level design of the engine.  

Calculations are executed in three steps: Validation, snapshot taking and proper calculation. Each step is discussed in the following subsections:

\subsection{Validation}
\label{sec:validation}

The framework needs to verify that the calculation tree is valid. It also needs to handle any missing tree data (array dimensions and data type) that can be predicted. If the framework has no sufficient data, or the \textit{calculation tree} is invalid, then the validation process returns an error in the form of a C++ exception.

Some of the validation logic has to be supplied by the \textit{operator} used in the \textit{calculation tree}. The \textit{operator} has to validate itself on details regarding its unique specifications, such as, the number of inputs and outputs.  It also has to determine the dimensions of the output \textit{arrays} when missing or incorrect.

\begin{figure}[h]
    \centerline{\includegraphics[clip=true, trim=1cm 1cm 1cm 1cm, scale=1]{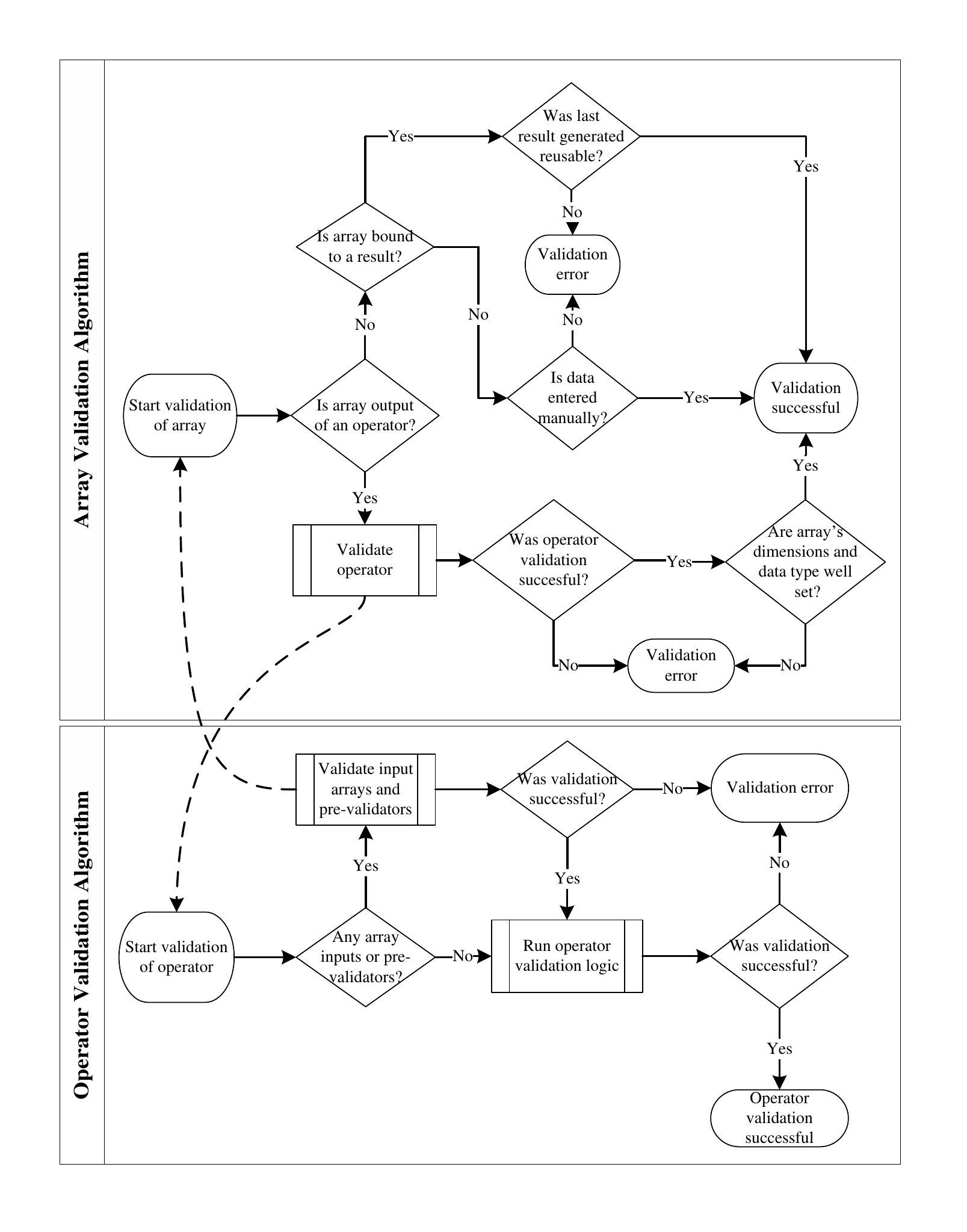}}
  \caption[Flow chart of the validation algorithm]{Flow chart of the validation algorithm. The algorithm is split in two sub-algorithms: one for validating \textit{arrays} and one for validating \textit{operators}. The starting point is the validation of the \textit{array} associated with the \textit{result} object through which the calculation has been requested. The algorithm iterates between the two sub-algorithms so that it traverses the \textit{calculation tree} until finding a non-output array or input-less operator. Once such objects are found it will reverse back while validating all objects.}
\label{fig:validation}
\end{figure}

The validation algorithm is depicted  in the flow chart shown in Figure \ref{fig:validation}.  The algorithm is split in two sub-algorithms, one for validation of \textit{arrays} and one for validation of \textit{operators}. These two sub-algorithms recur on each other. \textit{Array} validation in output mode requires the validation of the respective \textit{operator}, while \textit{operators} need the validation of  all input \textit{arrays} and pre-validators. In this way, the tree is traversed until non-output arrays or input-less operators are found. The algorithm then reverses back validating all objects.

\subsection{Snapshot taking}

Taking a snapshot means the copying of all relevant data regarding a calculation such that the proper calculation can be executed asynchronously. In this way, the proper calculation takes place in the background while the application can change trees, input new data, request other calculations or execute any other logic.  

To execute a calculation, the engine transforms the calculation tree into a stream of operations to run. Each element in the stream describes an operation in full, including transient data, such as state, locking mechanisms and memory allocation on GPU for input and output data. The engine relies exclusively on this data to perform a calculation.

\subsection{Proper calculation}

This step, computes the \textit{calculation tree} over GPUs. Simplistically speaking, it is a matter of memory management and kernel submission.  All the tasks are executed by the engine with the exception of kernel submission. The latter is executed by the \textit{operator} on request from the engine.

The engine is implemented using a multi-threaded approach. All the work done by the engine is divided into simple tasks, each handled by a separate thread. A blocking FIFO queue is used to communicate between threads. This approach enables the framework to monitor many events concurrently without the need of looping. It avoids busy waiting were events are continuously queried for status updates. It also helps the framework to act quickly on events over which a thread block waits. The waiting thread is immediately released as the event is fired.

\begin{figure}[h]
    \centerline{\includegraphics[scale=1,clip=true,trim=1cm 1cm 1cm 1cm]{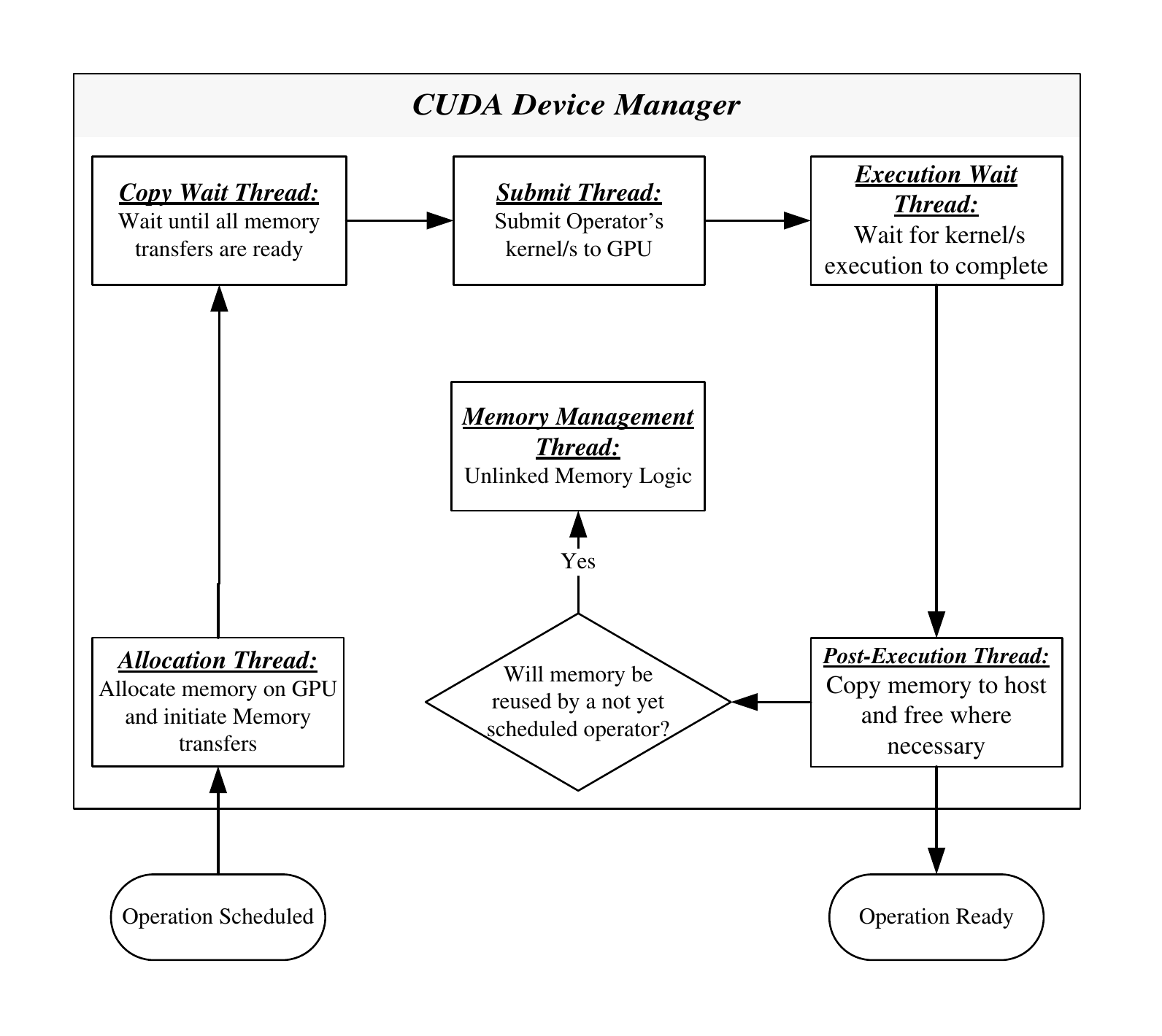}}
  \caption[The \textit{CUDA Device Manager}.] {A flow chart depicting the tasks applied to an operation once scheduled on the CUDA Device Manager. Each process block represents a task that gets executed in a separate thread. The memory management thread does not handle operations but instead administers memory allocated on GPU. The decision block at the middle of the diagram is executed by the \textit{post-execution thread}.}
\label{fig:devicemanager}
\end{figure}

The engine delegates the actual management of a device to a device manager. This simplifies functional support for multi-GPUs (and in the future other devices). A unique instance of the manager is brought up for every device supported. It has its own task threads as illustrated in Figure \ref{fig:devicemanager}.

The engine schedules an operation to be executed over a device by submitting it to the respective device manager. In systems having more than one GPU, a whole calculation request is submitted to one GPU while the next calculation request is submitted to the next GPU. This simple scheduling algorithm proved to be good enough for the imaging tool discussed in the next chapter.

GPU memory management works as follows: The \textit{allocation} thread continuously allocates memory for incoming operations until the memory is full or there are no operations in the pipeline. Once the GPU memory is full it waits until the \textit{post-execution} thread frees memory. Memory is freed once the computation related to an operator is ready and will not be reused for future operations. Memory that is not freed and is not set as input or output for any operation submitted to the device manager is managed by the \textit{memory management} thread. Such memory can halt the whole calculation process as it might not leave enough space for currently scheduled operators. In the case of such a scenario, the device manager caches the data on the host main memory (if not already done) and frees memory on the GPU.  Caching also takes place when data needs to be transferred from one GPU to another, or when de-fragmentation of memory is required. 

Unfortunately, the CUDA runtime API \cite{2011} is unable to allocate memory on a GPU while a kernel is being executed. This causes the \textit{allocation} thread to freeze up during kernel execution, reducing thread concurrency on the CPU. This has a direct impact on the overall performance since data transfers\footnote{By GPU best practices \cite{2011a} time to transfer data from host to GPU is hidden by doing memory transfers in parallel with kernel execution.} cannot be initialised prior to the allocation of the respective memory. In order to ease the problem, a locking mechanism is in place that denies concurrent allocation of memory and submission of threads. In this way,  the allocation process works in batch while kernel submission is locked.

\section{Collaboration}

In the chapter's introduction,  it was claimed that the framework is designed for collaboration. This section elaborates on this argument.
 
Framework related development of an application can be divided in two areas. The first area is the application layer. Development related to this area is easy and fast with no specialised expertise required. Whoever attempts development in this area, requires general knowledge on the behaviour of the calculation with no details on how it is deployed on the GPU. This area suits remarkably well to that scientist who has general expertise in programming and views programming as a means to reach scientific goals. 

Development of \textit{operators} is the second area. Developing an operator requires  specialised expertise in high-performance parallel computing over GPUs. Much less knowledge about the overall behaviour of the calculation is required, but in-depth and detailed knowledge of the \textit{operator} is a must. It is suited to a developer who has little interest on scientific goals and much more interest in writing high-performance code. 

The two areas are separate and the only commonalities are the framework itself and functional specification of the \textit{operator}. Therefore, it is is easy to promote collaboration between the scientist and the developer, and get the best out of the two. One notes that functional specification documents are a standard in the industrial software development world. It is the main tool with which developers communicate with their clients.  

\section{Standard Operators}

As part of the implementation, some standard operators have been developed. This section discusses the most noteworthy ones.

\subsection{The Fast Fourier Transform (FFT) operator}
\label{sec:FFTOperator}
This operator executes Fast Fourier Transform over arrays of any dimension. It is possible to divide the array into sub arrays of lesser dimensions and perform a Fourier transform over them. For example, in the case of a matrix, it is possible to request a 1-dimensional Fourier transform over each column. It is implemented using the CUDA cuFFT library \cite{Nvidia2007b} provided by NVIDIA. Unfortunately, the documentation \cite{Nvidia2007b} does not give substantial information on the buffer space required to execute an FFT. Documentation only states that it is between one and eight times the size of the input being FFTed. To be on the safe side, the operator has been set to request a buffer eight times larger than the input. 

\subsection{The all-prefix-sums operator}
\label{sec:all-prefix-sums}
Given a sequence $\{a_i\}^{n-1}_{i=0}$, the all-prefix-sums operator does an accumulated addition over the elements to return the sequence $\{b_i\}^{n-1}_{i=0}$ defined as follows:

\begin{equation}
b_i= \begin{cases}0 & if\ i=0\\b_{i-1}+a_{i-1} & otherwise \end{cases}
\end{equation}

This operation is also known as an \textit{exclusive scan}.

The implementation is based on the ideas presented in \cite{Nguyen2007}, which are based on the work of Blelloch \cite{Blelloch1990}. Balanced trees are used. This implementation computes lower levels of the balanced tree over registers giving a significant boost to performance.

\subsection{The general reduction template}
\label{sec:operatorreduction}
This is a generic \textit{operator}  (in C++ terms: a template). It provides general code for \textit{operators} that handle reductions of the form:

\begin{equation}
R=\bigoplus^{n-1}_{i=0}\left(\bigotimes^{k-1}_{j=0}(a_{j,i})\right)
\end{equation}

where
\begin{description}
\item{$\left\lbrace\left\lbrace a_{j,i} \right\rbrace^{n-1}_{i=0}\right\rbrace^{k-1}_{j=0}$} is a sequence of sequences, whereby each element sequence is an input to the GAFW \textit{operator}.
\item{$\bigoplus$} is a general mathematical operator that has to be associative.
\item{$\bigotimes$} is another general mathematical operator that does not need to be associative. 
\end{description}

Specialisation of the template needs to define the two mathematical operators and the value of $k$. The value of $n$ is determined from the size of the input \textit{arrays}. The \textit{operator} can reduce over dimensions. For example, in case of a matrix,  it is possible to reduce each column separately. The result will be a 1-dimensional \textit{array} of length equal to the number of rows of the matrix.

Reductions are simple to implement over GPUs. In this implementation,  a kernel is run with a configuration of maximum threads per block and a number of blocks high enough to reach nearly 100\% real occupancy. Work load is split evenly over all threads such that each thread reduces a subset of the $n$ elements in each sequence. Results are saved in shared memory so as to run $\bigoplus$ over all results produced by threads in a given block. This is stored in global memory in such a way that a second kernel re-applies $\bigoplus$ over the values calculated by each block.

\chapter{The Malta-imager}
\label{chap:mtimager}

The \textit{Malta-imager (mt-imager)} is a high performance image synthesiser for radio interferometry. It is implemented in C++ \cite{ISO2012} and CUDA \cite{2011}, and uses the GAFW infrastructure (refer to chapter \ref{chap:gafw}). It achieves unprecedented performance by means of GPUs. The CPU multi-threaded design for handling pre- and post-data processing is also a crucial ingredient in ensuring the best performance.   

The image is synthesised using \textit{w-projection} (refer to section \ref{sec:w-projection}). An independent multiple-Stokes dirty image is synthesised for each channel. Measurement data is expected to be available as a MeasurementSet \cite{Kemball2000} stored using the Casacore Table System \cite{Diepen2010}. Output images are stored as a 4-dimensional data cube FITS \cite{Pence2010} primary image. Most of the calculations are done in single-precision mode \cite{2008,Whitehead2011,Goldberg1991}.

\section{High level design and concepts}
\label{sec:highleveldesign}

\begin{figure}
\centerline{\includegraphics[scale=1, trim=1cm 1cm 1cm 1cm, clip=yes]{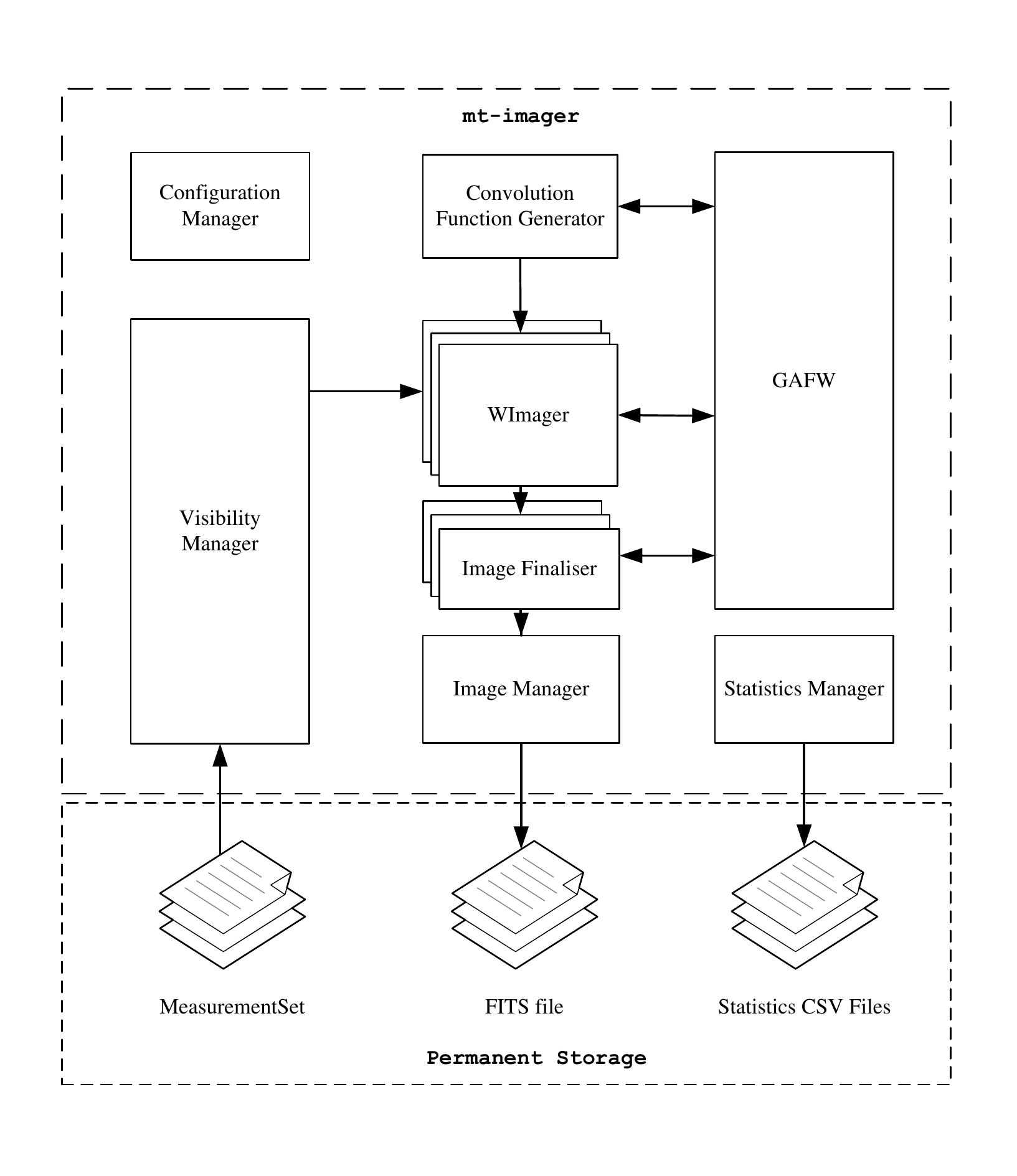}}
\caption[High-level design of the imaging tool]{High-level design of the imaging tool showing the various components and relevant data flow}

\label{fig:highlevelmtimager}
\end{figure}

The design is based on the \textit{General Array Framework} (GAFW) philosophy (refer to chapter \ref{chap:gafw}). The system is made up of seven autonomous components as depicted in Figure \ref{fig:highlevelmtimager}. None of them interact directly with each other, and with a few exceptions, they are unaware of each other. Data is communicated between each component using the GAFW, by means of \textit{result} objects (see section \ref{sec:GAFW:result}). The \textit{main()} function integrates all components together.

Each component is assigned a unique set of tasks. The \textit{Configuration Manager} takes care of producing configuration information for each component based on the local configuration and environment. The \textit{Visibility Manager} loads data from a \textit{MeasurementSet}, sorts, converts and inputs data in the GAFW infrastructure. The \textit{WImager} performs gridding over a multi-polarised grid, while the \textit{Image Finaliser} converts the grid to a multiple-Stokes dirty image. These components are implemented as GAFW modules, and all calculations are made over GPUs. Channels are processed independently, and, for each channel, separate instances of the two components are set up. Numerical representations of the convolution functions $\tilde{C}_w(u,v)$ (defined in equation \ref{equ:C_W}) are generated by the \textit{Convolution Function Generator}. The \textit{Image Manager} stores output images in FITS files, while the \textit{Statistics Manager} processes statistics, generated by each component, including the GAFW. It then reports them in various CSV (Comma-Separated Values) files.

Data is processed in chunks to exploit parallel mechanisms available on the hardware. This is essential to ensure high-performance. A GPU is by itself a parallel device which can only achieve high-performance through parallel methods. Presenting the GPU with a suitably sized chunk of data ensures best gridding performance. In case of a multi-GPU system, by virtue of the GAFW, the imaging tool grids independent channels over different GPUs so as to achieve concurrency over GPUs.

Fast gridding on GPUs is the imaging tool's strong suit. Nevertheless CPU bound pre-processing and post-processing of data is a necessity, since, by design GPUs are limited. For example, GPUs cannot load data from hard-disk, neither do they recognise C++ objects. Also, they cannot save images to permanent storage. If these pre- and post-processing steps are not well handled on the CPU, then, they can severely compromise performance. 

Execution time of pre- and post- processing steps is hidden by having the CPU, GPU and permanent storage IO running in parallel. Most components perform their tasks asynchronously to each other. Since data is processed in chunks, the \textit{Visibility Manager} prepares the chunks in the background through a multi-threaded mechanism. Once the first chunk is available for gridding the \textit{WImager} component (by virtue of the GAFW) grids the data over the GPU while the \textit{Visibility Manager} continues its task of preparing other chunks. This achieves concurrent use of the CPU and GPUs. 

Channels are processed one after each other\footnote{For multi-GPU systems, channels are processed in parallel by an amount equal to the number of GPUs.}. Once a channel dirty image is finalised the \textit{Image Manager} saves the image to disk while subsequent channels are being processed. This ensures concurrent use of the GPU and permanent storage. The \textit{Visibility Manager}, by virtue of its multi-threaded design ensures concurrency between CPU processing and permanent storage IO and exploits the multi-core infrastructure of modern CPUs.

The core of gridding is based on Romein's algorithm \cite{Romein2012} which is enhanced, implemented and adapted for the necessities of this imaging tool. The algorithm requires that data is grouped by baseline and sorted in time. The ordering is done by the \textit{Visibility Manager}.

The imaging tool supports single, dual or quad polarised\footnote{Polarisation can be linear or circular. Both are supported.} data. The term multi-polarisation is used as to describe any number of polarisations. It should be noted tha the imaging tool handles each multi-polarised visibility record as one entity.

Output images are converted in the \textit{I,Q,U,V} stokes format or a subset of, depending on the polarisations available.

Visibility data is gridded using the natural weighting scheme (see section \ref{sec:weighting}). The required variance is read from the \textit{MeasurementSet}.

Flagging is also supported. During pre-processing phases of data, not handled by the imaging tool, some visibility records might be flagged for various reasons. These include erroneous readings. The imaging tool does not grid any such flagged data. Note that flagging is done per polarisation, and a flag has value of 1 when the respective polarised data is not to be gridded and 0 otherwise.

The final point in this section is about channel frequencies. Since, in the general case, the interferometer resides on the Earth, channel frequencies are normally given in the topocentric frame of reference. However, to make corrections for Doppler effects caused by the motion of the Earth, the imaging tool grids using the \textit{Local Standard of Rest Kinematic} (LSRK) frame of reference. Frequency values in this frame of reference are not given directly by the \textit{MeasurementSet}, so a conversion is required. This is taken care of by the \textit{Visibility Manager} using the CPU. It is the only calculation done in double-precision mode \cite{2008}.

\section{Runtime configuration and the \textit{Configuration Manager}}

Runtime configuration of the \textit{mt-imager} is done via command line arguments and configuration files. The configuration is a set of parameters defined by keyword and value. This is similar to the \textit{Common Astronomy Software Applications(CASA)} \cite{McMullin2007} and the standalone \textit{lwimager} tool based on the casacore libraries \cite{casacore}. The keyword and its respective value are separated by the equals (\textit{=}) character. For example, \textit{nx=100} defines a parameter named \textit{nx} with its value set to $100$. A type such as a string, boolean or integer is attributed to a parameter value. Boolean parameters can be set to true by putting, a minus (\textit{-}) just before the parameter name. For example, \textit{-test} and \textit{test=true} are equivalent.

Parameters can be set within a manually specified configuration file. In principle, all parameters can be specified as command line arguments to \textit{mt-imager}. This makes the command quite long, and subject to errors (an issue in \textit{lwimager}). The use of a configuration file solves the problem. It is a simple text file where parameters are each listed on a line of their own. Empty lines are allowed, and lines beginning with the number sign (\textit{\#}) are assumed as comments and ignored. Parameters set through the command line have precedence over those defined in the configuration file. This enables the user to have a default configuration stored in a  file and partially overridden by command line arguments.   

The \textit{mt-imager} uses a logging system based on \textit{Apache log4cxx} API \cite{log4cxx}. It requires an extra configuration with a format defined by the API. Its location is configurable through a parameter.

The \textit{Configuration Manager} is the holder of all configuration logic. It is a passive component as all the logic runs in the main thread. It does not interact with any components. It produces a component specific configuration that is passed by the main thread to the component during initialisation. Due to its nature, the \textit{Configuration Manager} is code-wise dependent on the other components. This dependency is one-way since the other components are neither aware nor dependent on the \textit{Configuration Manager}.

\section{The Convolution Function Generator}
\label{sec:confuncgen}
The \textit{Convolution Function Generator} is entrusted in calculating the \textit{w-projection} oversampled convolution functions $\tilde{C}_w(u,v)$\footnote{$\tilde{C}_w(u,v)$ is defined in section \ref{equ:C_W}.}. It is not practical to calculate the convolution functions during gridding, instead, they are numerically evaluated and presented to the \textit{WImager} component in an oversampled format. 

\subsection{General points}

This component can generate convolution functions in two modes: \textit{normal} and \textit{w-projection} mode. The mode  used is configurable at runtime.

In normal mode, the generator only evaluates $\tilde{C}_w(u,v)$ at $w=0$ and \textit{w-term} is ignored (refer to section \ref{sec:imagesynthesis}). The term "\textit{normal}" is used by other imaging tools such as CASA \cite{McMullin2007} and \textit{lwimager}. In these tools,  it describes the same behaviour as in the \textit{mt-imager}. In this mode, the support of the convolution function is runtime configurable. The system  will zero pad or trim the function so as to produce the desired support. This feature is useful for executing performance tests, and it was pivotal in many of the experiments described in chapter \ref{chap:results}.

In \textit{w-projection} mode, $\tilde{C}_w(u,v)$ is evaluated over various \textit{w-planes} depending on the runtime configuration. Each convolution function is trimmed to its support size. Support is evaluated after examining the generated data. As already discussed in section \ref{sec:w-projection}, support of $\tilde{C}_w(u,v)$ is a function of $w$ that increases with increasing $w$. 

In the two modes,  the choice of the tapering function is run-time configurable. Tapering functions are implemented as GAFW operators, and the choice is defined by specifying the name of the operator. In this thesis,  only one tapering function operator has been developed called \textit{casagridsf}. This implements the same prolate spheroidal function used in the casacore API \cite{casacore}. It is based on work presented by Schwab \cite{SchwabOct1980} and has been adapted to work over GPUs.

The oversampling factor is a run-time configurable variable. Since memory is limited it is suggested to keep it to a low value of 4, which is the value proposed by Cornwell \etal\ \cite{Cornwell2008}. This value is hard-coded in \textit{lwimager} and CASA. Zero-padding is used as an interpolation scheme. 

The generator samples in $w$. As recommended by Cornwell \etal\ \cite{Cornwell2008}, $w$ is scaled in $\sqrt{w}$ rather than linearly. No necessity exists to calculate for $w<0$  since the convolution functions are symmetrical around $w=0$. The maximum $w$ to consider is runtime configurable. 

\subsection{Mathematical treatment and outputs}
\label{sec:gen-out}

The \textit{Convolution Function Generator} outputs three GAFW \textit{results} that contain all convolution functions data required for \textit{WImager} to do its job. Some simple mathematical treatment is given here to help describe the content of the outputs. 

When a record is gridded, only the numerical data of one convolution function that falls on the pixels of the UV-grid is used. Any oversampled point that does not fall on the UV-grid is not considered. Define $M(w,u,v)$ as the function describing the operation that chooses data. $(u,v,w)$ are the baseline components of the record being gridded expressed in number of wavelengths. The function $M(w,u,v)$ returns the chosen data in a matrix with dimension $S_{w}\times S_{w}$, where $S_{w}$ is the support of the convolution function $\tilde{C}_{w}(u,v)$. Define the half-width $h_w=(S_w+1)/2$ and let $m_{i,j}(w,u,v)$ be an element of $M(w,u,v)$, where $i$ is the 1-based index of the row, and $j$ is the 1-based index of the column. Then:

\begin{equation}
\label{equ:choose}
m_{i,j}(w,u,v)=\tilde{C}_w((j-h_w)\Delta v - \delta v,(i-h_w)\Delta u - \delta u)  
\end{equation}     

where $\delta u$  and $\delta v$ satisfy the simultaneous  equations:
\begin{subequations}
\label{equ:sat}
\begin{equation}
-\frac{\Delta u}{2} < \delta u \leq \frac{\Delta u}{2}
\end{equation}
\begin{equation}
-\frac{\Delta v}{2} < \delta v \leq \frac{\Delta v}{2}
\end{equation}

\begin{equation} 
\delta u=u+ k_1 \Delta u  \quad k_1\in \mathbb{Z} 
\end{equation}
\begin{equation}
\delta v=v+ k_2 \Delta v  \quad k_2\in \mathbb{Z} 
\end{equation}
\end{subequations}

Note that the image\footnote{The image of an arbitrary function $f(X)$, is defined as the set $\{f(x): x \in X\}$.} of $M(w,u,v)$ is finite since $\tilde{C}_w(u,v)$ is oversampled in $u$ and $v$, and sampled in $w$.   

As defined in equation \ref{equ:normalizer}, the \textit{WImager} also calculates a normaliser. There  are substantial computational savings if the summation of the real parts of the elements of $M(w,u,v)$ are pre-calculated. Let $\zeta(w,u,v)$ define such summation such that:

\begin{equation}
\label{equ:sum}
\zeta(w,u,v)=\sum^{S_w}_{i=1} \sum^{S_w}_{j=1} \Re(m_{i,j}(w,u,v))
\end{equation}

where $\Re$ is the real operator.

The three outputs containing all data related to $\tilde{C}_w(u,v)$ are now explained. The first output is referred to as the \textit{Convolution Functions Numeric Data} output. It contains all numerical data of the oversampled $\tilde{C}_w(u,v)$. The data is laid down as follows: The data is first ordered by convolution function in increasing $w$. Further ordering for each convolution function is in such a way that elements of each matrix member of the image of $M(w,u,v)$ is coalesced in memory, sorted in the row-major form. This is in accordance with GPU best practices since when a record gets gridded, the convolution function data selected, is accessed in parallel. All matrix members of the image of $M(w,u,v)$ are ordered in increasing order by $\delta u$ and $\delta v$ . 

The second output is referred to as the \textit{Convolution Function Data Position and Support} output. It contains two elements for each convolution function. The first element is the function's support. The second is an index pointing to the first element of the first output that describe the convolution function. This index is vital to search for the  right data to use for gridding. 

The third output contains the image of $\zeta(w,u,v)$. The layout is similar to the layout of the first output.

\subsection{Data Generation}

A CPU/GPU hybrid algorithm is used to generate the three outputs described. CPU work is done using the imaging tool's main thread. The payload on the CPU is negligible, and the thread spends most of time waiting for some GAFW result to be available. The main thread is not available to spawn any new work while this component is calculating. This does not degrade performance since any new work to spawn after the generation of the convolution functions is dependent on the generated data. The \textit{Visibility Manager} component is initialised before the generator in such a way that convolution functions are generated by the \textit{Convolution Function Generator} component at the same time that the \textit{Visibility Manager} prepares data for gridding.

\setlength{\algomargin}{2em}
\begin{algorithm}
\SetAlgoLined
\KwData{\textit{w-planes} to consider}
\KwResult{The three outputs defined before}
\ForAll{w-planes}{
\textit{On GPU through the GAFW}

\Begin{
Calculate the sampled  $\mathcal{F}^{-1}(\tilde{C}_w(u,v))$ with zero padding\;
Apply FFT\;
Normalise the output as to get $\tilde{C}_w(u,v)$\;		
Find the support of $\tilde{C}_w(u,v)$\;
}

\textit{On CPU in the main thread}

\Begin{
Wait for support info to be available\;
Manually fill up the second output\;
}

\textit{On GPU through GAFW}

\Begin
{
Trim and Re-order numerical data of $\tilde{C}_w(u,v)$
}

}
\textit{On GPU through GAFW}

\Begin
{
Amalgamate all reordered $\tilde{C}_w(u,v)$ in one sequence as to finalise the first output\;
Calculate sums as to generate the third output\;
}

\caption[The \textit{Convolution Function Generator} component algorithm]{The algorithm used by the \textit{Convolution Function Generator} component to generate the three outputs.}
\label{algo:convgen} 
\end{algorithm}

The three outputs are generated using Algorithm \ref{algo:convgen}. All functions are normalised using a constant normaliser value such that $C_0(0,0)=1$. Support is calculated by detecting the pixel nearest to the edge that has a  magnitude larger than $10^{-3}$. The GPU thread configuration is set such that, for most of the steps, one thread handles one pixel.

\section{The \textit{WImager} Component}
\label{sec:Wimager}

The \textit{WImager} component executes the \textit{w-projection} algorithm (refer to section \ref{sec:w-projection}) over GPUs, through the General Array Framework.

An instance of the component handles one multi-polarised UV-grid. Since the imaging tool treats each channel separately, an instance of the \textit{WImager} component is created for every channel to grid. 

The component is designed to be flexible such that it can be used in other configurations. For example, channel data could be grouped by baseline and gridded over independent instances of the \textit{WImager} component. The grids would be added up on finalisation.  The component can be used as it is for \textit{w-stacking} (discussed in section \ref{sec:w-projection}) where each \textit{w-plane} is gridded separately using independent instances of the component.

The \textit{WImager} component is a GAFW module (refer to section \ref{sec:gafw-modules}). All inputs and outputs are GAFW \textit{result} objects (see section \ref{sec:GAFW:result}). The whole algorithm is implemented as one calculation tree and is fully GPU based. It is free from CPU calculations to ensure the asynchronous behaviour explained in section \ref{sec:highleveldesign}.

\begin{figure}[H]
\centerline{\includegraphics[scale=1, trim=1cm 1cm 1cm 1cm, clip=yes]{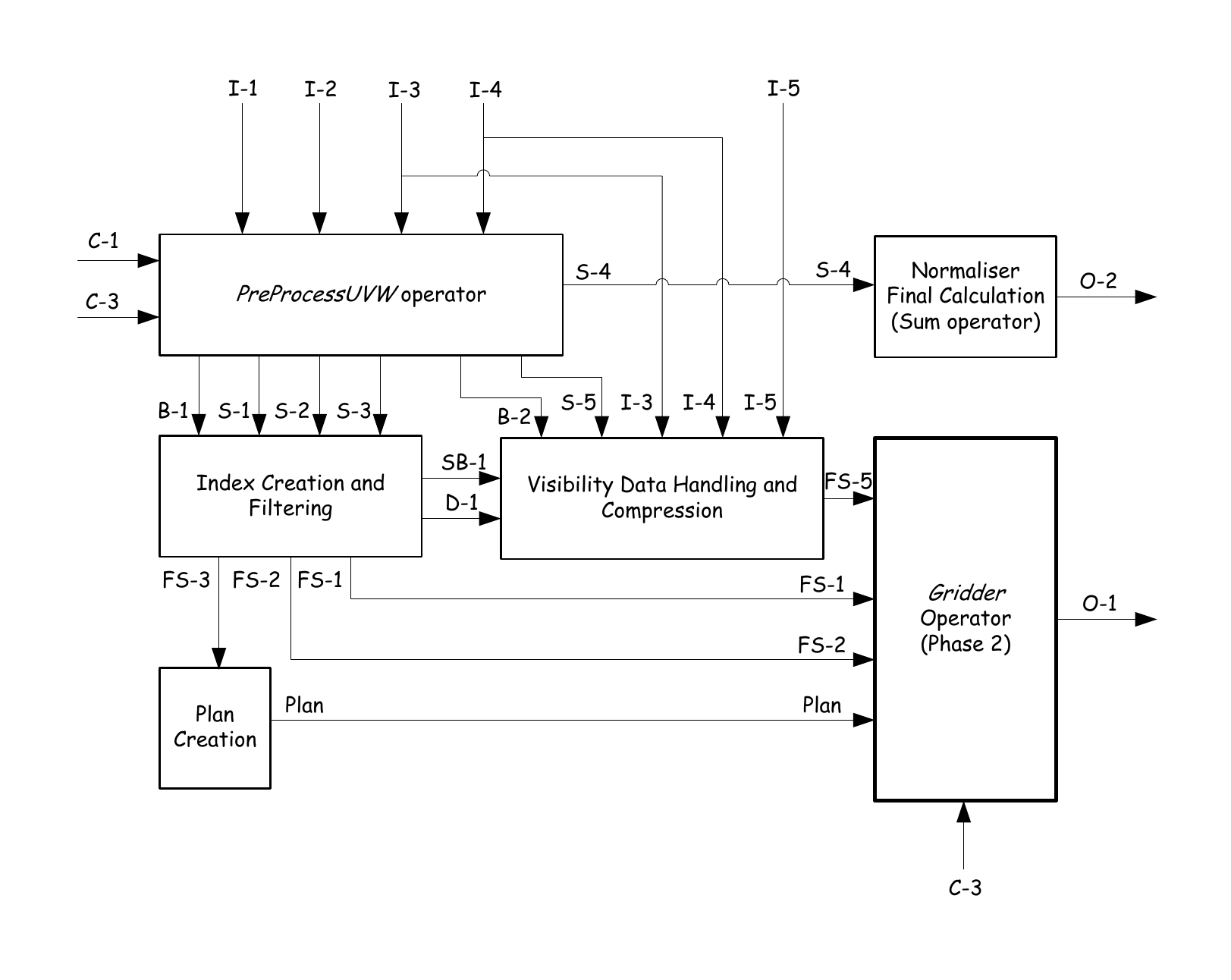}}
\caption{\textit{Calculation tree} of the \textit{WImager} algorithm }
\label{fig:wimageralgo}

\end{figure}

\begin{figure}[H]
\centerline{\includegraphics[scale=1,trim=1cm 1.2cm 1cm 1.5cm, clip=yes]{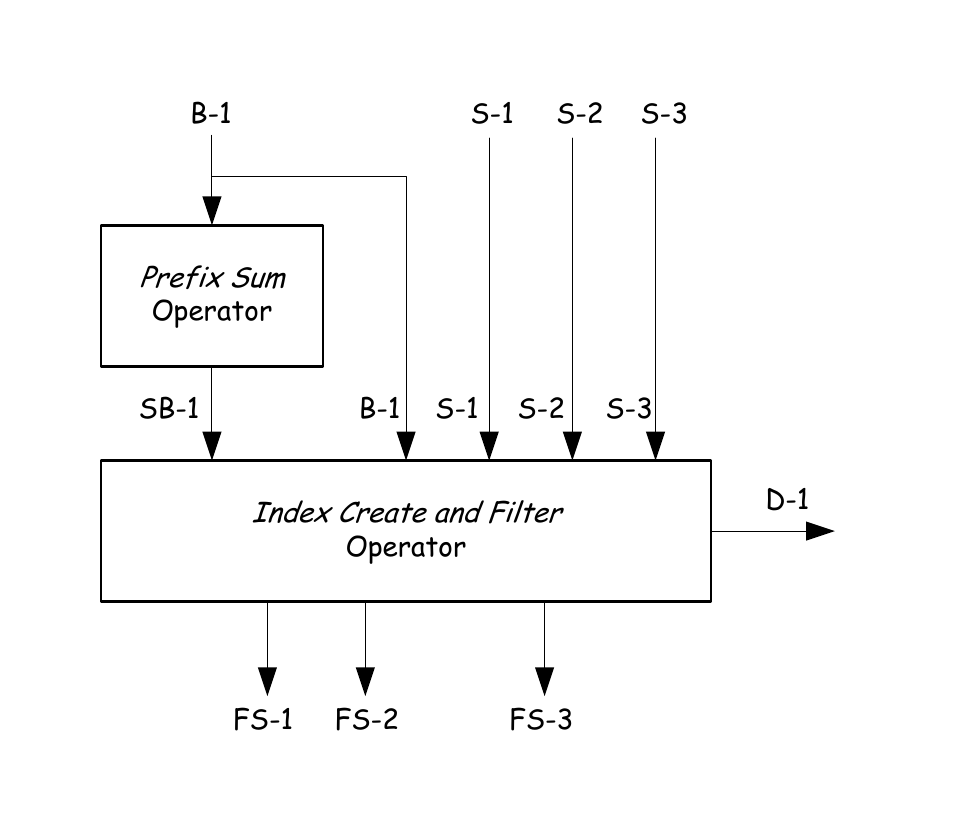}}
\caption[\textit{Calculation tree} of the \textit{Index Creation and Filtering} block]{\textit{Calculation tree} of the \textit{Index Creation and Filtering} block defined in Figure \ref{fig:wimageralgo}}
\label{fig:indexcreation}
\end{figure}

\begin{figure}[H]
\centerline{\includegraphics[scale=1,trim=1cm 1cm 1cm 1cm, clip=yes]{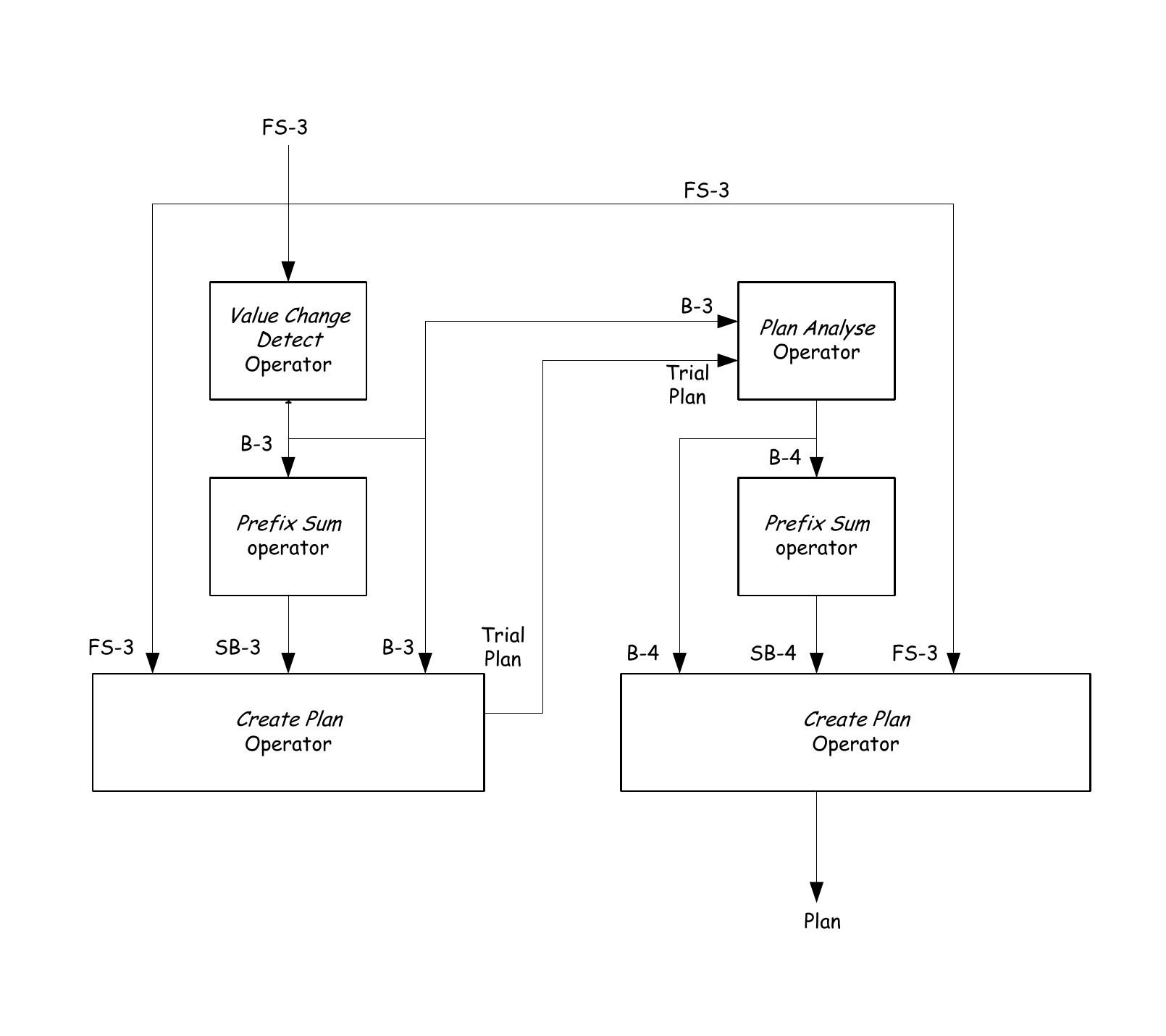}}
\caption[\textit{Calculation tree} of the \textit{Plan Creation} block]{\textit{Calculation tree} of the \textit{Plan Creation} block defined in Figure \ref{fig:wimageralgo}}
\label{fig:plancreation}
\end{figure}

\begin{figure}[H]
\centerline{\includegraphics[scale=1,trim=1cm 1cm 1cm 1cm, clip=yes]{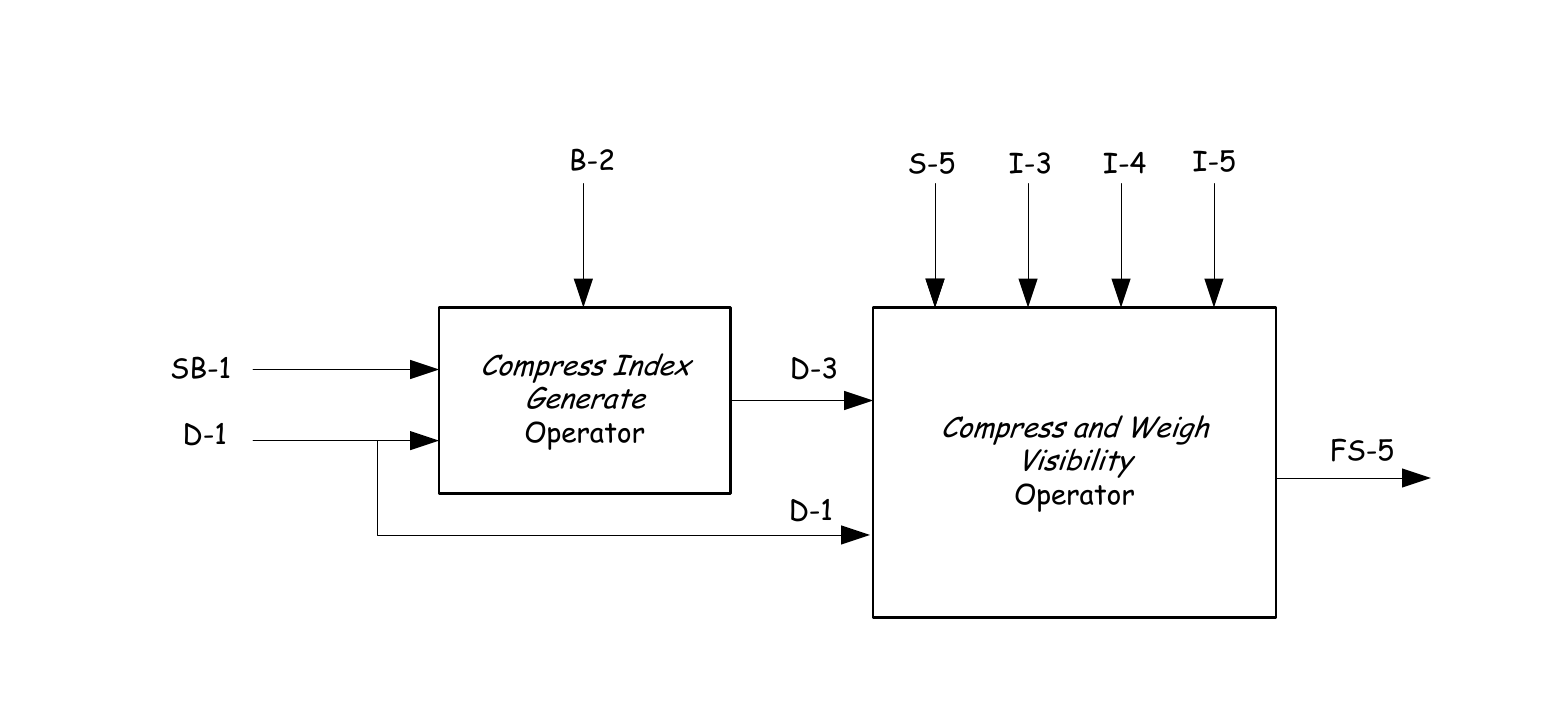}}
\caption[\textit{Calculation tree} of the \textit{Visibility Data Handling and Compression} block]{\textit{Calculation tree} of the \textit{Visibility Data Handling and Compression} block defined in Figure \ref{fig:wimageralgo}}
\label{fig:visibilityhandling}
\end{figure}

The implemented algorithm is, from here onwards, referred to as the \textit{WImager} algorithm. The calculation tree is depicted in Figure \ref{fig:wimageralgo}. Some GAFW operators are grouped up in one block and then expanded in subsequent Figures \ref{fig:indexcreation}, \ref{fig:plancreation} and \ref{fig:visibilityhandling}. Arrays are denoted with keys, tabulated in Table \ref{tab:legend}. The table includes the mathematical reference used to represent each sequence and a reference to the section, where the sequence is discussed.

The \textit{WImager} algorithm is based on Romein's algorithm, reviewed in section \ref{sec:romein}. It is enhanced and adapted to suit for the requirements of the \textit{mt-imager}. One of the main enhancements made over Romein's algorithm is \textit{compression}. It is discussed in section \ref{sec:compression}.

Records that are input to the \textit{WImager} algorithm are expected to be grouped by baseline and sorted by time. If this requirement is not respected, the algorithm will still work, but with a heavy penalty in performance. Note that there is no restriction on the amount of baselines in an input chunk of data and records per baseline can vary. Input visibility records do not need to be on the grid and flagging is supported. The algorithm adapts to the varying support of the convolution functions. It is perfectly fine that records in one input chunk require different sized convolution functions to be gridded. 

The \textit{WImager} algorithm is conveniently split up in two phases: the preparation phase and the gridding phase. The main difference between the two phases is the thread configuration. In the first phase,  each multi-polarised visibility record is processed by one thread or in some parts with a ratio of threads per record that goes below unity. In the gridding phase,  a record is processed by a block of threads. In this case,  the threads per record ratio goes in the order of thousands and millions and thus expensive. The gridding phase can hold all preparation phase logic but, for best performance, the gridder is stripped from any logic that can be implemented in the preparation phase. Note that the gridding phase is implemented in one GAFW \textit{operator} and thus all the other \textit{operators} in the calculation tree shown in  Figure \ref{fig:wimageralgo} are part of the preparation phase.

The rest of this section explains in detail the \textit{WImager} algorithm. The subject is tackled as follows: First, nomenclature and terminology used is defined and explained (section \ref{sec:terminology}). Inputs, outputs and mathematical equations describing the algorithm are given in section \ref{sec:wimager-inputs}. In the subsequent section, the gridding phase is discussed, whereby all logic that is delegated to the preparation phase is identified. The discussion on the \textit{WImager} component is concluded by explaining in detail the preparation phase in section \ref{sec:wimager-preparation}.


\begin{table}
  \centering
    \begin{tabular}{p{1.9cm}p{9cm}p{2.3cm}}
    \hline
	    
    \textit{Key} & \textit{Sequence name} & \textit{Mathematical Symbol} \\
    \hline
	\multicolumn{3}{c}{\textit{Visibility Records Inputs (defined in section \ref{sec:wimager-inputs}})}\\
	\hline
	I-1   & Baseline in meters & $\{(u'_i,v'_i,w'_i)\}$\\ 
	I-2  & Channel Frequency & $\{\lambda_i^{-1}\}$  \\ 
    I-3   & Weight & $\{\varepsilon_i\}_p$ \\  
    I-4   & Flags & $\{f_i\}_p$ \\ 
    I-5   & Visibility & $\{\vis_{i}\}_p$ \\
    \hline
	\multicolumn{3}{c}{\textit{Convolution Functions Input Data (defined in section \ref{sec:gen-out}) }}\\
	\hline
	C-1   & Convolution Functions Numeric Data & N/A  \\
	C-2   & Convolution Functions Data Position and Support & N/A \\ 
	C-3   & Convolution Functions Data Sum & N/A \\ 
	\hline
	\multicolumn{3}{c}{\textit{Outputs (defined in section \ref{sec:wimager-inputs})}}\\
	\hline
	O-1   & UV-Grid & $\mathcal{V}_{grid,p}$ \\ 
	O-2   & Normaliser & $N_p$ \\ 
	\hline
	\multicolumn{3}{c}{\textit{Binary Sequences (defined in section \ref{sec:wimager-preparation})}}\\
	\hline
	B-1   & Gridding Indicator & $\{g^b_i\}$ \\ 
	B-2   & Compression Indicator & $\{c^b_i\}$ \\ 
	B-3   & Support Change Indicator & $\{\chi^{bf}_i\}$ \\ 
	B-4   & Manipulated Support Change Indicator & N/A\\ 
	\hline
	\multicolumn{3}{c}{\textit{Scanned Sequences (defined in section \ref{sec:wimager-preparation})}}\\
	\hline
	SB-1 &  Prefix Sum of Gridding Indicator & $\{g^s_i\}$ \\ 
	SB-3   & Prefix Sum of Support Change Indicator & $\{\chi^{sf}_i\}$\\ 
	SB-4   & Prefix Sum of Manipulated Support Change Indicator & N/A\\
	\hline
	\multicolumn{3}{c}{\textit{Index Sequences (defined in section \ref{sec:wimager-preparation})}}\\
	\hline
	D-1  & Gridding Index & $\{g^d_i\}$ \\ 
	D-2  & Last Compressed Entry Index & $\{h_i^d\}$ \\
	Trial Plan & Trial Plan & \\
	Plan & Plan & \\  
	\hline
	\multicolumn{3}{c}{\textit{Non-Filtered Sequences (defined in section \ref{sec:wimager-preparation})}}\\
	\hline
	S-1  & Position & $\{a_i,b_i,c_i,d_i\}$  \\ 
	S-2  & Convolution Data Pointer & $\{z_i\}$ \\ 
	S-3  & Support & $\{s_i\}$ \\
	S-4  & Normaliser & $\{\varrho_i\}_p$ \\ 
	S-5  & Take Conjugate & $\{\psi_i\}$ \\ 
	\hline
	\multicolumn{3}{c}{\textit{Filtered Sequences (defined in section \ref{sec:wimager-preparation})}}\\
	\hline
	FS-1  & Position & $\{a^f_i,b^f_i,c^f_i,d^f_i\}$  \\
	FS-2  & Convolution Data Pointer & $\{z^f_i\}$ \\ 
	FS-3  & Support & $\{s^f_i\}$ \\ 
	FS-5  & Weighted and Compressed Visibilities & $\{\gamma^f_{i,p}\}$ \\
	\hline
    \end{tabular}%
    \caption{Legend for Figures \ref{fig:wimageralgo}, \ref{fig:indexcreation}, \ref{fig:plancreation}, \ref{fig:visibilityhandling}}
	\label{tab:legend}
\end{table}%
\FloatBarrier
\subsection{Nomenclature and Terminology}
\label{sec:terminology}
The various terminologies and nomenclatures used in this section are defined and explained. As pointed out in section \ref{sec:highleveldesign}, visibility records are input to the component in chunks. As indicated in Figure \ref{fig:wimageralgo} the record data chunk is split in different \textit{arrays}. It is reasonable to refer and represent GAFW arrays related to record data as mathematical sequences. 

The usual "curly brackets" format is used to define mathematical sequences (for example: $\{q_i\}$). It is to be assumed that the sequence is finite, and that the first element is at index 0, that is $\{q_i\}=\{q_0,q_1,...,q_{n-1}\}$. The letter $n$ implicitly defines the length of the sequence. 

As record data is processed by the \textit{WImager} algorithm, some of it is removed. The process of removal is referred to as filtering. Sequences that have records removed from are referred to as \textit{filtered sequences}. Notation-wise, filtered sequences are distinguished from their non-filter counterpart by having the letter $f$ as a superscript to the element. For example, $\{q^f_i\}$ is the filtered version of the sequence $\{q_i\}$.

An element in any sequence gives data about one record. Elements in different sequences that are either all non-filtered or all filtered sequences, describe the same record only if they are at the same position. 

The algorithm makes use of binary sequences. The term \textit{indicator} is used to refer to elements in such sequences. The letter $b$ is set as superscript of the sequence's element to show that the sequence is a binary sequence. For example, $\{g^b_i\}$ is a binary sequence. Note that if the binary sequence is also filtered, the letter $f$ is also retained as a superscript. For example,  $\{g^{bf}_i\}$ is a filtered binary sequence.   

Sometimes a prefix sum (exclusive scan) is run over a binary sequence to produce a \textit{scanned} sequence. The resulting scanned sequence is denoted by the letter $s$ set as superscript of the sequence's element. For example, $\{g^s_i\}$ is the resulting scanned sequence of $\{g^b_i\}$. If the binary sequence is a filtered one, the superscript letter $f$ is retained and the resultant filtered scanned sequence is denoted by $\{g^{sf}_i\}$. 

\textit{Index sequences} are defined as sequences containing integer index elements pointing to a record. These are generated by using binary sequences as predicates. \textit{Index sequences} are denoted by using the predicate sequence. The superscript letter $d$ is used to denote that the sequence is an \textit{index sequence}. For example, $\{g^d_i\}$ is the \textit{index sequence} generated using $\{g^b_i\}$ as predicate. If the \textit{index sequence} points to records in a filtered sequence, the superscript letter $f$ is also used (for example $\{g^{bf}_i\}$).         

To represent sequences made up of tuples, the usual notation of a comma separated list of sub-elements enclosed in a parenthesis, is used. For example, $\{(a_i, b_i, c_i, d_i)\}$ represent a finite sequence of 4-tuple elements. Most sequences are directly linked with the input data that define in full a visibility record.

Various sequences include polarisation dependent data. For such sequences, the letter $p$ will distinguish between each polarisation and is set as a subscript after the twisted brackets. For example, $\{v_i\}_p$ contains polarisation dependent data. When referring to an element, the notation $v_{i,p}$ will be used.

\subsection{Input, outputs and mathematical equations}
\label{sec:wimager-inputs}
The \textit{WImager} algorithm receives visibility record data in five inputs. Another three extra inputs contain data of the convolution functions in numerical form. These extra inputs are supplied by the \textit{Convolution Function Generator} and discussed in section \ref{sec:confuncgen}. There are two outputs for this algorithm, which are: a multi-polarised UV-grid and a normaliser. Details on the inputs and outputs of the algorithm follow.

The five visibility records related inputs are:

\begin{enumerate}
\item baseline in meters as projected on the UVW axis - $\{(u'_i,v'_i,w'_i)\}$. The apostrophe indicates that values are in meters and not in number of wavelengths.
\item channel frequency (per speed of light) - $\lbrace\lambda_i^{-1}\}$. where $\lambda_i$ represents the channel wavelength, and which is the inverse of the channel frequency divided by the speed of light. 
\item flags\footnote{Flagging was discussed in the penultimate paragraph of section \ref{sec:highleveldesign}.} - $\{f_i\}_p$
\item weight - $\{\varepsilon_i\}_p$
\item visibility - $\{\vis_{i}\}_p$
\end{enumerate}

The first output of the algorithm is the multi-polarised UV-grid $\mathcal{V}_{grid,p}(u,v)$ sampled at regular intervals $\Delta u$ and $\Delta v$. The pair of equations \ref{equ:gridding} gives a mathematical representation of how the algorithm calculates this output. It is based on the \textit{w-projection} gridding mathematical treatment given in sections \ref{sec:convgriddingmath} and \ref{sec:wprojmathtreatmnet} with some adaptations and added features.

\begin{subequations}
\label{equ:gridding}
\begin{equation}
\label{equ:gridding1}
Q_i(u,v)=\varepsilon_{i,p}(g_i+c_i)(1-f_{i,p}) \cdot v_{i,p} \cdot \tilde{C}_{\left({w'_i}/{\lambda_i}\right)}\left(u-\frac{u'_i}{\lambda_i},v-\frac{v'_i}{\lambda_i} \right)
\end{equation}
\begin{equation}
\label{equ:gridding2}
\mathcal{V}_{grid,p}(u,v)=\sum^{n-1}_{i=0}\left[ Q_i(u,v) \cdot \shah\left(\frac{u}{\Delta u},\frac{v}{\Delta v}\right)\right]
\end{equation}
\end{subequations}

$\tilde{C}_w(u,v)$ represents the oversampled w-projection convolution functions, defined in equation \ref{equ:C_W}.

$g_i$ is the gridding indicator. It is calculated within the algorithm and defines if a record is to be gridded by the gridding phase. When $g_i$ is set to 1, the respective record is gridded by the gridding phase while if set to 0 it is not. 

$g_i$ is set to 0 if and only if one of the following three conditions is met: The first condition is when gridding the record requires  updates to pixel outside the grid. The second condition is when all polarisations of the record are flagged (that is, $\forall p\ f_{i,p}=1$). The last condition which sets $g_i$ to 0 is when the record is compressed\footnote{Compression is discussed in section \ref{sec:compression}.}. 

When a record is compressed, the compression indicator $c_i$ is set to 1. Otherwise, the compression indicator is set to 0. Note that for a given record $g_i$ and $c_i$ cannot both be 1. Hence $(g_i+c_i)$ can have values of 0 or 1. When $(g_i+c_i)$  equates to 0, then the given record has no effect on the output grid. If $g_i=0$ and $c_i=1$ then the record will still be gridded and will affect the output, but not through the gridding phase. This is explained in detail in section \ref{sec:compression}.

It should be noted that when a flag $f_{i,p}$ is set to 1, $(1-f_{i,p})=0$ and the respective polarised visibility is ignored, since $Q_{i,p}(u,v)$ will acquire a value of 0 over the whole grid. The gridding phase will still grid the polarised visibility, but with no effect since it will grid zeros.

The second output of the \textit{WImager} is the polarisation dependent normalisers $N_p$. These are required by the \textit{Image Finaliser} component to output images in units of $Jy/beam$. The following equation gives a mathematical representation.

\begin{equation}
\label{equ:normalizer}
N_p=\sum^{n-1}_{i=0} \varepsilon_{i,p} (g_i+c_i)(1-f_{i,p})\cdot \zeta\left(\frac{w'_i}{\lambda_i},\frac{u'_i}{\lambda_i},\frac{v'_i}{\lambda_i}\right) 
\end{equation}
$\zeta(w,u,v)$ is defined in equation \ref{equ:sum} and its image is given in the \textit{Convolution Function Data Sum} input.

\subsection{Compression}
\label{sec:compression}

Romein's algorithm exploits the behaviour of the baseline trajectory, and in the \textit{WImager} algorithm the behaviour of the baseline trajectory is exploited further. There is a probability that for some consecutive records sampled on the trajectory, the change of $(u,v,w)$ is so small that the convolution function values and the position of gridding do not change. In mathematical terms there is a probability that for two consecutive records at position $i$ and $i+1$ the equality shown below holds:

\begin{dmath}
\tilde{C}_{\left({w'_i}/{\lambda_i}\right)}\left(u-\frac{u'_i}{\lambda_i},v-\frac{v'_i}{\lambda_i} \right)\cdot \shah\left(\frac{u}{\Delta u},\frac{v}{\Delta v} \right)=\tilde{C}_{\left({w'_{i+1}}/{\lambda_{i+1}}\right)}\left(u-\frac{u'_{i+1}}{\lambda_{i+1}},v-\frac{v'_{i+1}}{\lambda_{i+1}} \right)\cdot \shah\left(\frac{u}{\Delta u},\frac{v}{\Delta v} \right)
\label{equ:compression}
\end{dmath}

In this case by virtue of equation \ref{equ:gridding}, the weighted and flagged visibilities of the two entries, that is, $\varepsilon_{i,p} v_{i,p} (1-f_{i,p})$ and $\varepsilon_{i+1,p} v_{i+1,p} (1-f_{i+1,p})$, are summed together in the preparation phase and gridded as one record. This thesis terms the process as \textit{compression} in view that some records are "compressed" into one.  

The effectiveness of \textit{compression} is depended on many parameters. For example, \textit{compression} depends on the observation integration time, the grid configuration $(\Delta u,\Delta v)$, and the oversampling factor of the convolution functions. A short integration time causes the interferometer to sample records at a faster rate than longer integration times. This means that the Earth would have rotated less from one sampled record to another over a trajectory. Therefore, there will be smaller changes in $(u,v,w)$ which imply a higher probability of compression. Short baselines tend to have a shorter trajectory than long baselines. Within a given time interval, sampled records over short baselines travel shorter distances on the UV-grid than over longer baselines. This leads to a higher probability of compression for shorter baselines.

\subsection{The gridding phase}
\label{sec:gridphase}
The gridding phase does the final gridding. It is implemented by one operator named the \textit{gridder} and is based on Romein's algorithm. This section discusses implementation details, together with adaptations made to Romein's algorithm so as to suit the imaging tool. The discussion focuses on the required inputs that the preparation phase needs to generate for the \textit{gridder} to be as fast as possible.

\subsubsection{Filtering}

Referring to equation \ref{equ:gridding}, whenever $g_i=0$, then, the record is not gridded and is ignored by the gridding phase. The calculation for $g_i$ is delegated to the preparation phase. In order to evade any selection logic based on the value of $g_i$, the preparation phase filters out any records with $g_i=0$. All input sequences to the gridding phase, related to visibility records, are thus filtered sequences. 

As pointed out earlier, if a polarised visibility value is flagged, then, the gridder will grid zeros instead of ignoring the polarised value. The gridder does not handle flags and does not have any flag data as input. The zeroing of the visibility value is done in the preparation phase. Such a strategy evades logic in the gridding phase related to flagging and reduces the shared memory usage of a record in the gridder kernels. As will be explained later on, the gridder loads chunks of records in shared memory prior to gridding them. Shared memory is a limited resource, and there is an impact on performance when few records are loaded. Not loading flags reduces the record footprint in shared memory and thus increases the amount of records that can be loaded at one go.

\subsubsection{\textit{Convolution Position} sequence $\{(a_i^f,b_i^f,c_i^f,d_i^f)\}$}

Part of the calculation required to find the position of the convolution function on the UV-grid adheres well to the thread configuration of the preparation phase and is delegated to it. Instead of the $\{(u'_i,v'_i,w'_i)\}$ and $\{\lambda_i^{-1}\}$ sequences, the gridding phase is presented with a 4-integer tuple sequence $\{(a_i^f,b_i^f,c_i^f,d_i^f)\}$. The tuple, which proposed in Romein's algorithm, defines the position of the convolution function as follows: Each multi-polarised pixel on the grid is 0-based indexed by two integers $(y,x)$ representing the $v$ and $u$ value of the pixel respectively. The pixel with $v$ and $u$ values of 0 is in the middle of the grid and the pixel indexed as $(0,0)$ has the most negative $v$ and $u$ values\footnote{Note that this indexing scheme is the same as that of C++ indexing method of arrays.}.

If the record's position on the grid is on the pixel indexed by $(y_i,x_i)$ and the grid is $l$ pixels long on the u-axis then: 

\begin{subequations}
\begin{equation}
a_i=x_i-h_{w_i}+1
\end{equation}
\begin{equation}
b_i=(y_i-h_{w_i}+1) \times l + a_i
\end{equation}
\begin{equation}
c_i=-a_i \mod S_{w_i}
\end{equation}
\begin{equation}
d_i=-(y_i-h_{w_i}+1) \mod S_{w_i} 
\end{equation}
\end{subequations}

$h_w$ and $S_w$ are the half width and support of the convolution function that will be used to grid the record. They were defined in section \ref{sec:gen-out}

\subsubsection{\textit{Convolution Data Pointer} sequence $\{z_i\}$}

Similar to the generation of the \textit{Convolution Position} sequence, the logic that chooses the numeric data in the \textit{Convolution Functions Numeric Data} input is handled by the preparation phase. The \textit{gridder} is presented with an index pointing to the first element of the convolution function data that will be used to grid. Note that thanks to this sequence (the \textit{Convolution Data Pointer} sequence), the \textit{gridder} does not need the \textit{Convolution Functions Data Position} input.

\subsubsection{\textit{Weighted and Compressed Visibilities} sequence $\{\gamma^f_{i,p}\}$ }

The calculation of the weighted and flagged visibilities $(1-f_{i,p})\epsilon_i v_i$ is done in the preparation phase and the answer is presented in the \textit{Weighted and Compressed Visibilities} sequence. Each element in the sequence includes the compressed visibilities which are summed up with the record to be gridded.

\subsubsection{The Plan and local chunks}
\label{sec:localchunks}

Similar to Romain's algorithm a CUDA block grids a sub-chunk of the input data, which is here referred to as the \textit{local chunk}. It is first loaded on shared memory for fast access. The number of records in the \textit{local chunk} is let to vary. Whenever support changes, the \textit{gridder} needs to commit all the grid point updates to the global memory and reconfigure itself. It makes sense to constrain the \textit{local chunk} to entries of the same support, forcing it to be of a variable length. Note that there is a constraint on the maximum length since shared memory is a limited resource. This limit is dependent on the convolution function support that the \textit{local chunk} will grid. This is because records with small convolution functions are gridded using a different kernel than the other records. For performance reasons, the kernel gridding small convolution functions is configured with less shared memory than the other kernel.

Since \textit{local chunks} are variable in length, a plan is required to avoid fetching logic during the loading in shared memory. The plan is a sequence of \textit{local chunks} defined by the index of the first record and the common convolution function support.

\begin{equation}
Plan=\{(j^{df}_0,s_0^f),(j^{df}_1,s_1^f),...(j^{df}_{n-1},s^f_{n-1}),(j^{df}_n,0)\}
\end{equation}

where 
\begin{description}
\item[$Plan$] is the plan sequence
\item[$j^{df}_i$] is an index of the \textit{local chunk}'s first record in the respective filtered sequences.
\item[$s^f_i$] is the 1-dimensional support of the convolution functions that the chunk will use.
\item[$n$] is the number of chunks. 
\item The last element $(j^{df}_n,0)$ points to the record after last in the filtered sequences.
\end{description}
Note that the element of the $k^{th}$ \textit{local chunk} starts from the entry pointed out by $j^{df}_k$ and ends at $(j^{df}_{k+1}-1)$ in the filtered sequences.  

\subsubsection{Implementation details}

Gridding is done using two kernels. One kernel grids convolution functions with support less than $31\times 31$ while the second kernel handles larger convolution functions. Each kernel scans through the plan to see which \textit{local chunks} to grid. Expense of such a scan is insignificant as the memory consumed by plans is normally small. 

Two different kernels are used because small convolution functions need a different configuration to be gridded efficiently. This issue is also pointed out by Romein \cite{Romein2012}. The large convolution function kernel is configured with 1024 threads per block. Shared memory is configured to the maximum possible that lets the kernel run at $100\%$ theoretical occupancy. Configuring the small convolution function kernel with 1024 threads per block does not make any sense as the convolution functions are smaller in size than 1024 pixels. A configuration of 256 threads per block is therefore used instead. This reduces the theoretical occupancy and performance. Reducing shared memory usage (by reducing the maximum \textit{local chunk} length) makes up for the loss in performance. Note that the small convolution function kernel has been tuned so as to get the best performance in gridding $15\times 15$ convolution functions.

Different numbers of polarisations are handled by means of C++ templates. The mentioned kernels are implemented as C++ templates, and different specialisations are created to handle single-, dual- and quad-polarisations.

Convolution function data is loaded from textures similar to the test case presented in Romein's algorithm.
\subsection{The Preparation Phase}
\label{sec:wimager-preparation}
The main property of the preparation phase is that execution over GPU is configured with a thread to records ratio near unity. As shown in Figure \ref{fig:wimageralgo}, the preparation phase is divided into five blocks made up of one or a group of GAFW operators. In this section,  all the blocks and operators are discussed in detail.

\subsubsection{The \textit{PreProcessUVW} operator}

The \textit{PreProcessUVW} operator is the decision maker of the preparation phase. The operator decides whether a record is to be gridded or compressed, and is configured in such a way that one thread processes one record. The operator does not do any filtering of non-griddable data, and it does not handle any mathematics related to visibilities. It just instructs the preparation phase logic which are the records to filter out or compress by outputting binary sequences.

The operator prepares most of the inputs for the gridding phase. It does most of the work required to calculate the normaliser as per equation \ref{equ:normalizer} and performs other calculations.

Figure \ref{fig:wimageralgo} show all inputs and outputs of this operator and hereunder details of the operator are discussed by focusing on the outputs.          

\begin{description}
\item[\textit{Gridding Indicator} sequence $\lbrace g_i\rbrace$] \mbox{}\\This indicator is explained in section \ref{sec:gridphase}. It indicates if the record will be handled by the gridding phase or not. The respective element in the sequence is set to 1 if the record should be processed by the gridding phase. Otherwise it is set to 0. The preparation phase will subsequently filter out any records which have their respective gridding indicator set to 0. 

\item[\textit{Position} sequence $\{(a_i,b_i,c_i,d_i)\}$]\mbox{}\\ This sequence is explained in the gridding phase section \ref{sec:gridphase}, and defines the position on the grid were the convolution function will be added when gridding the respective record.

\item[\textit{Convolution Data Pointer} sequence $\{z_i\}$]\mbox{}\\ This sequence is explained in section \ref{sec:gridphase}. Elements of this sequence point to the first element in the \textit{Convolution Functions Numeric Data} input, that will be used as to grid the respective record.

\item[\textit{Compression Indicator} sequence $\{c_i\}$]\mbox{}\\This is a binary sequence indicating whether a record should be compressed with its predecessor. The element is set to 1 if the respective entry is to be compressed and 0 if not. $a_i$ (part of the \textit{Position} sequence)  and $z_i$ are checked if they are equal to $a_{i-1}$ and $z_{i-1}$ respectively to verify the compression equality defined in equation \ref{equ:compression}. The following equation gives details:
\begin{equation}
\label{equ:ci}
c_i=\left\lbrace \begin{array}{l l}
1 & \quad \text{if\ } i\ne 0\ \&\ z_i=z_{i-1}\ \&\ e_i=e_{i-1} \&\ \text{record is griddable}\\
0 & \quad \text{otherwise}
\end{array} \right.
\end{equation}

Note that a record is griddable if it does not require to update any grid points outside the grid boundaries, and at least there is one polarised visibility value that is not flagged.

\item[\textit{Take Conjugate} sequence $\{\psi_i\}$]\mbox{}\\This sequence instructs the visibility handling logic to take the conjugate of the complex visibility rather than the inputted one. The convolution functions are only calculated for $w_i \ge 0$, since symmetry exists around the plane $w=0$. It is deduced, from the measurement equation \ref{equ:measurmentrewritten}, that  $\mathcal{V}(u,v,w)$ is equal to the conjugate of $\mathcal{V}(-u,-v,-w)$, and thus all records with $w'_i< 0$ are gridded by considering a baseline of $(-u'_i,-v'_i,w'_i)$ and taking the conjugate of visibility. 

The value of each element in the sequence is set up according to the following equation:
\begin{equation}
\psi_i=\left\lbrace \begin{array}{l l}
1 & \quad \text{if\ } w'_i \ge 0 \\
-1 & \quad \text{otherwise}
\end{array} \right.
\label{equ:conj}
\end{equation}

The choice of values shown in equation \ref{equ:conj} makes the maths for handling visibility data easy as shown in equation \ref{equ:conjugatetake}. 

\item[\textit{Support} sequence $\{s_i\}$]\mbox{}\\This sequence stores the support of the selected convolution function, for gridding of the record. The elements of the sequence equate to: 
\begin{equation}
	s_i=S_{w'_i/\lambda_i}
\end{equation}
Data is available from the \textit{Convolution Functions Data Position and Support} input.

\item[\textit{Normaliser} Sequence $\{\varrho_i\}_p$]\mbox{}\\The \textit{PreProcessUVW} operator does most of the calculations required for the normaliser. Elements in this sequence are set to:

\begin{equation}
\varrho_{i,p} =\varepsilon_{i,p} (g_i+c_i)(1-f_{i,p})\cdot \zeta\left(\frac{w'_i}{\lambda_i},\frac{u'_i}{\lambda_i},\frac{v'_i}{\lambda_i}\right)
\end{equation}

To complete the calculation of the normaliser given in equation \ref{equ:normalizer}, the sequence is simply summed up.
\end{description}

\subsubsection{Index Creation and Filtering}

The \textit{Index Creation and Filtering} logic, filters out some of the record data generated by the \textit{PreProcessUVW} operator. All records that are not to be handled by the gridding phase (that is the records with $g_i=0$) are removed from the output sequences, and at the same time the \textit{Gridding Index} sequence is generated. The elements of the \textit{Gridding Index} sequence point to record data in the non-filtered sequences that will be processed by the gridding phase. This index is required for subsequent processing of visibility data in the preparation phase.

Figure \ref{fig:indexcreation} details the part of the calculation tree that filters various sequences\footnote{The sequences filtered are: \textit{Position} sequence, \textit{Convolution Data Pointer} sequence and \textit{Support} sequence.} and generates the index.  The \textit{Gridding Indicator} sequence acts as a predicate and an exclusive scan\footnote{Refer to section \ref{sec:all-prefix-sums} for operator details.} is run over it. The output sequence of the exclusive scan is represented by $\{g^s_i\}$  

\begin{eqnarray}
g^s_0 &=&0  \\
g^s_i &=&g^s_{i-1}+g_{i-1}=\sum_{i=0}^{n-1}g_i 
\end{eqnarray}

For any record with $g_i=1$  then $g^s_i$ gives the position of the record in the filtered sequences. Algorithm \ref{algo:index} is thus applied to create the index sequence and filtered sequences.

\setlength{\algomargin}{2em}
\begin{algorithm}
\SetAlgoLined
\KwData{\textit{Gridding Indicator} Sequence: $\{g_i\}$, The exclusive scan of $\{g_i\}$: $\{g^s_i\}$,  Non-filtered sequences: $\{\Theta_i\}$}
\KwResult{\textit{Gridding Index} Sequence: $\lbrace g^d_i\rbrace$, Filtered sequences: $\{\Theta_i^f\}$}
\ForAllP{elements in $\{g_i\}$}{
\If{$g_i$ is true}{$g^d_{g^s_i}=i$\; 
$\Theta_{g^s_i}^f=\Theta_i$\;}
\Else { do nothing\; }
}
\caption{Algorithm of the \textit{Index Create and Filter} operator}
\label{algo:index}
\end{algorithm}

The algorithm is implemented as one kernel in the \textit{Index Create and Filter Sequences} operator. The kernel is run with a thread per element of $\{g_i\}$ and does the filtering and index creation at one go. 

Note that what is here referred to as filtering is also known as \textit{stream compaction} and the algorithm described is a well known technique in the world of parallel computing. Some good reviews on this subject are found in  \cite{Nguyen2007,Horn2005}

\subsubsection{Plan Creation}

\textit{Plan creation} is depicted in Figure \ref{fig:plancreation}. The plan is a sort of index sequence as each element points to an element in the filtered sequences\footnote{The plan also contains the value of the support of the convolution functions to be used to grid the \textit{local chunk}.}. The same technique as Algorithm \ref{algo:index} is used to create the plan and is explained in the next paragraph. 

First, a binary sequence showing which elements need to be indexed in the plan is generated. This is done by the \textit{Value Change Detect} operator. Using a thread per element configuration, the filtered \textit{Support} Sequence $\{s^f_i\}$ is analysed for changes in support. Such a change delimits \textit{local chunks}. The results are saved in the \textit{Support Change Indicator} sequence $\{\chi^{bf}_i\}$ using the equation below: 

\begin{equation}
\chi^{bf}_i=\left\lbrace \begin{array}{l l}
1 & \quad \text{if\ } i=0\ \text{or\ } s^f_i\ne s^f_{i-1} \\
0 & \quad \text{otherwise}
\end{array} \right.
\end{equation}       

Running an exclusive scan over the \textit{Support Change Indicator} sequence and applying Algorithm \ref{algo:plan} generates the trial plan. Algorithm \ref{algo:plan} is implemented in the \textit{Plan Create} operator using the usual thread per element configuration.

\setlength{\algomargin}{2em}
\begin{algorithm}
\SetAlgoLined
\KwData{Filtered \textit{Support} Sequence: $\{s^f_i\}$,  \textit{Support Change Indicator} Sequence: $\{\chi^{bf}_i\}$, Exclusive scan of $\{\chi^{bf}_i\}$: $\{\chi^{sf}_i\}$ }
\KwResult{Plan $\{j^{df}_i,t^f_i\}$}
\ForAllP{Elements of $\{\chi^{bf}_i\}$}{
\If{$\chi^{bf}_i$ is true}{$j^d_{\chi^{sf}_i}=i$\; 
$t_i=s^f_i$\;}
\Else { do nothing\; }
}
\caption{Algorithm for the \textit{Plan Create} operator}
\label{algo:plan}
\end{algorithm}
    
The trial plan generated is not fully compliant with what is required by the gridding phase, since the length of a \textit{local chunk} cannot exceed a defined limit dependent on the support of the convolution function (refer to section \ref{sec:gridphase}). The trail plan generated is analysed by the \textit{Plan Analyse} operator that detects which \textit{local chunks} are too large. The \textit{Support Change Indicator} is manipulated, and some of its elements are set to 1 so as to split large chunks into smaller ones. The process of the plan creation is then re-executed to generate the final plan.

\subsubsection{Visibility Handling Logic}

The last input for the gridding phase to be generated by the preparation phase is the \textit{Weighted and Compressed Visibility} sequence $\{\gamma_{i}^f\}_p$. Two operators generate this input as shown in Figure \ref{fig:visibilityhandling}. The main calculations are done by \textit{Compress and Weigh Visibility} operator.

An index sequence called \textit{Last Compressed Entry Index} sequence$\{k^d_i\}$ is generated by the \textit{Compress Index Generate} operator. Elements in this index sequence point to the last record in the non-filtered sequences that are to be compressed with the record indexed by the Gridding Index sequence $\{g^d_i\}$. That is, $\forall i$ records from $g^d_i$ to $k^d_i$ are to be compressed together. If there are no records to be compressed to a record pointed to by $g^d_i$ then $k^d_i=g^d_i$.

Algorithm \ref{algo:compressindex} is used to generate the \textit{Last Compress Entry Index} sequence. It works as follows: The \textit{Compress Index Generate} operator inspects the non-filtered \textit{Compressed Indicator} sequence $\{c_i\}$ using a thread configuration of one thread per element. A record, that is last in a subsequence of records to be compressed together, is characterised by having the subsequent record not set for compression. The operator detects the position of such records and writes their position in the \textit{Last Compress Entry Index} sequence. The position where the index element is saved in the \textit{Last Compress Entry Index} sequence, is calculated by subtracting 1 from the respective element in the exclusive scan of the \textit{Gridding Indicator} sequence. Copying the \textit{Gridding Index} sequence on initialisation caters for those records with no subsequent records to compress.

\setlength{\algomargin}{2em}
\begin{algorithm}
\SetAlgoLined
\KwData{\textit{Gridding Index} sequence: $\{g^d_i\}$,  \textit{Compress Indicator} sequence: $\{c_i\}$, Exclusive scan of $\{g^d_i\}$: $\{g^s_i\}$}
\KwResult{\textit{Last Compressed Entry Index} sequence: $\{k^d_i\}$}

$\{k^d_i\}=\{g^d_i\}$

\ForAllP{ elements of $\{c_i\}$}{
\If{$c_i=1$}{
\If{$c_i$ is the last element in the sequence \textbf{or} $c_{i+1}=0$}{
	$h^d_{(g^s_i-1)}=i$\;
}
\Else { do nothing\; }
}
\Else { do nothing\; }
}
\caption{Algorithm of the \textit{Compress Index Generate} operator}
\label{algo:compressindex}
\end{algorithm}            

The \textit{Compress and Weigh Visibilities} operator generates the \textit{Filtered Weighted Visibility and Compressed} sequence $\{\gamma_{i}^f\}_p$ at one go, by means of the \textit{Last Compress Entry Index} sequence and the \textit{Gridding Index} Sequence. Assigning a thread per element in $\{\gamma_{i}^f\}_p$, equations \ref{equ:compress1} and \ref{equ:conjugatetake}, are executed to generate the output.
 
\begin{subequations}
\begin{equation}
\Re \gamma_{i,p}^f=\sum_{x=g^d_i}^{k^d_i}(1-f_{x,p})\varepsilon_{x,p}\Re \varphi_{x,p}
\label{equ:compress1}
\end{equation}
\begin{equation}
\Im \gamma_{i,p}^f=\sum_{x=g^d_i}^{k^d_i}\psi_x(1-f_{x,p})\varepsilon_{x,p}\Im \varphi_{x,p}
\label{equ:conjugatetake}
\end{equation}
\label{equ:viscal}
\end{subequations}

where $\Re$ and $\Im$ are operators returning the real and imaginary part respectively of a complex value. $\{\psi_i\}$ is the \textit{Take Conjugate} sequence. The sequences  $\{f_{i,p}\}$, $\{\varphi_{i,p}\}$ and $\{\varepsilon_{i,p}\}$ are the flags, visibility and weight sequences respectively and are all inputs of the \textit{WImager} algorithm.

\subsubsection{Statistics logic}
\label{sec:wimagerstats}

The \textit{WImager} component creates four different statistics about the gridding of data. They are all generated from the outputs of the \textit{PreProcessUVW} operator and are processed further by the \textit{Statistics Manager} component.  Following are the equations:

\begin{subequations}
\begin{equation}
\text{Total multi-polarised records gridded}=\sum_{i=0}^{n-1} g_i
\end{equation}
\begin{equation}
\text{Total multi-polarised records compressed}=\sum_{i=0}^{n-1} c_i
\end{equation}
\begin{equation}
\text{Total grid point updates executed by the gridder }=\sum_{i=0}^{n-1}\left( g_i\times {s_i}^2\right)
\end{equation}
\begin{equation}
\text{Total grid point updates saved because of compression}=\sum_{i=0}^{n-1}\left(c_i\times {s_i}^2 \right)
\end{equation}
\end{subequations}

A grid point update is defined as an update made to one point on the UV-grid. If a single polarised record is gridded with a $7 \times 7$ convolution function, 49 grid points are updated. Note that for the calculations of these statistics, the \textit{General Reduction} template operator is used (refer to section \ref{sec:operatorreduction}). GPUs' execution of the above equations is efficient, and the time consumed to generate such statistics is negligible.  

\section{The \textit{Visibility Manager} Component}
\label{sec:visibilitymanager}
The \textit{Visibility Manager} component does all the necessary work to prepare data for consumption by the \textit{WImager} component. The following are the main tasks entrusted to the \textit{Visibility Manager}:

\begin{enumerate}
\item Loading of measurement data from permanent storage.
\item Grouping of data by baseline. 
\item Conversion of channel frequencies from the topocentric frame of reference to LSRK.
\item Loading of data in the \textit{General Array Framework}.
\end{enumerate} 

The \textit{Visibility Manager} component does all its work on the CPU since its main purpose is to prepare data to be used by the GPU. It is CPU intensive, and by means of a multi-threaded approach, most of the resources of the modern multi-core CPU are used. All work is done in the background, as to ensure the asynchronous behaviour argued in section \ref{fig:highlevelmtimager}. 

\begin{figure}
\centerline{\includegraphics[scale=0.95]{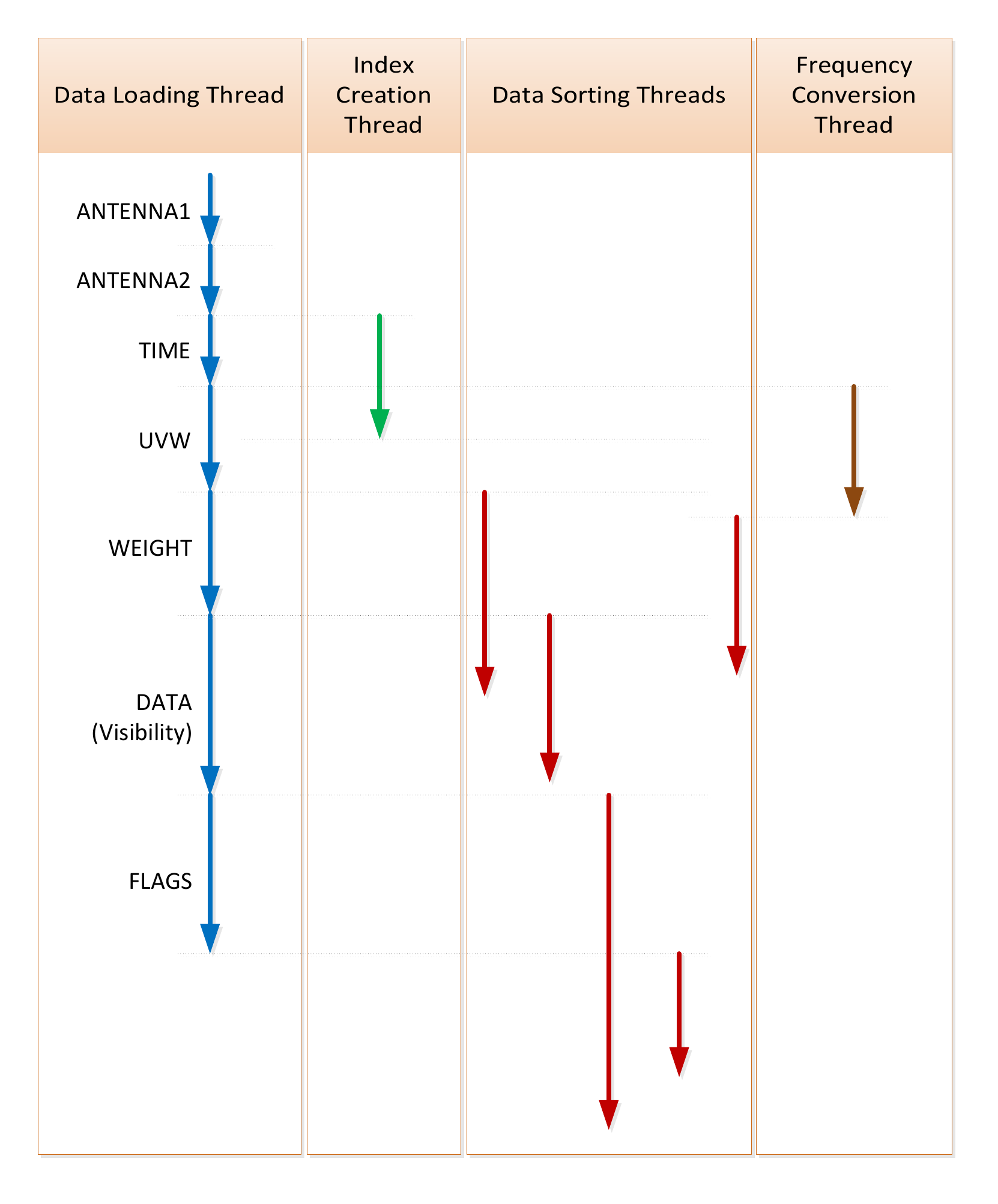}}
\caption[CPU thread concurrency of the \textit{Visibility Manager} ]{CPU thread concurrency of the \textit{Visibility Manager}. The blue thread loads data from a MeasurementSet one column at a time. The thread entrusted to create the sorting index waits until ANTENNA1 and ANTENNA2 are loaded from the storage device, and commences it's task. The frequency conversion thread waits until the TIME column is loaded. All sorting threads will wait until the data to be sorted is available, including the sorting index.}
\label{fig:vismultithread}
\end{figure}

Figure \ref{fig:vismultithread} depicts the way multi-threading has been designed to work for the \textit{Visibility Manager}. All work that is done by the \textit{Visibility Manager} is divided into simple tasks, and a thread is assigned to do one of these tasks. Only one thread is assigned for loading data from disk. A task is executed by a thread as soon as all required data  is present in memory. Note that there is also a dependency on the sorting index which is generated once the first two columns are loaded. In most cases,  the creation of the index is fast enough not to have threads waiting for the index after the necessary data is loaded from disk. Sorting of frequency is only commenced once conversion is ready. Synchronisation between threads is brought out using block waiting. In this way,  no CPU resources are wasted while a thread is waiting for all data to be available.

The \textit{Visibility Manager} is expected to load large amounts of data (in order of Gigabytes). The loading process is expected to take its time, and this has a direct effect on performance since the gridding process can only commence after all data is loaded in memory. The impact on performance is dependent on the I/O bandwidth of the storage device. It will be shown in chapter \ref{chap:results} that the loading phase is the main limiting factor of the \textit{mt-imager}. 

Having so many data to load directly implies that there is an equal amount of data to prepare. This takes time, but thanks to the multi-threaded design a substantial amount of preparation work is done in parallel with loading. This reduces the \textit{Visibility Manager}'s overall impact on performance. 

\subsection{Loading of Measurement Data from Permanent Storage}

The \textit{Visibility Manager} loads data from one \textit{MeasurementSet} \cite{Kemball2000} in the \textit{Casacore Table System} format \cite{Diepen2010}. It has been designed and coded in such a way that it can easily support multiple MeasurementSets and different formats such as HDF5 \cite{TheHDFGroup2000-2010}. Most of the C++ code, already assumes multiple MeasurementSets, but this feature was not finalised due to time constraints.   

The \textit{Visibility Manager} relies on the casacore \cite{casacore} ms API to load data from disk to memory. As already pointed out, only one thread is assigned to the loading process. This is because the casacore ms API is not well suited for multi-threading, and the underlying storage device might not perform well when accessed in parallel. The thread loads data column by column.

It is worthwhile to point out the "problems" that the \textit{Visibility Manager} has to cope with because of the casacore ms API and MeasurementSets. Most of the data is loaded in formats which need conversion prior to loading in the \textit{General Array Framework}. For example, complex visibilities are retrieved as C++ \textit{std:complex<float>} objects and GPUs do not understand such objects. They only know about simple structures of two consecutive floats. The raw data in a MeasurementSet is normally sorted in time but rarely (rather never) grouped by baseline. Channel frequency is given in the topocentric frame of reference, but the imaging tool requires it to be in the LSRK frame of reference. The conversion is variable over time, and its calculation is computationally intensive.

\subsection{Data sorting and preparation}
\label{sec:vissort}

A tailor-made solution has been implemented to group records by baseline, with records in each group sorted in time. The casacore ms API provides sorting functionality, but it was found to be too slow for the requirements of the imaging tool. The \textit{Visibility Manager} component loads data from a MeasurementSet, in the same order saved on the disk. In most cases,  it is already sorted in time, and the \textit{Visibility Manager} relies on this fact. If the MeasurementSet is not sorted in time, the imaging tool will still work, potentially with less performance.

All data sorting is done using a generated index. The generation is done by a purposely created thread that analyses ANTENNA1 and ANTENNA2 columns to group all data by baseline. Algorithm \ref{algo:sortindex} is used to generate the index. An amount of C++ vectors equal to the amount of possible baselines is set up. Each vector represents a baseline and is populated by index elements pointing to records of the baseline they represent. All the vectors are at the end amalgamated together serially to generate the final index. Note that auto-correlations (visibility measurements using the same antenna) are not gridded, and removed immediately. Once the sorting index is created, all other sorting threads can initiate their job. Most likely they will need to wait for the data that will be sorted, to be available in memory.

\setlength{\algomargin}{2em}
\begin{algorithm}
\SetAlgoLined
\KwData{ANTENNA1 and ANTENNA2 columns}
\KwResult{Sorting Index}
\Begin{
Create Array of C++ vectors equal to the number of baselines\;
\textbf{Define} integer i:=0\;
\ForAll{records in MeasurementSet}{
\If{record's ANTENNA1==ANTENNA2}{ ignore\;}
\Else{
	Push i in related baseline vector \;
}
	i++\;
}
Sorting Index=All vectors amalgamated together \;
}
\caption{Algorithm for the creation of sorting index}
\label{algo:sortindex}
\end{algorithm}

The work to do for sorting, conversion of data (except frequency conversions) and loading in the \textit{General Array Framework} are handled at one go by the sorting threads. A sorting thread works over one column and prepares data chunk by chunk. It processes the whole data for a channel and then proceeds to another channel. For single-GPU systems,  one sorting thread per column is assigned. In multi-GPU systems,  each column of data is handled by an amount of sorting threads equal to the number of GPUs available. Too many active sorting threads on the CPU can hinder performance and thus unnecessary sorting threads need to be avoided. The thread configuration is reasonably good since the imaging tool synthesises a channel in full over one GPU before proceeding to the next. If multiple channels are synthesised together over the same GPU, but on different grids, then it is quite likely that the \textit{General Array Framework} will be forced to cache out grids and re-load them when necessary. Caching is expected to occur since GPU memory is limited, and it is likely to be too small to hold many grids at the same time. Caching consumes time, so it has to be avoided. 
  
\subsection{Conversion of frequency}

In general, channel frequency is provided in the topocentric frame of reference while it is desirable to grid using frequency values in the LSRK frame of reference. The conversion of frequency from the topocentric to the LSRK frame of reference is a function of the time. The casacore measures \cite{casacore} API is used to make the conversion. Unfortunately, the API is not fully thread-safe, implying that channels cannot be processed in parallel. To top it all, the conversion is computationally expensive.

In order to have the best performance the conversion of frequencies is done as follows: One single thread is assigned to do the conversions for all channels. Channel frequencies for a given record are processed together. In this way, some computation is avoided since part of the conversion is only a function of time. As the thread loops over records and makes conversions, it checks if the previous record has the same time of observation and if so it does not go through the whole conversion process. Instead it copies the results obtained from the previous record. Records are processed in the same natural order they get loaded,  which is expected to be sorted in time, thus saving a lot of computational power.

Once conversion of frequency is ready, sorting threads do their job in the exact way as explained in section \ref{sec:vissort}.

\section{The \textit{Image Finaliser} Component}

The \textit{Image Finaliser} Component is a simple GAFW module that converts a multi-polarised UV-grid to a dirty multiple-Stokes intensity image. The output of this module is a finalised dirty image ready to be saved to disk by the \textit{Image Manager} (refer to section \ref{sec:imagemanager}). A separate instance of the \textit{Image Finaliser} is set up for each channel, similar to the \textit{WImager} component. The \textit{WImager}'s GAFW \textit{result} outputs are set as inputs to the \textit{Image Finaliser} component.  

The outputted stokes are dependent on the polarisations available. If the grid is quad-polarised, I,Q,U and V stokes are outputted. If the grid is dual-polarised or single polarised, only the stokes that can be calculated from the available polarisations are outputted. 

The algorithm implemented is GPU based using one GAFW \textit{calculation tree}. The step-by-step procedure is documented in Algorithm \ref{algo:finalizer}.

\setlength{\algomargin}{2em}
\begin{algorithm}
\SetAlgoLined
\KwData{multi-polarised UV-grid and normaliser value for each polarisation}
\KwResult{Multiple-Stokes dirty image }
\Begin{
Step 1: Inverse FFT of each polarised grid\;
Step 2: Convert to multiple-Stokes image and normalise\;
Step 3: Correct for the the tapering function as explained in sub section \ref{sec:convgriddingmath}\;
Step 4: Correct for convolution function sampling effects shown by equation \ref{equ:coversampled}\;
}
\caption{Algorithm used for image finalisation}
\label{algo:finalizer}
\end{algorithm}

The implementation of Algorithm \ref{algo:finalizer} is simple. Step 1 is done using the FFT operator discussed in section \ref{sec:FFTOperator}. Steps 2-4 are all element by element operations and thus a thread is assigned to each element.

\section{The \textit{Image Manager} component}
\label{sec:imagemanager}

The \textit{Image Manager} component is responsible to write the resultant dirty images to a FITS \cite{Pence2010} file whose location is configurable.  All multiple-Stokes dirty images are saved in a primary 4-dimensional image cube using the world coordinate system (WCS) \cite{Greisen2009,Calabretta2002,Greisen2002}.

The axes, saved in ascending order are: right ascension, declination, stokes polarisation and channel frequency. The output is compatible with other popular imaging tools like \textit{lwimager} and CASA. The CFITSIO API \cite{Pence1999} together with casacore \cite{casacore} are used to generate the axes.

As per the general design of the imaging tool, all logic is executed in the background. All work is done in one thread. The thread begins its execution by creating the FITS file and then block waits over GAFW \textit{result} objects until data is available. Channel images are saved one-by-one and in order, as soon as the image data is available. This strategy ensures that most of the writing to disk of a channel image is done in parallel with the gridding over the GPU of the next channel. This hides most of the time consumed in writing images to disk. In order to evade possible I/O contention between the \textit{Visibility Manager} and the \textit{Image Manager}, the latter is only initialised once the finalisation of the first channel has been requested to the GAFW. The finalisation request can only happen after loading of measurement data from disk has ended.

\section{The \textit{Statistics Manager} component}

The \textit{Statistics Manager} is entrusted in handling all statistics generated by the imaging tool and to save such data in CSV (Comma-separated values) files. 

The imaging tool generates statistics of various types. The \textit{Visibility Manager} measures timings of the various activities it does. This includes the time an activity started and when it ended. It uses the \textit{Boost.Chrono} API \cite{boostchrono} to do so. Other components and the \textit{main()} function also measure the time taken for the execution of some of their activities. The \textit{WImager} component generates the statistics listed in section \ref{sec:wimagerstats} while the GAFW engine measures the operators' execution time over GPU using CUDA events. 

All generated statistics are processed by the component and saved in three different CSV files. The first file contains all statistics generated by the engine. The second file contains all timings generated by the imaging tool's components and the \textit{main()} function. The third file contains performance data related to gridding. Note that these three files have been vital in extracting performance measurements for the results reported in chapter \ref{chap:results}.

The first and second files are straight forward to generate. The component just adds a new row to the file once a particular related statistic is received. More challenging is the generation of the third file since each row is generated from many statistics. The \textit{Statistics Manager} saves in memory any statistic that it needs for the third file and waits until it has a complete set to generate a row in the third file.

The statistics system of the imaging tool is built over the GAFW statistics system discussed in section \ref{sec:gafw-statistics}. All statistics are objects whose class inherit a general base class defined by the GAFW. The \textit{Statistics Manager} implements the interface defined by the GAFW, through which statistics are pushed to the component via a thread-safe FIFO queue. The \textit{Statistics Manager} does all work in a single background thread, and the FIFO queue has block waiting capabilities such that the thread can wait over the queue when it has no work to do.
 
\chapter{Performance analysis}
\label{chap:results}

This chapter reports performance results obtained by \textit{mt-imager}. The \textit{WImager} algorithm (refer to section \ref{sec:Wimager}) is analysed in depth since it is the tool's strong suit. Analyses on the implemented Romein's algorithm, builds up knowledge on the algorithm with data not published by Romein \cite{Romein2012}. Other analysis focus on the overall behaviour of the imaging tool whereby limitations are identified. A comparative study with a standard CPU based imaging tool known as \textit{lwimager} is presented. It measures the performance advantage of the \textit{mt-imager} over \textit{lwimager} to show that the thesis main objective is achieved.

\section{Hardware used}
\label{sec:hardware}

All experiments are done over a high-end desktop PC. Specifications are listed in Table \ref{tab:specs}.:

\begin{table}[htbp]
  \centering
    \begin{tabularx}{\textwidth}{ ll }
    \hline
    \textit{Component}  & \textit{Specification} \\
    \hline
    Mother Board & MSI Z77 MPower \\
    CPU   & Intel(R) Core(TM) i7-3770K CPU @ 3.50GHz \\
    GPU   & MSI GeForce GTX 680 \\
    Memory & DDR3 32gb / 1600mhz Corsair Vengeance [4x8GB] \\
    Storage 1  & HD 3,5" SATAIII  1TB SEAGATE ST1000DM003 7200rpm 64MB\\
	Storage 2 & OCZ RevoDrive 3 X2 PCI-Express SSD 240 GB\\
    OS & Ubuntu 12.04.1 LTS \\
    \hline
    \end{tabularx}%
    \caption{Hardware and software specifications of the PC used for analysis}
    \label{tab:specs}
\end{table}%
\FloatBarrier

This PC was the best option within the budget available at the time of purchase. Note that a new series of GeForce cards have been recently released. 

For experiments that required two GPUs, an extra VEVO GeForce GTX 670 is used. It was the best GPU available at the University, at the time of testing. It performs less than the GTX 680 but still suitable for the analysis presented in this chapter.

The second storage device is a fast PCIe Solid State Device (SSD). The operating system and \textit{mt-imager} binaries are stored on the SATA magnetic storage device while the Data Sets are stored on the SSD card. As it will be evident from the reported results, the loading of data from storage is a limiting factor for performance, and thus the use of the SSD device has considerable weight on the results obtained.

\section{Data Sets}

Three data sets are used for analysis. Details are tabulated in Table \ref{tab:datasets}. 

\begin{table}[htbp]
  \centering
    \begin{tabular}{llll}
    \hline
     \textit{Data Set Number} & \textit{Data Set 0} & \textit{Data Set 1} & \textit{Data Set 2} \\
    \hline
    Telescope & LOFAR & LOFAR & LOFAR \\
    Integration time &   10    & 10.139 & 3.00024 \\
    Size on disk & N/A & 1.3 Gigabytes & 5.3 Gigabytes \\  
    Number of channels & 1 or 16 & 1     & 16 \\
    Records per channel & 2138400 & 6578850 & 6501600 \\
    Polarisations per record & 1,2,4 & 4 &4 \\
    Data format & C structure & \textit{MeasurementSet} & \textit{MeasurementSet} \\
    \hline
    \end{tabular}%
    \caption{Data sets used for analysis}
    \label{tab:datasets}
\end{table}%

In all experiments, all channels are imaged. Some data is also flagged, and some records are out of the grid. This gives a realistic scenario to these experiments.

Data Set 1 and Data Set 2 have been carefully selected such that the overall performance of the \textit{mt-imager} for different scenarios  can be analysed in depth. Data Set 2 is approximately 4$\times$ larger in size then Data Set 1, so the imaging tool requires to load much more data, prior to gridding. Two data sets contain nearly the same amount of record per channel but Data Set 2 has 16 channels while Data Set 1 contains only 1 channel. This means that there is much more records to grid for Data Set 2 than for Data Set 1. In conclusion, Data Set 2 is much more difficult to image than Data Set 1. 

Data Set 0 is the same data set used in the analysis presented by Romein \cite{Romein2012}. It is used as a point of departure for the analysis whereby an environment is created to be as similar as possible to Romein's test scenario (that is, the test case presented in \cite{Romein2012}). Some differences do exist since the \textit{mt-imager} is implemented using different assumptions. 

Since Data Set 0 is stored in a C structure, a \textit{special Visibility Manager} has been developed to cater for it. Only $(u,v,w)$ data and the channel topocentric frequencies are supplied. Time of observation is not included, denying the possibility of a frequency conversion. Auto-correlations (that is records resulting from the correlation of the same signal from the same antenna) are flagged. The \textit{special Visibility Manager} sets visibilities to unity and mimics scenarios of 1, 2 or 4 polarisations. In Romein's test scenario, channel data was interleaved and all records gridded on one grid. This feature has been implemented in the \textit{special Visibility Manager} but not in the proper \textit{Visibility Manager}. Here, it is referred to as \textit{channel interleaving}.

\section{Performance measurement metrics}

In this chapter, performance is measured using different metrics discussed in the following subsections:

\subsection{Gridding rate \textit{(Giga grid point updates per second)} } 
\label{sec:updatemetric}
This metric is used to describe performance of the \textit{WImager} gridding phase. The unit G shall stand for \textit{Giga grid point updates per second}. 

A grid point update is defined as an update made to one point on the UV-grid. If a single polarised record is gridded with a $7\times7$ convolution function, the gridder updates 49 grid points. The metric, takes the number of polarisations into account and a quad-polarised record gridded using a $7\times7$ convolution function updates $49\times4=196$ grid points. The time interval considered is the total execution time of the gridder kernels. It is measured by the \textit{General Array Framework} engine using CUDA events. 

The metric can be measured in two ways to get the total rate and/or the real rate. For the total rate, compressed records are included in the metric as if they were gridded independently. For the real rate, only genuinely gridded records are considered. In either case, records not gridded because of reasons beyond \textit{compression} are dismissed.

\subsection{Record gridding rate \textit{(Mega records/second)}} 

This is similar to the previous metric. Instead of counting grid point updates, multi-polarised records are considered. Polarisation is not factored in, and a gridded multi-polarised record is counted as one. Similarly to the previous metric, this metric can be measured as a total rate or real rate.

\subsection{Compression ratio} 

This ratio is given in terms of either records or grid point updates. It defines the ratio of compressed records or grid point updates against genuinely gridded records or grid point updates. A value of 0 indicates that no records were compressed.

\subsection{Record preparation rate (\textit{Mega records/second})}

This metric is used to quantify the performance of the \textit{WImager} preparation phase. This is the total number of records presented to the WImager component against the total execution time on the GPU of all preparation phase kernels. The execution time is measured by the \textit{General Array Framework} engine using CUDA events. 

\section{Experiments}
This section describes all the experiments reported in this chapter. Results are analysed in the next section. 

Some experiments are intended to analyse the performance of the \textit{WImager} component. The system is configured to work in \textit{normal} mode whereby only one \textit{w-plane} is considered (that is one convolution function is used over the whole $w$ range). Experiments are repeated several times using the following differently sized convolution functions (in pixels):\footnote{Note that performance data for convolution functions of size $7\times7$ and $511\times511$ pixels have not been published by Romein \cite{Romein2012}} $7\times7$, $15\times15$, $23\times23$, $31\times31$, $63\times63$, $95\times95$, $127\times127$, $255\times255$, $511\times511$. Proper \textit{w-projection} is only enabled when examining the overall performance.

Experiments are divided into separate batches. Each batch is numbered and each experiment is referred to using its batch number and another number. The two numbers are divided using a dot (.) such as $1.1$.

Unless otherwise stated, it is to be assumed that \textit{channel interleaving} is disabled and \textit{compression} enabled. It is also to be assumed that quad-polarisation is used when imaging Data Set 0.

All experiments presented is this chapter have square dimensions and the intensity image pixel length and width are always equal.

The following subsections describe all experiments batch by batch.

\subsection{Experiment batch 1: \textit{Compression}}
This batch of experiments is intended to comprehend the behaviour of \textit{compression}. Data Set 0 is used. Channel interleaving and \textit{compression} are enabled or disabled as necessary. The \textit{special Visibility Manager} outputs quad-polarised records. The image dimensions and pixel length are set to the same values as Romein's test scenario. Table \ref{tab:batch1} describes each experiment. The first experiment sets the environment nearest to Romein's test scenario.

\begin{table}[htbp]
  \centering
    \begin{tabularx}{\textwidth}{ P{1.6cm}XXXXP{2.4cm} }
  
    \hline
    \textit{Ref No} & \textit{Image dimensions (pixels)} & \textit{Pixel length (arcsec)} & \textit{Over- sampling factor} & \textit{Channel interleaving} & \textit{Compression} \\
    \hline
    1.1 & $2048\times2048$ & 14.5711 & 8 & enabled & disabled \\
    1.2 & $2048\times2048$ & 14.5711 & 8 & disabled & disabled \\
    1.3 & $2048\times2048$ & 14.5711 & 8 & enabled & enabled \\
    1.4 & $2048\times2048$ & 14.5711 & 8 & disabled & enabled \\
    \hline
    \end{tabularx}%
    \caption{Experiment details of batch 1}
    \label{tab:batch1}
\end{table}%

\FloatBarrier

\begin{figure}[H]
   \centerline{\includegraphics[scale=1]{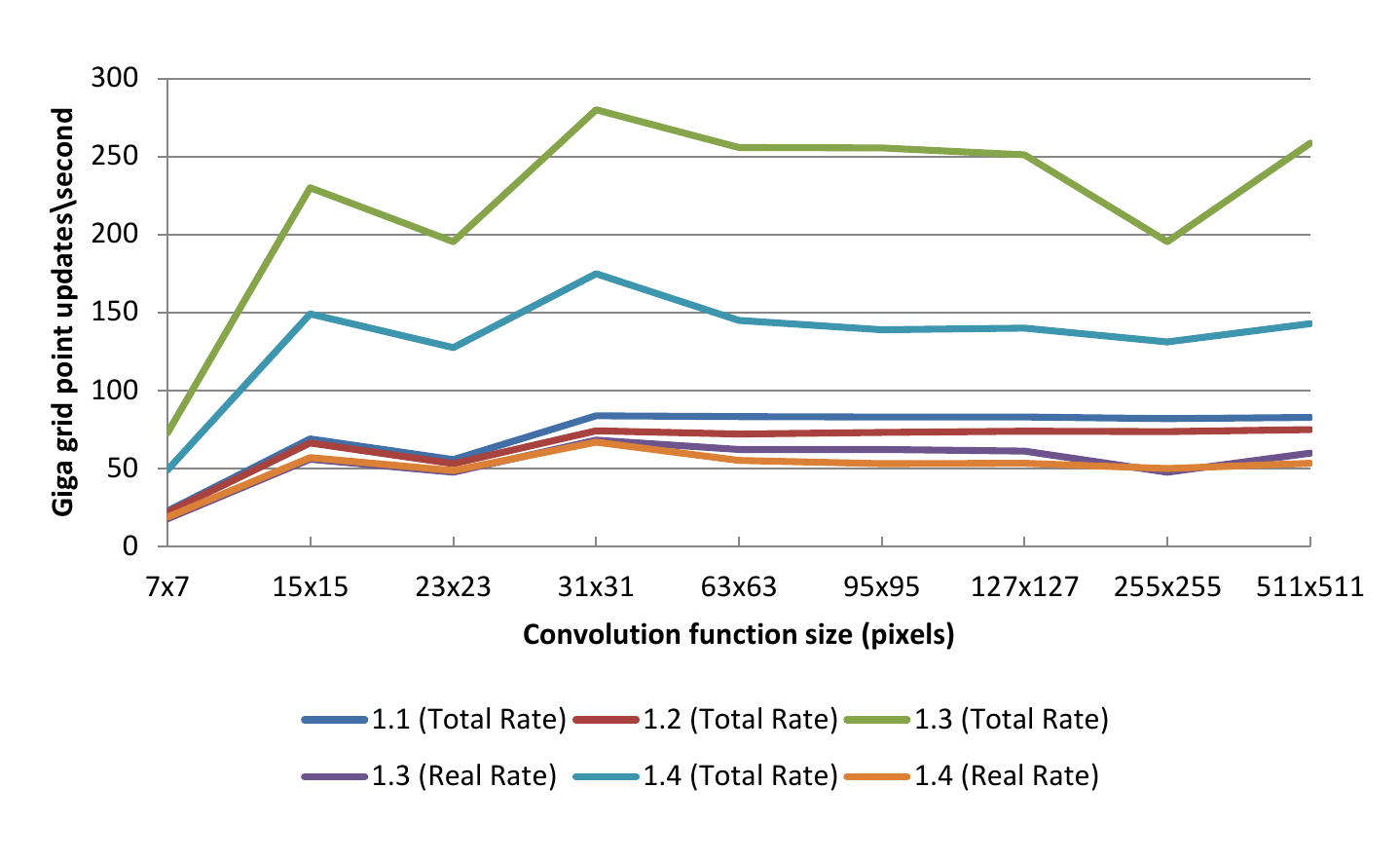}}
   \caption[Real and total gridding rate results for experiment batch 1]{Real and total gridding rate results for experiment batch 1. The numbers in the legend refer to specific experiments as tabulate in Table \ref{tab:batch1}.}
\label{fig:res-batch1}
\end{figure}

Gridding Rates are reported in Figure \ref{fig:res-batch1}.

\FloatBarrier

\subsection{Experiment batch no 2: Effects of UV-grid pixel length and width}

This batch of experiments is intended to study how the gridder real performance is affected on varying the UV-grid sampling interval. This is achieved by changing the image dimensions. Details are listed in Table \ref{tab:batch2}.
 
\begin{table}[htbp]
  \centering
    \begin{tabularx}{\textwidth}{P{1.6cm}LLLL }
    \hline
    \textit{Ref No} & \textit{Data Set No} & \textit{Image dimensions (pixels)} & \textit{Pixel length (arcsec)} & \textit{Oversampling factor} \\
    \hline
    2.1   & 0     & $4096\times4096$ & 14.5711 & 8 \\
    2.2   & 0     & $2048\times2048$ & 14.5711 & 8 \\
    2.3   & 0     & $1024\times1024$ & 14.5711 & 8 \\
    2.4   & 1     & $4096\times4096$ & 14.5711 & 8 \\
    2.5   & 1     & $2048\times2048$ & 14.5711 & 8 \\
    2.6   & 1     & $1024\times1024$ & 14.5711 & 8 \\
    2.7   & 2     & $4096\times4096$ & 14.5711 & 8 \\
    2.8   & 2     & $2048\times2048$ & 14.5711 & 8 \\
    2.9   & 2     & $1024\times1024$ & 14.5711 & 8 \\
    \hline
    \end{tabularx}%
  \caption{Experiment details of batch 2}
    \label{tab:batch2}
\end{table}
\FloatBarrier

\begin{figure}[H]
    \centerline{\includegraphics[scale=1]{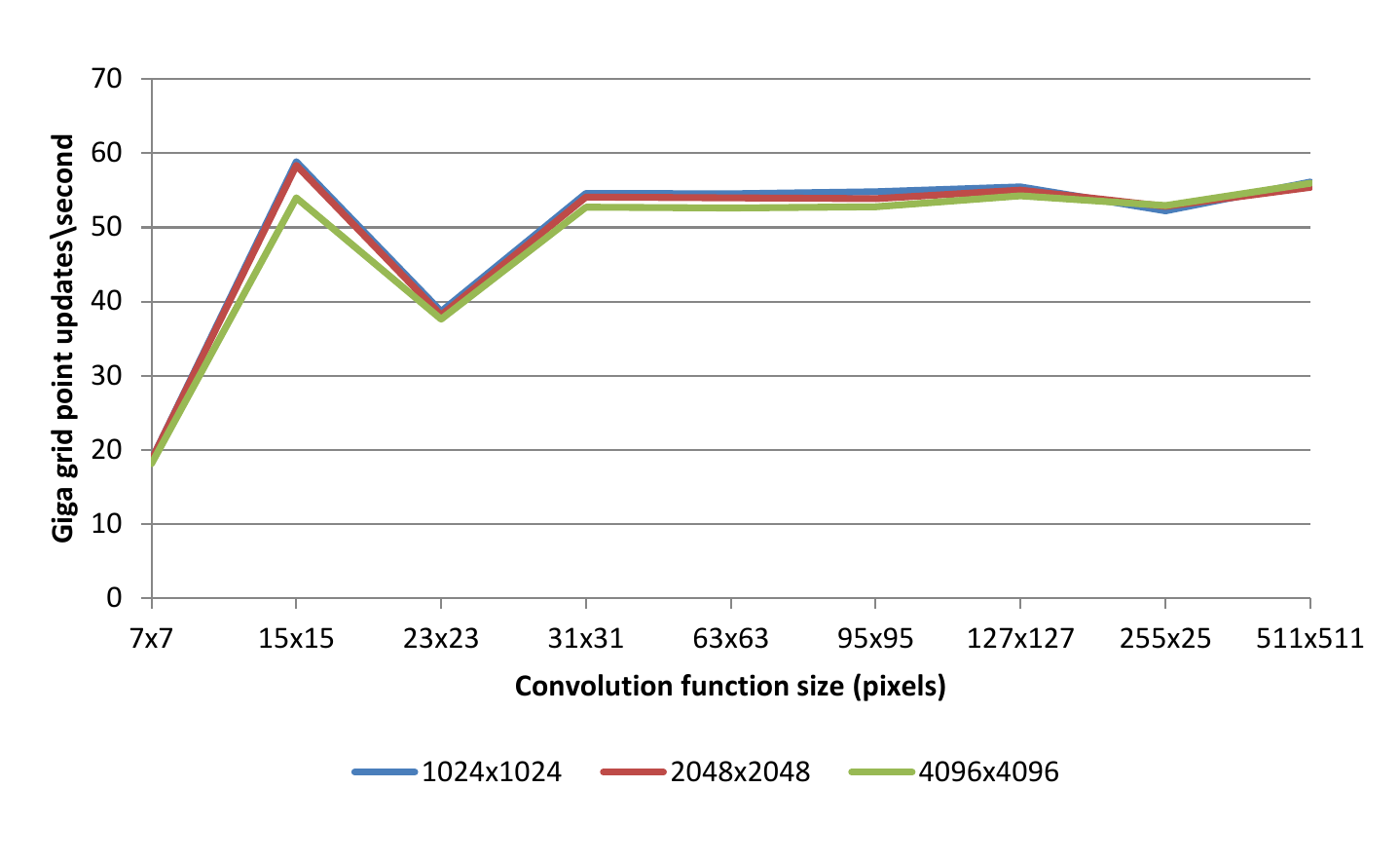}}
    \caption[Real gridding rate results of experiment batch 2 (First Data Set)]{Real gridding rate results per convolution function size for experiments 2.1 to 2.3 (Data Set 0)}
\label{fig:res-deltau-set0}
\end{figure}

\begin{figure}[H]
    \centerline{\includegraphics[scale=0.995]{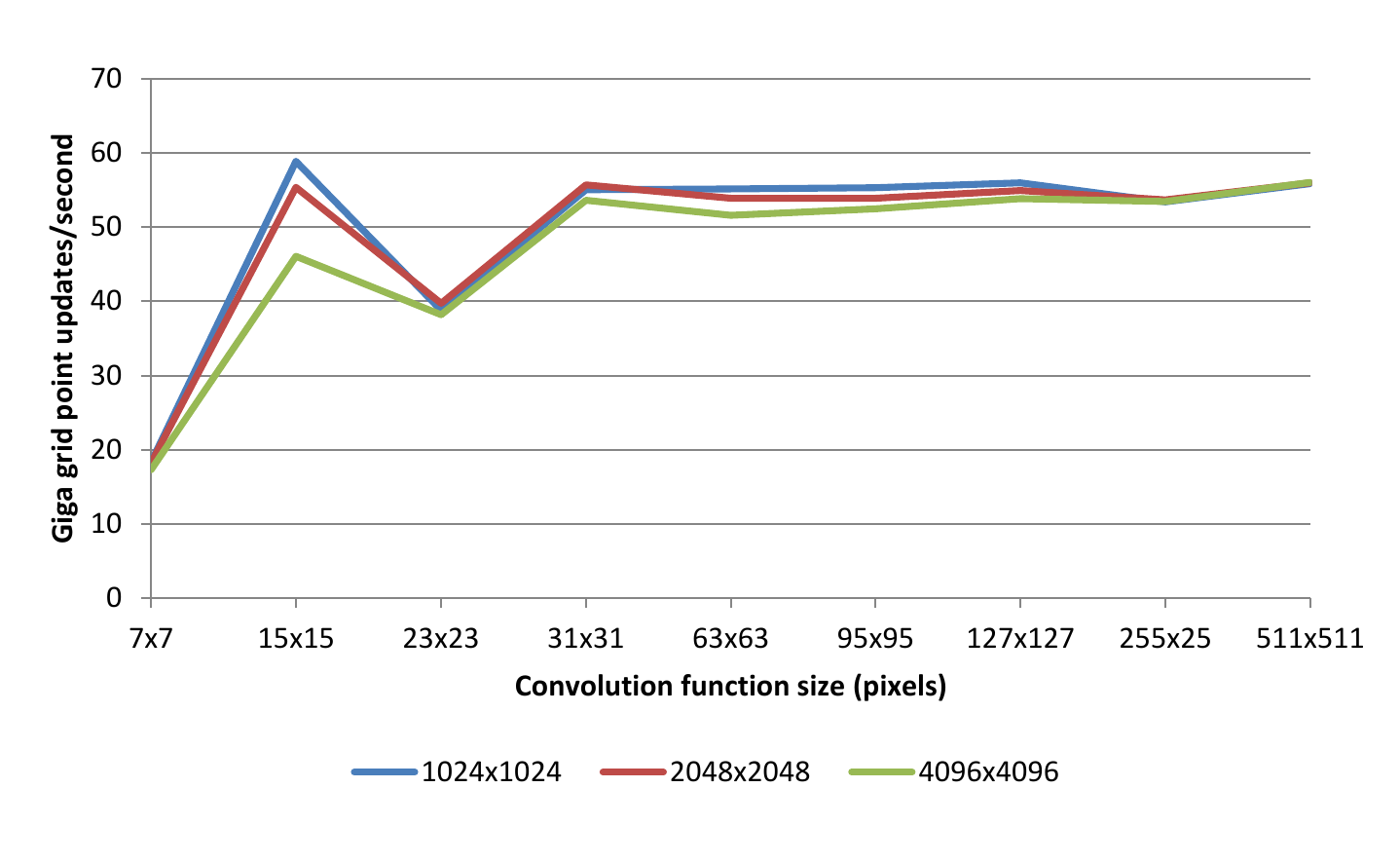}}
    \caption[Real gridding rate results of experiment batch 2 (Second Data Set)]{Real gridding rate results per convolution function size for experiments 2.4 to 2.6 (Data Set 1)}
\label{fig:res-deltau-set1}
\end{figure}

\begin{figure}[H]
    \centerline{\includegraphics[scale=0.995]{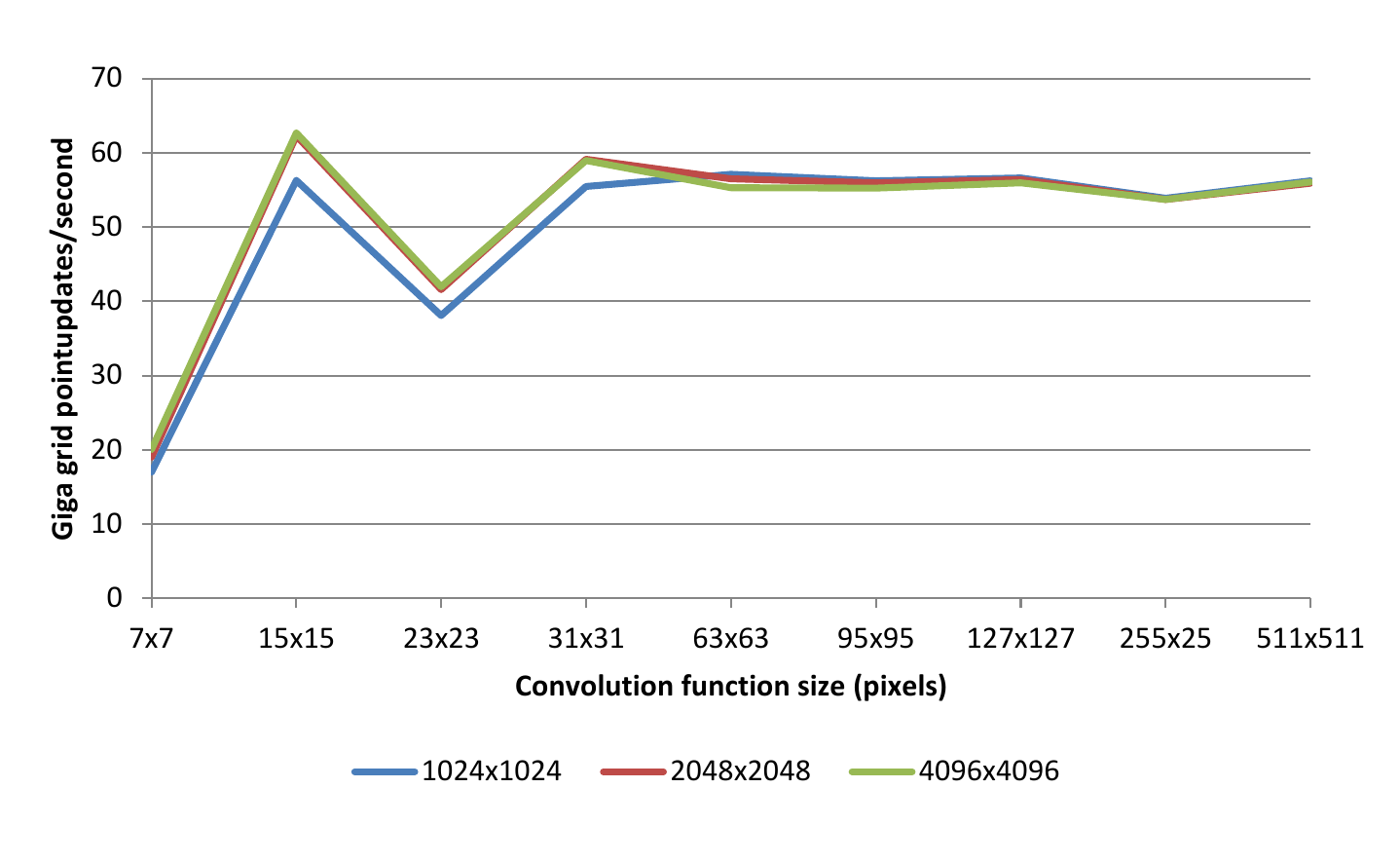}}
	\caption[Real gridding rate results of experiment batch 2 (Third Data Set)]{Real gridding rate results per convolution function size for experiments 2.7 to 2.9 (Data Set 2)}
\label{fig:res-deltau-set2}
\end{figure}

The real gridding rates obtained from these experiments are shown in Figures \ref{fig:res-deltau-set0}, \ref{fig:res-deltau-set1} and \ref{fig:res-deltau-set2}.

\FloatBarrier
\subsection{Experiment batch no 3: Effects of oversampling factor}

In this batch,  the oversampling factor is varied to analyse its impact on the real gridder performance. Note that an oversampling factor of 2 is generally not enough for accurate imaging. Experiment details are given in Table \ref{tab:batch3}.
 
\begin{table}[htbp]
  \centering
    \begin{tabularx}{\textwidth}{P{1.6cm}LLLL }
    \hline
    \textit{Ref No} & \textit{Data Set No} & \textit{Image dimensions (pixels)} & \textit{Pixel length (arcsec)} & \textit{Oversampling factor} \\
    \hline
    3.1   & 0     & $2048\times2048$ & 14.5711 & 2 \\
    3.2   & 0     & $2048\times2048$ & 14.5711 & 4 \\
    3.3   & 0     & $2048\times2048$ & 14.5711 & 8 \\
    3.4   & 1     & $2048\times2048$ & 14.5711 & 2 \\
    3.5   & 1     & $2048\times2048$ & 14.5711 & 4 \\
    3.6   & 1     & $2048\times2048$ & 14.5711 & 8 \\
    3.7   & 2     & $2048\times2048$ & 14.5711 & 2 \\
    3.8   & 2     & $2048\times2048$ & 14.5711 & 4 \\
    3.9   & 2     & $2048\times2048$ & 14.5711 & 8 \\
    \hline
    \end{tabularx}%
    \caption{Experiment details of batch 3}
    \label{tab:batch3}
\end{table}%

\begin{figure}[H]
    \centerline{\includegraphics[scale=0.995]{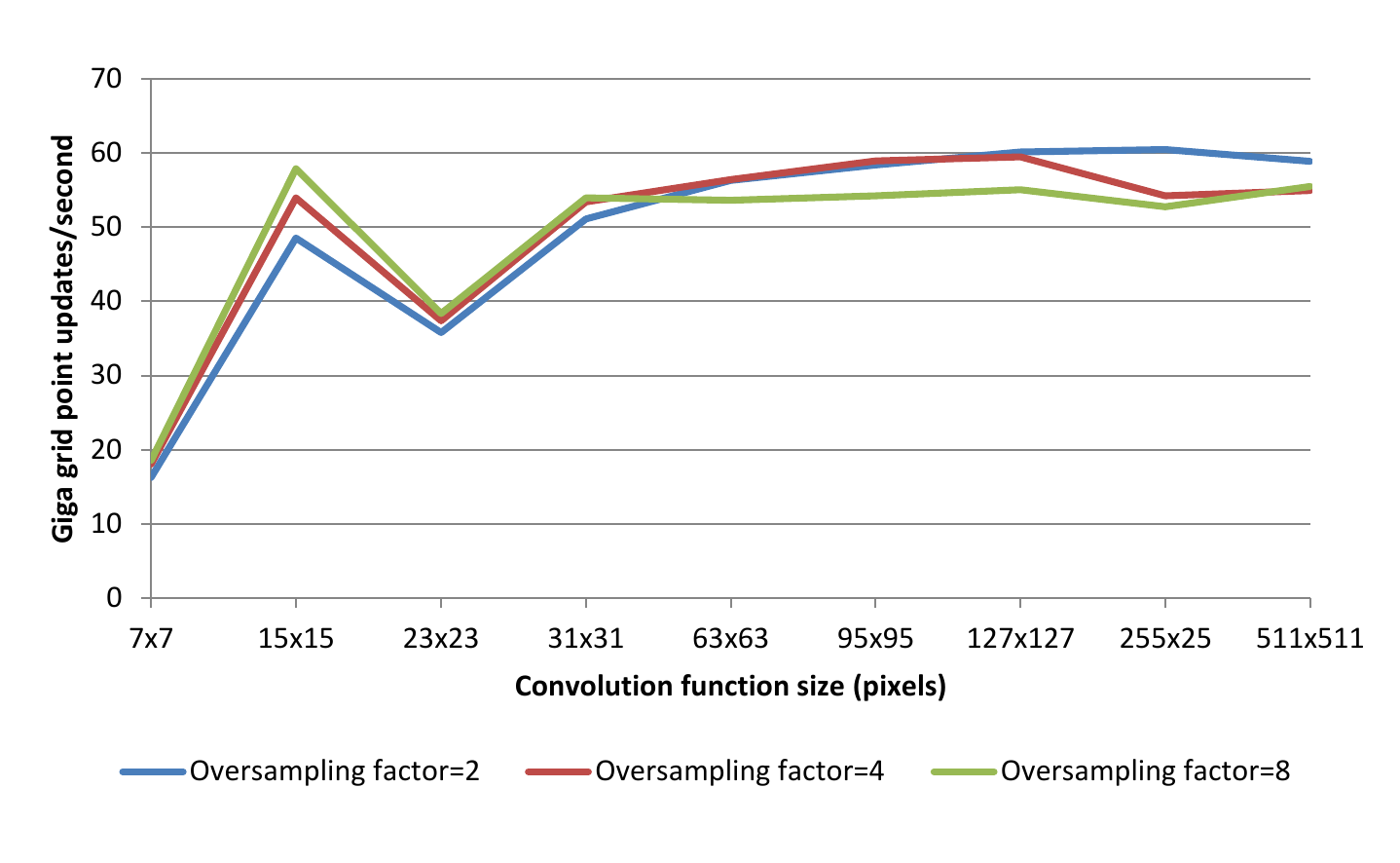}}
    \caption[Real gridding rate results of experiment batch 3 (First Data Set)]{Real gridding rate results per convolution function size for experiments 3.1 to 3.3 (Data Set 0)}
\label{fig:res-sam-set0}
\end{figure}

\begin{figure}[h]
  \centerline{\includegraphics[scale=0.995]{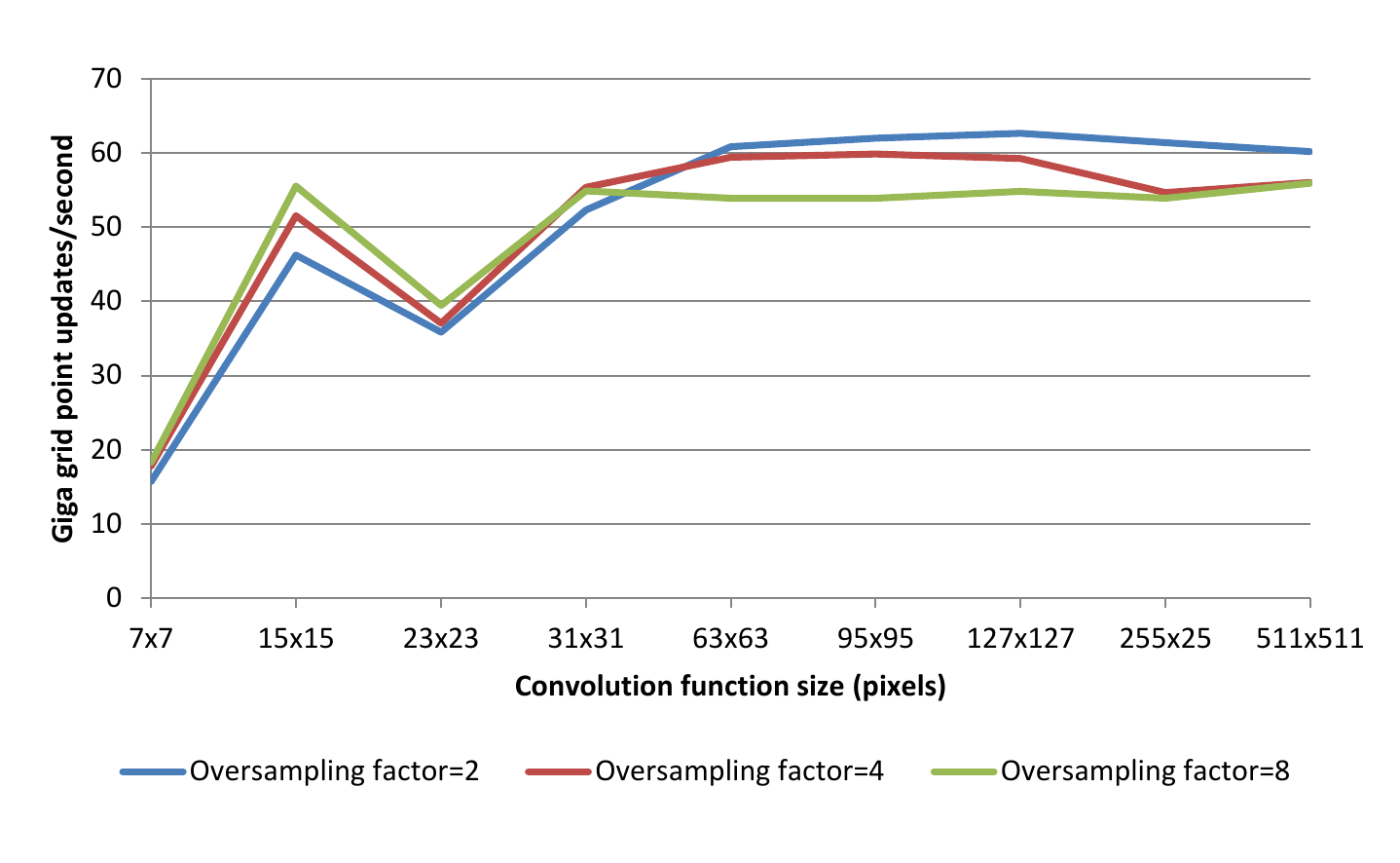}}
  \caption[Real gridding rate results of experiment batch 3 (Second Data Set)]{Real gridding rate results per convolution function size for experiments 3.4 to 3.6 (Data Set 1)}
\label{fig:res-sam-set1}
\end{figure}

\begin{figure}[h]
    \centerline{\includegraphics[scale=0.995]{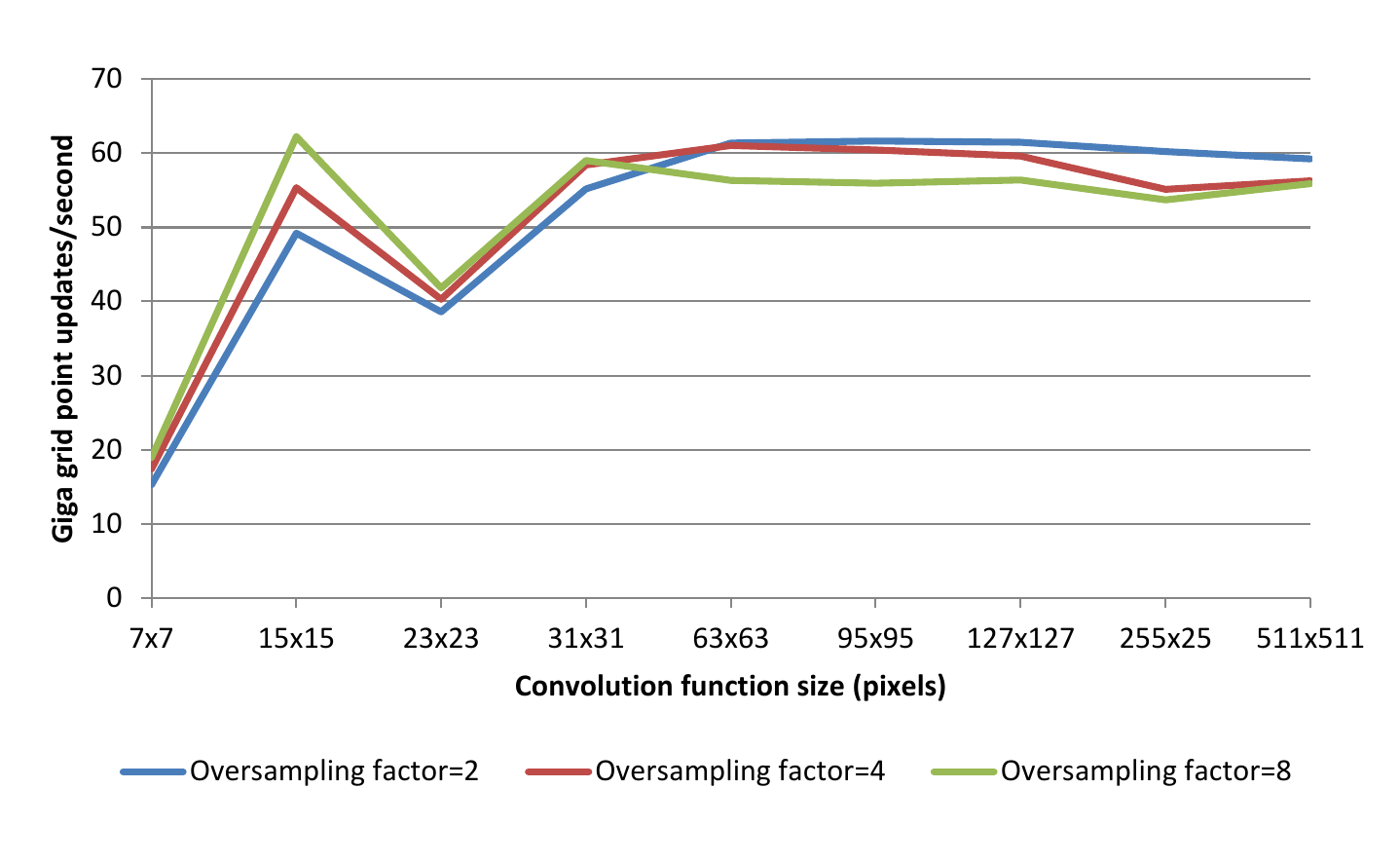}}
  \caption[Real gridding rate results of experiment batch 3 (Third Data Set)]{Real gridding rate results per convolution function size for experiments 3.7 to 3.9 (Data Set 2)}
\label{fig:res-sam-set2}
\end{figure}

\begin{figure}[h]
    \centerline{\includegraphics[scale=0.995]{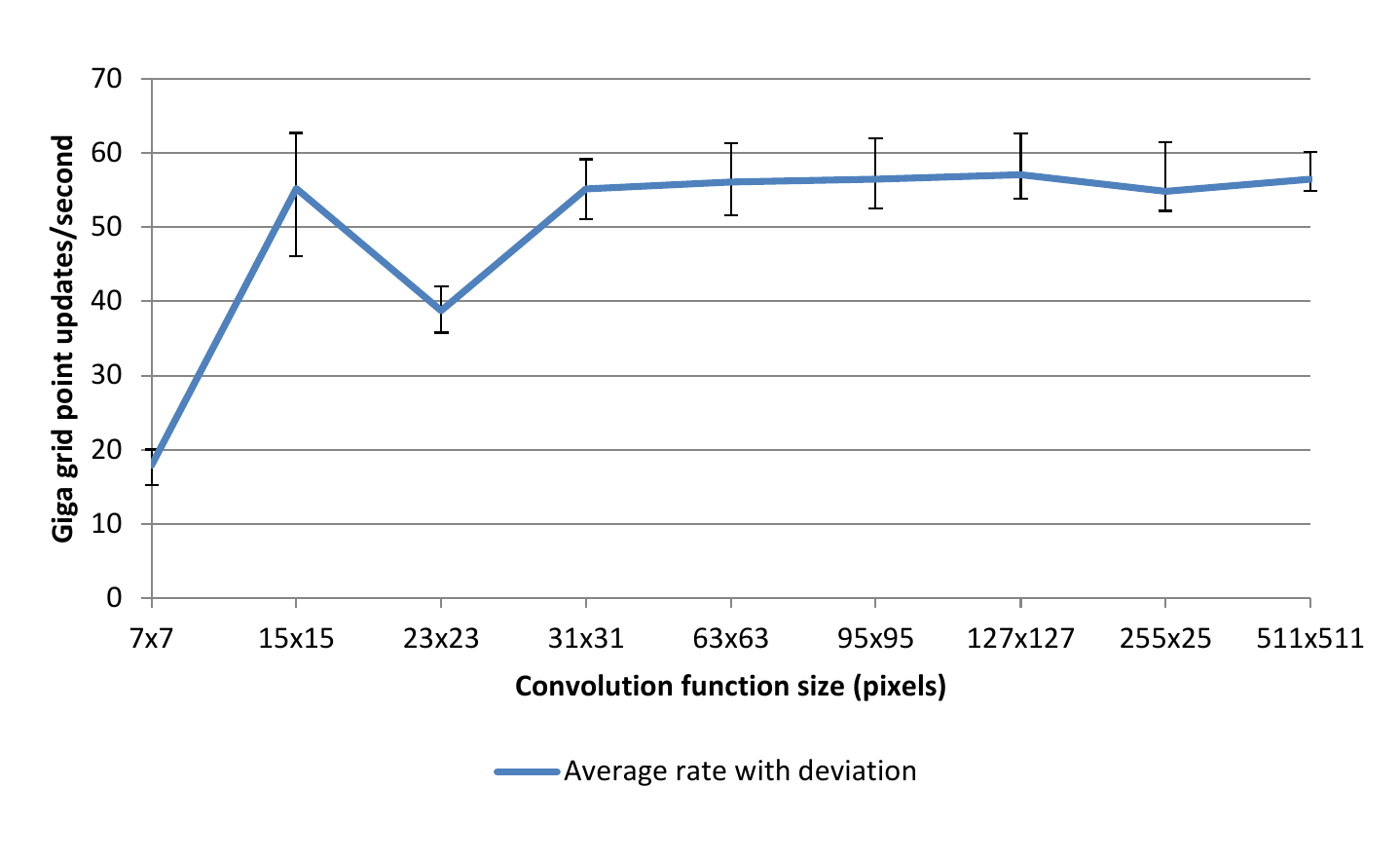}}
    \caption[Average gridder performance results]{Plot showing average real gridding rate for experiments of batch 2 and batch 3, together with total variation. The total variation is represented by the error bar.} 
\label{fig:res-real}
\end{figure}

Figures \ref{fig:res-sam-set0}, \ref{fig:res-sam-set1} and \ref{fig:res-sam-set2} plot the real gridding rate obtained. Figure \ref{fig:res-real} plots the average rate obtained from the second and third batch (this batch) of experiments, together with the total variation. 
 

\FloatBarrier

\subsection{Experiment batch no 4: Number of polarisations}
The aim of this batch is to analyse the gridder's real performance vis-a-vi number of polarisations gridded. Only Data Set 0 is used. Table \ref{tab:batch4} gives details.

\begin{table}[htbp]
  \centering
    \begin{tabularx}{\textwidth}{ P{1.6cm}LLLLL }
    \hline
    \textit{Ref No} & \textit{Data Set No} & \textit{Image dimensions (pixels)} & \textit{Pixel length (arcsec)} & \textit{Over- sampling factor} & \textit{No of polarisations}\\
    \hline
    4.1   & 0     & $2048\times2048$ & 14.5711 & 8 & 1\\
    4.2   & 0     & $2048\times2048$ & 14.5711 & 8 & 2\\
    4.3   & 0     & $2048\times2048$ & 14.5711 & 8 & 4\\
     \hline
    \end{tabularx}%
\caption{Experiment details of batch 4}
    \label{tab:batch4}

\end{table}%

\begin{figure}[H]
    \centerline{\includegraphics[scale=0.995]{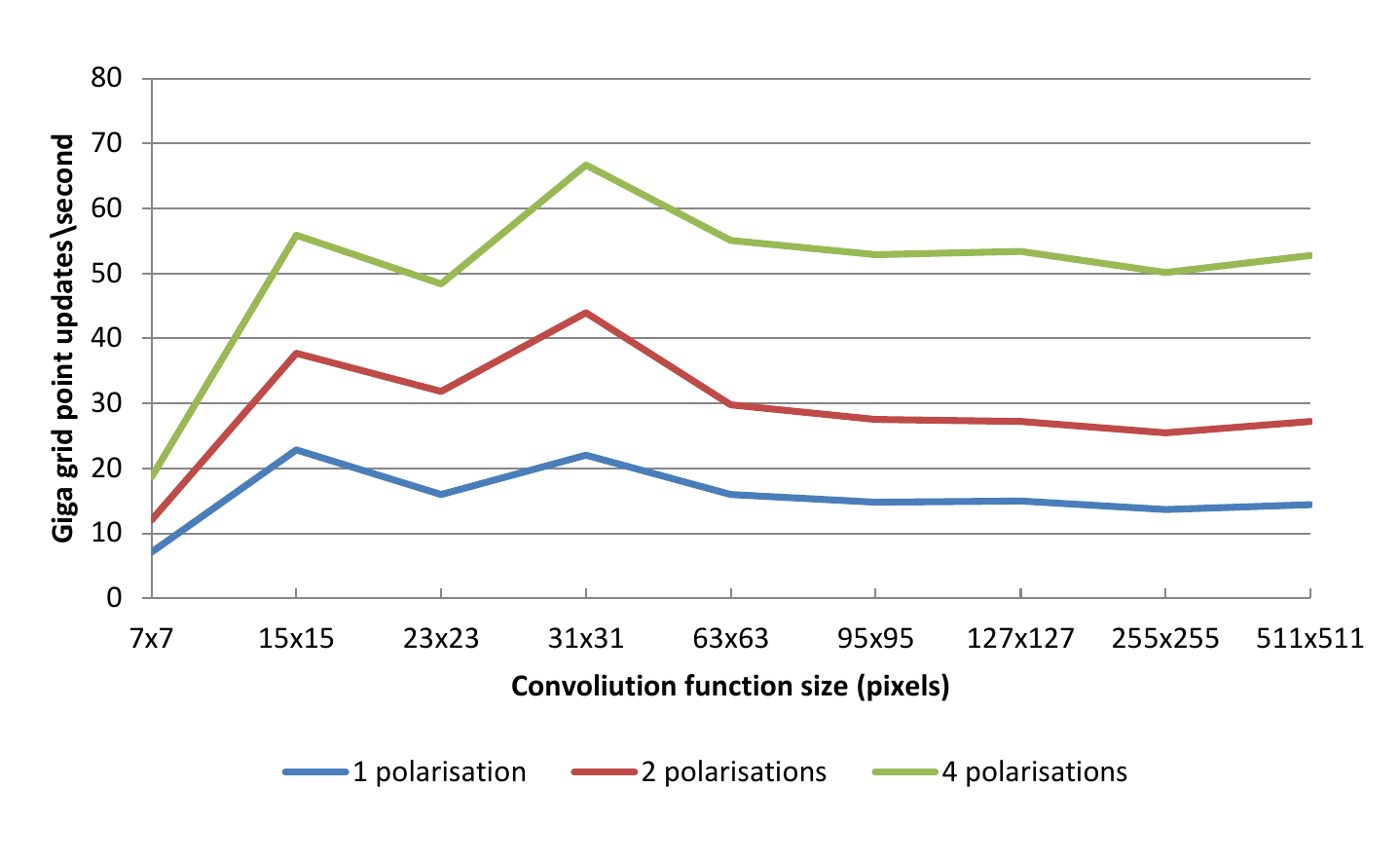}}
  \caption[Real gridding rate results of experiment batch 4]{Real gridding rate per convolution function size for experiment batch 4}

\label{fig:res-polarizations}
\end{figure}

\begin{figure}[H]
    \centerline{\includegraphics[scale=0.995]{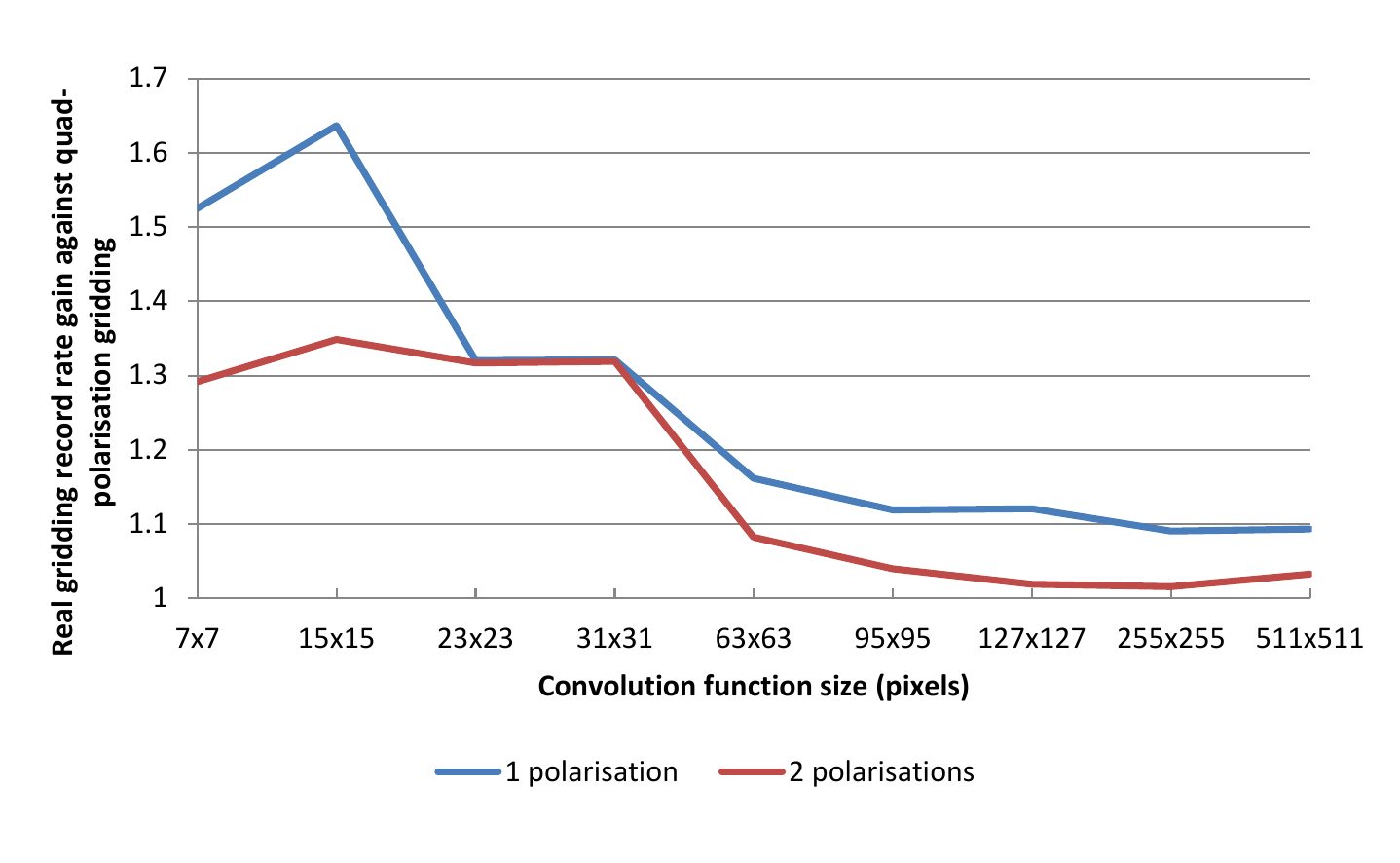}}
  \caption[Multiple increase in real gridding rate results of experiment batch 4]  {Multiple increase in real record gridding per convolution function size obtained against quad-polarisation imaging for experiment batch 4}

\label{fig:res-pol-comp}
\end{figure}

The following two figures report the results obtained. Figure \ref{fig:res-polarizations} plots the real gridding rate for each experiment. Figure \ref{fig:res-pol-comp} plots the multiple increase in record gridding rate when reducing the number of polarisations from 4 to 2 and 1. Reducing the number of polarisations, reduces the grid point updates generated by each record and thus an increase in real record gridding rate is expected.


\subsection{Experiment batch no 5: Overall performance analysis}

Proper \textit{w-projection} is enabled for this batch and Data Sets 1 and 2 are imaged using the two configurations described in Table \ref{tab:confbatch5}.

\begin{table}[htbp]
  \centering
    \begin{tabularx}{\textwidth}{XXX}
    \hline
    \textit{Configuration parameter} & \textit{Configuration 1} & \textit{Configuration 2} \\
    \hline
    Image dimensions & $1024\times1024$ & $4500\times4500$ \\
    Pixel length/width & 20    & 18.95 \\
    No of $w$-planes & 200   & 16 \\
    Oversampling Factor & 4 & 4 \\
    \hline
    \end{tabularx}%
    \caption{Imaging configuration for experiment batch 5}
    \label{tab:confbatch5}
\end{table}%

These experiments are repeated using \textit{lwimager} for a comparative analysis. Note that, for Data Set 2, \textit{lwimager} uses single precision while for Data Set 1 \textit{lwimager} uses double precision. It is an automated feature of the tool.

The two configurations have been carefully selected to show how the \textit{mt-imager} performs in different scenarios. Configuration 1 is a lightweight configuration in the sense that a relatively inexpensive computation is required. Convolution functions are not larger than $29\times29$ pixels and therefore a maximum of 841 grid point updates are required to grid one polarisation for a record. Configuration 2 is a heavy weight configuration requiring expensive computation. The largest convolution function is $349\times349$ pixels implying a maximum of 121801 grid point updates to grid a single polarisation of a record (approximately 145$\times$ more than Configuration 1). Configuration 2 images a much larger field of view than Configuration 1, implying smaller sampling intervals in the UV-grid. This reduces the likelihood of \textit{compression} being effective, and making it more difficult for \textit{mt-imager} to image Configuration 2. 

Since the two configurations are applied over the "small" Data Set 1 and the "large" Data Set 2, the scenarios presented here are a combination of small/large data sets with lightweight/heavyweight configurations. The most important run is the imaging of Data Set 2 using Configuration 2 since it is the most difficult and where high-performance is mostly required.   

\begin{figure}[H]
\centerline{\includegraphics[scale=0.995]{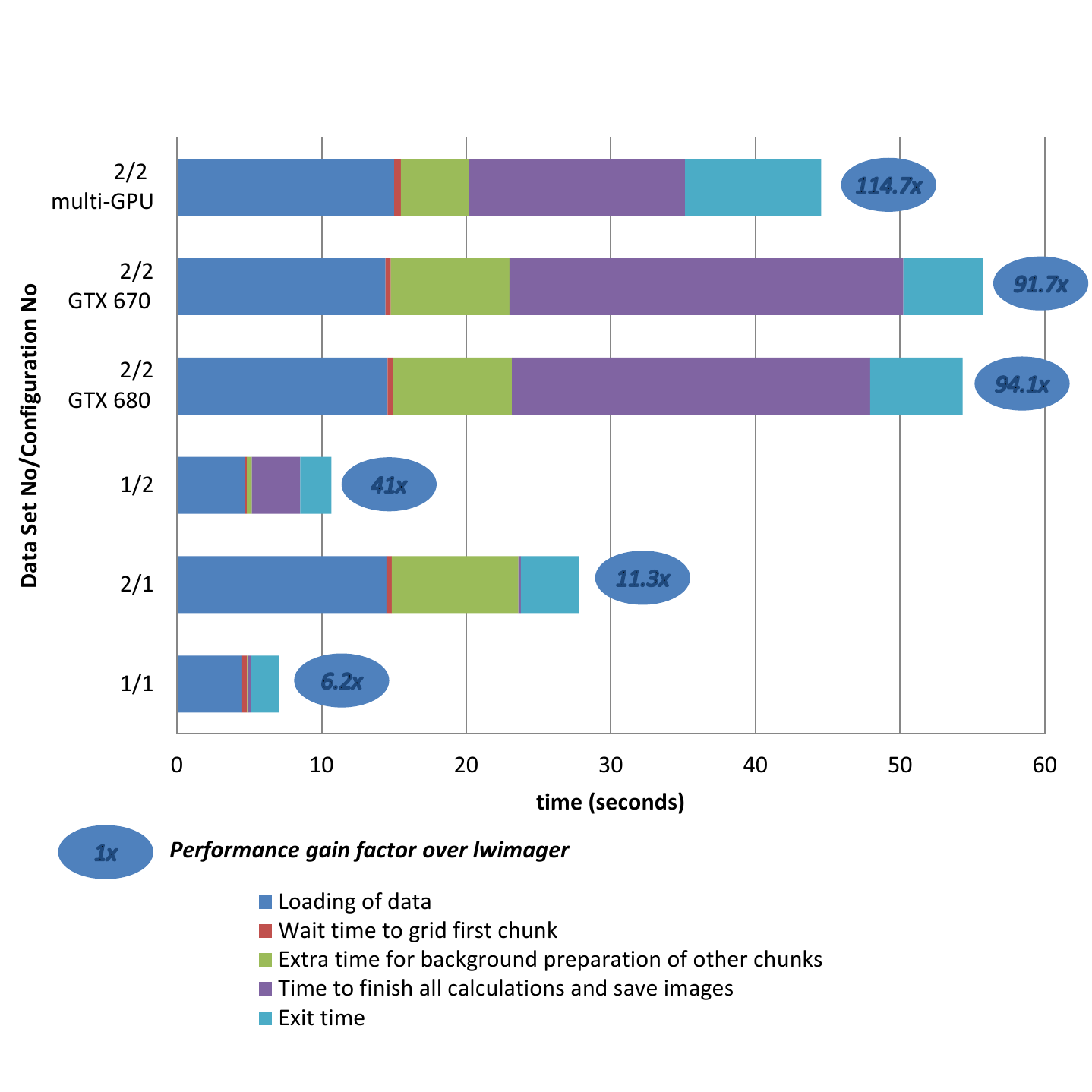}}
\caption[Execution timelines of \textit{mt-imager} for experiment batch 5 and 6]{Execution timelines of \textit{mt-imager} for experiment batch 5 and 6. The performance gain over lwimager is also stated.}
\label{fig:timeline}
\end{figure}

\begin{figure}[H]
    \centerline{\includegraphics[scale=0.995]{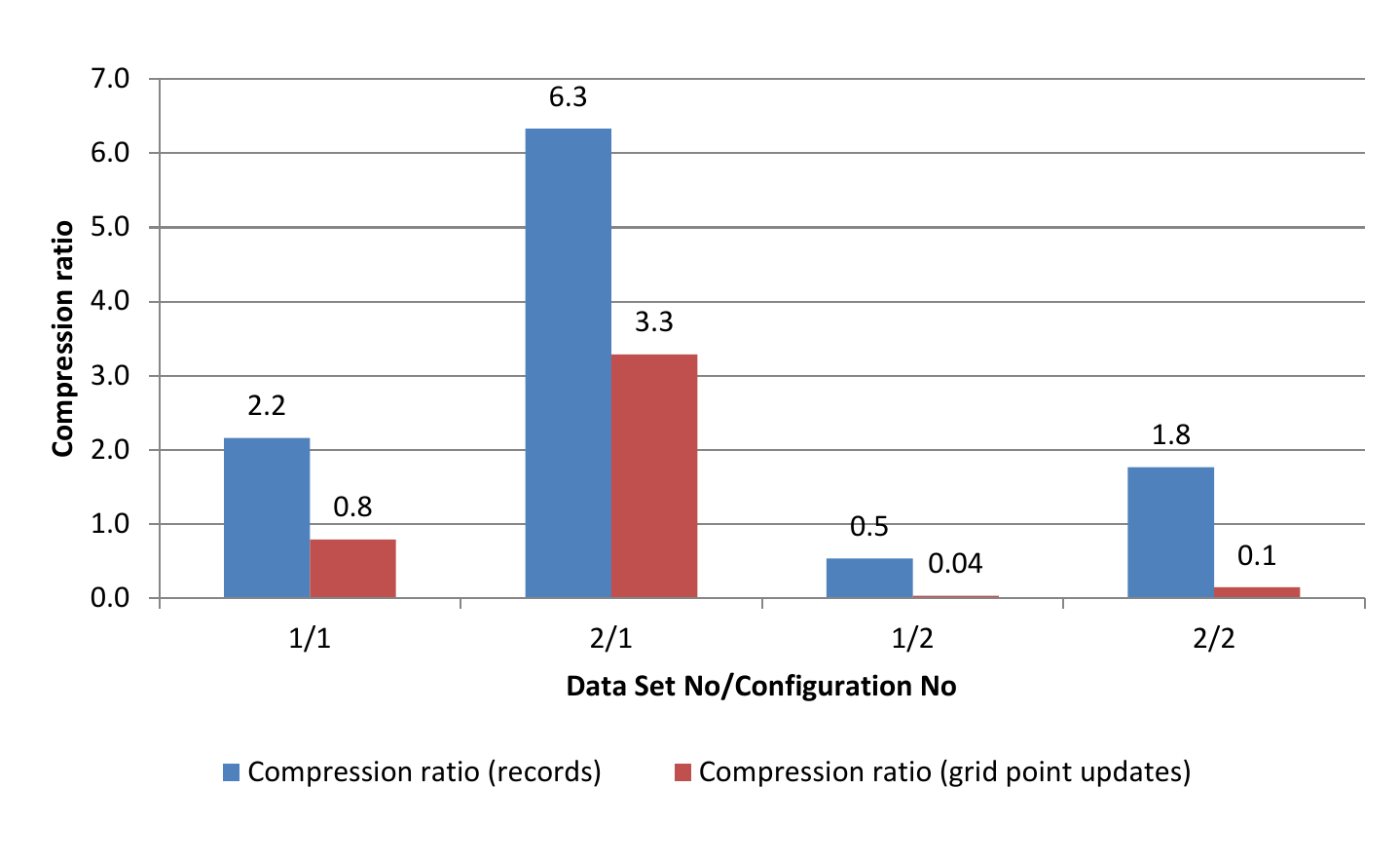}}
  \caption{\textit{Compression} ratios obtained for experiment batch 5.}

\label{fig:res-compressionratio}

\end{figure}

\begin{figure}[H]
    \centerline{\includegraphics[scale=0.995]{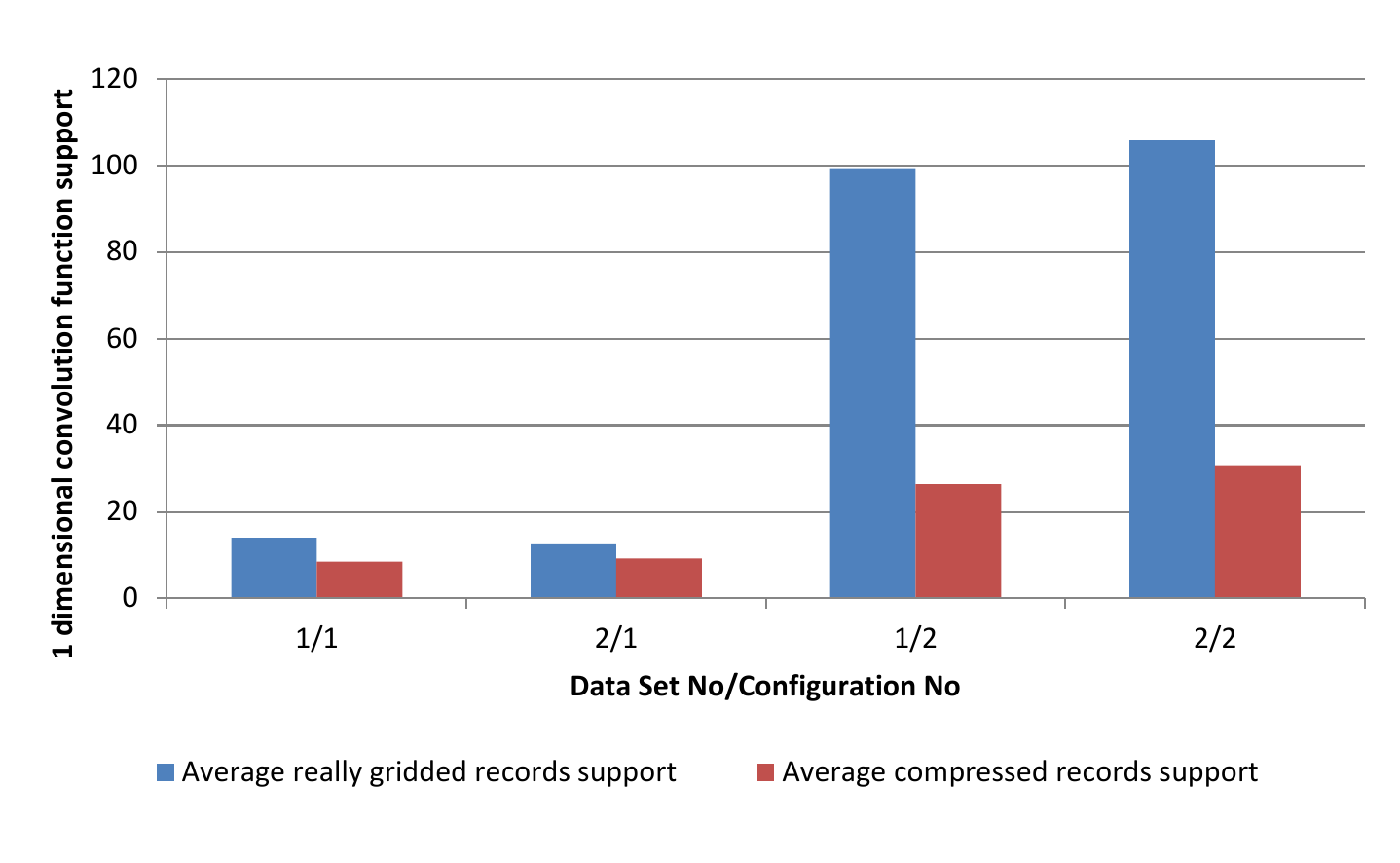}}
 \caption{Average convolution function support for experiment batch 5}
\label{fig:res-avg-support}
\end{figure}

\begin{figure}[H]
    \centerline{\includegraphics[scale=0.995]{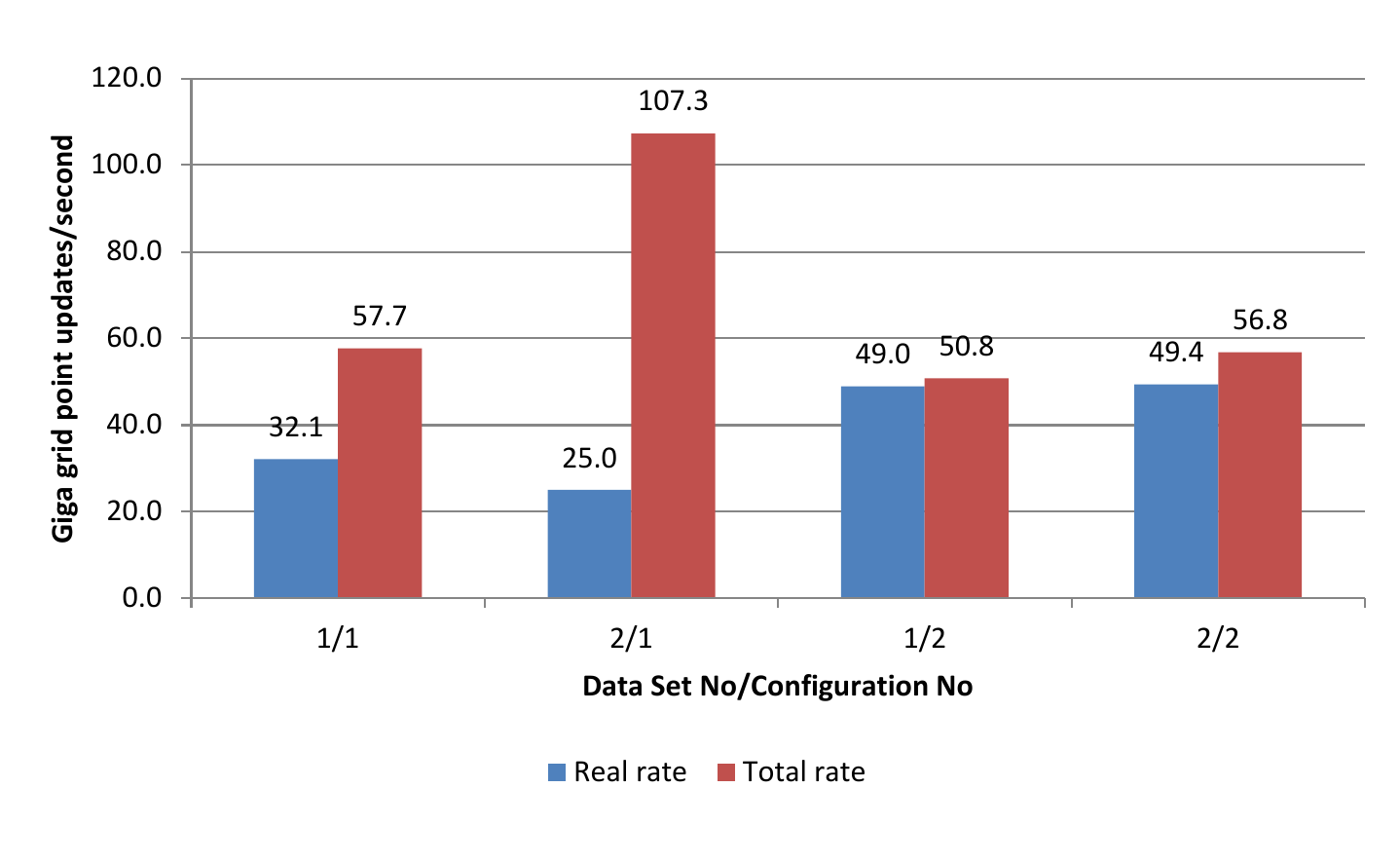}}
  \caption{Gridding rate results for experiment batch 5}
\label{fig:res-overall-gridderrate}
\end{figure}

\begin{figure}[H]
    \centerline{\includegraphics[scale=0.995]{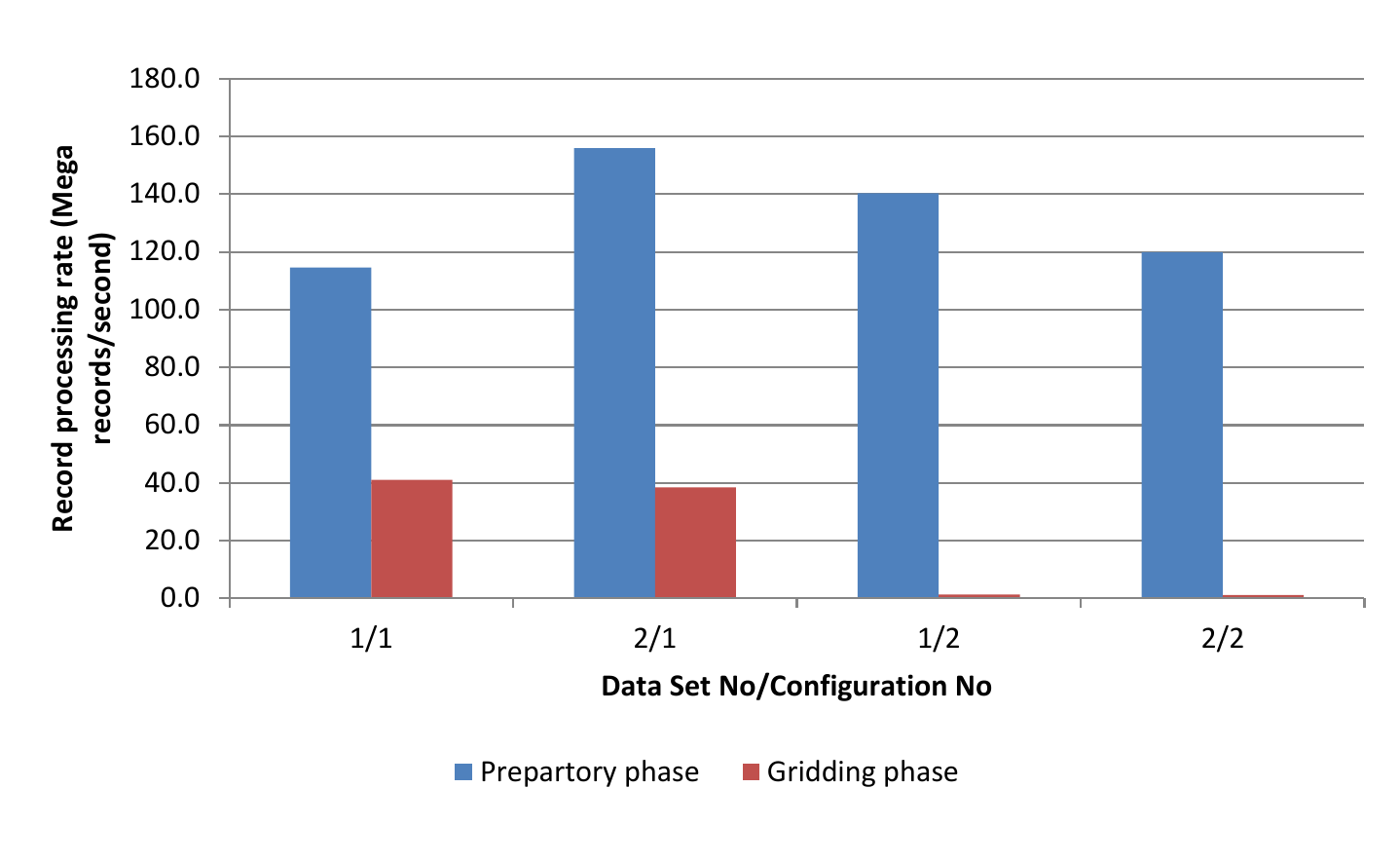}}
  \caption[Preparation and gridding phase record rate results for experiment batch 5]{Plot of preparation and gridding phase record rate for experiment batch 5 . All records presented to the \textit{Wimager} are considered for the preparation phase rate. As for the gridding rate, only records that are truly gridded are considered.}

\label{fig:res-overall-rate}
\end{figure}

\begin{figure}[H]
  
    \centerline{\includegraphics[scale=0.995]{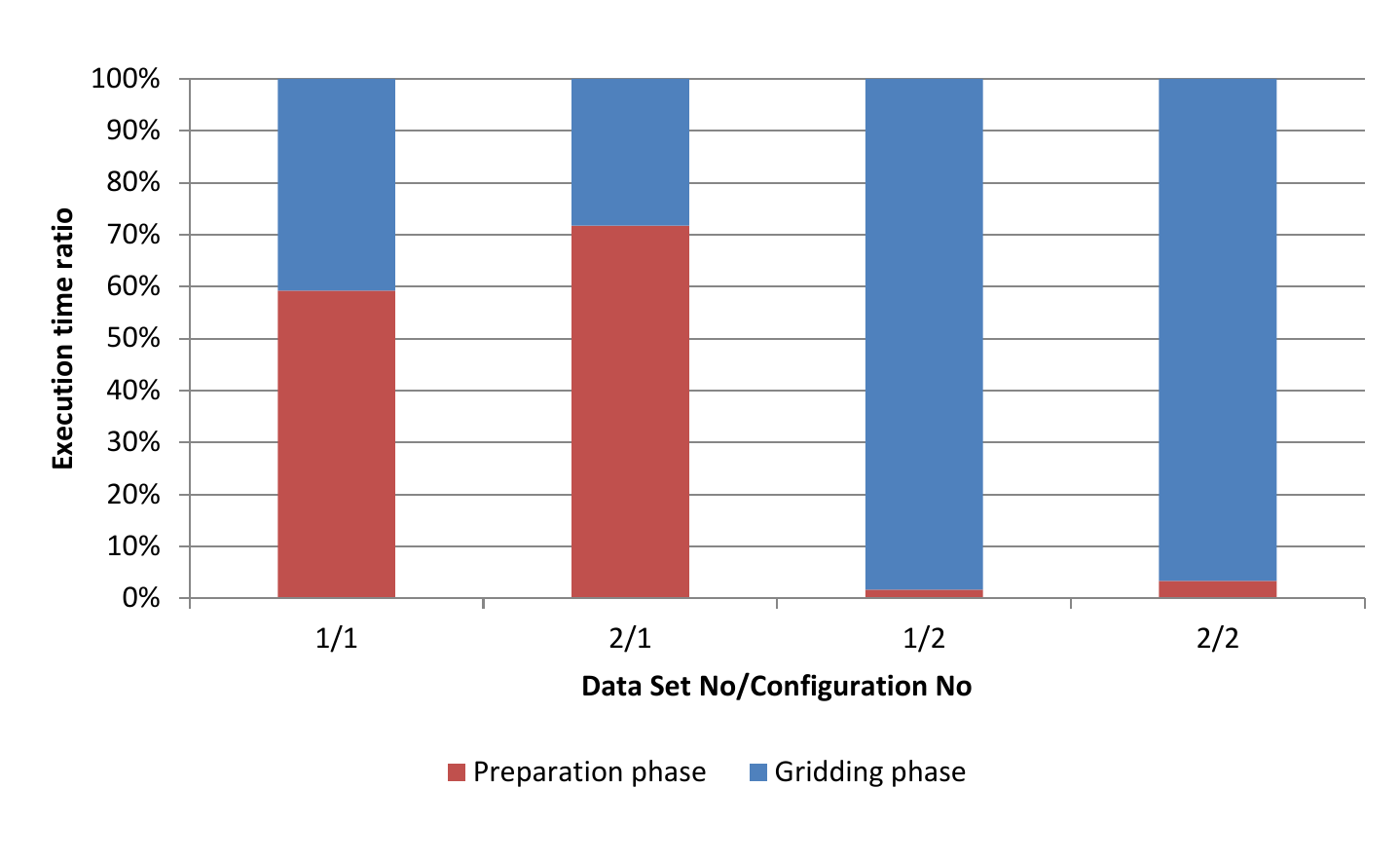}}
\caption{Plot giving the ratio of \textit{WImager} preparation phase execution time against gridder execution time for experiments in batch 5}
\label{fig:res-kernel-time}
\end{figure}

Figures \ref{fig:timeline} to \ref{fig:res-kernel-time} report the results obtained. Figure \ref{fig:timeline} gives a timeline of main events occurring during a run of the imaging tool. It must be kept in mind that \textit{mt-imager} does work in parallel, so during a defined time interval it would be doing more than one activity. During loading of data, the \textit{Visibility Manager} component prepares data chunks. After loading, GPUs grid any ready data chunks, while the \textit{Visibility Manager} component prepares other data chunks. Convolution functions are generated over the GPU, at the same time when the \textit{Visibility Manager} component is loading data from disk.

Figures \ref{fig:res-compressionratio} and \ref{fig:res-avg-support} report \textit{compression} related data, to give more insights on how \textit{compression} works.

The \textit{WImager} algorithm performance is reported in Figures \ref{fig:res-overall-gridderrate}, \ref{fig:res-overall-rate}\ and \ref{fig:res-kernel-time}.


\subsection{Experiment batch no 6: Multi-GPU analysis}

Experiments in this batch aims in analysing the scalability of \textit{mt-imager} over a number of GPUs. Two experiments are defined and in the two of them Data Set 2 is imaged over Configuration 2 (defined in Table \ref{tab:confbatch5}). In the first experiment \textit{mt-imager} images over the GTX670 card only. In the second experiment \textit{mt-imager} images over the two GPUs, that is the GTX680 and GTX670 (refer to section \ref{sec:hardware} for further details). Timeline results are shown in Figure \ref{fig:timeline}.

\section{Analysis and discussion}

\subsection{\textit{Compression}} 

In the first batch of experiments (refer to Figure \ref{fig:res-batch1}), switching off \textit{channel interleaving} while compression is disabled (experiment 1.2) reduces the gridder performance for the large convolution functions (31$\times$31 and wider),  by around 10G. No change in performance resulted for smaller convolution functions. On enabling compression, the total rate for interleaved channels goes to a maximum of 280G and for non-interleaved channels goes to 175G. This implies a 3-fold performance increase for \textit{interleaved channels} and 2-fold increase for \textit{non-interleaved channels}. The real gridding rate goes down as compression is enabled, but as argued in section \ref{sec:realrate} it goes to a steady value.

From these results,  it can be deduced that performance of Romein's algorithm without compression has some dependence on the records that would be compressed if compression is enabled. The methods implemented in this thesis tackle such "compressible" records in a way to obtain higher performance.

Unfortunately, the claimed total rates are not likely to be achieved in realistic scenarios where proper \textit{w-projection} is used. In section \ref{sec:compression} it was discussed that the probability of compression is not evenly distributed over the UV-grid but is more prominent for short baselines. Short baselines tend to have small convolution functions, and thus the highest probability for compression is for records with the least computational exigencies. This argument is well supported by the reported results of experiment batch 5. Figures \ref{fig:res-compressionratio} and \ref{fig:res-avg-support} show that for all the runs made using proper \textit{w-projection}, the average convolution function size for compressed records is smaller than that for really gridded records. Consequently, the \textit{compression} ratio in terms of grid point updates\footnote{This controls the total gridding rate.} is smaller than the respective record \textit{compression} ratio. This reduction reached a factor of 10 for some runs. This implies that performance gains delivered by \textit{compression} can be severely degraded. This does not necessarily make \textit{compression} ineffective since as shown in Figure \ref{fig:res-overall-gridderrate}, Data Set 2 is gridded using Configuration 1 at a total rate of 107G. One notes that when there are no records to compress there is no performance loss against a scenario with \textit{compression} disabled. \textit{Compression} can either not affect performance or enhance it.

Integration time effects \textit{compression}. Data Set 2 has an integration time roughly 3$\times$ shorter than Data Set 1, and results depicted in Figure \ref{fig:res-compressionratio} reveal the highly different \textit{compression} ratios between Data Set 1 and Data Set 2. This is in-line with the arguments given in section \ref{sec:compression}.

A final note on \textit{compression} regards \textit{channel interleaving}. From the results reported on the first batch of experiments, it is clear that \textit{compression} occurred over channels. This shows that \textit{compression} can also deliver performance in multi-frequency image synthesis. In such synthesis, channels are gridded over the same grid. The developed imaging tool does not support such a feature, but it is a good point to note for the future. The good news about \textit{compression} over channels is that the probability of occurrence should be distributed evenly over the grid. It is likely to be dependent on channel frequency bandwidth, whereby the narrower the channel frequency bandwidth, the higher is the probability for \textit{compression}. Worth recalling that, for a wide field of view, channel bandwidth has to be as narrow as possible (refer to section \ref{sec:channelbadwidth}).

\subsection{Real gridding rate and gridder scalability}
\label{sec:realrate}
It is deduced from the second and third batch of results (Figures \ref{fig:res-deltau-set0}-
\ref{fig:res-real}) that there is little effect on the gridder real performance when varying the oversampling factor or the UV-grid sampling interval. Figure \ref{fig:res-real} gives the average rate obtained from these experiments per convolution function size and reports the total variations in performance. The biggest fluctuations happened for the $15\times15$ convolution function whereby a maximum rate variation of 16G resulted. For all the others,  the maximum variation was less than 10G. 

It is also deduced that the gridder is scalable in terms of convolution function size. The gridder, grids convolution functions of size larger or equal to $31\times31$ at a nearly constant rate of around 50G. There is clearly  a degradation in performance for convolution function of size less than $31\times31$. However, this does not constitute a real issue since they are relatively light in terms of computation.

Rates are nearly maintained when proper \textit{w-projection} is enabled. Figure \ref{fig:res-overall-gridderrate} reports that, for the heavy-weight Configuration 2, the gridder rate was around 49G for the two data sets. It is slightly less than the gridding rate achieved by convolution functions larger or equal to $31\times31$, but much higher than the gridding rate achieved for some of the small convolution functions ($7\times7$ and $23\times23$). In view that proper \textit{w-projection} uses many convolution function sizes and that the local chunk length (refer to section \ref{sec:localchunks}) is variable, then, this rate can be considered as acceptable. 

Performance achieved using the light-weight Configuration 1 is also acceptable though much less than Configuration 2. Given that, in this configuration, all convolution sizes are less than $31\times31$ pixels, then such a drop is expected since these sizes tend to be gridded at a lower rate.      

\subsection{Number of polarisations}
\label{sec:res-polarizations}
Results of experiment batch 4 reported in Figures \ref{fig:res-polarizations} and \ref{fig:res-pol-comp}, show that the gridder does not down-scale with the number of polarisations being gridded. Gridding dual or single polarised records require  half or  quarter of grid point updates respectively than gridding quad-polarised records. If the gridding rate (in grid point updates/sec) remains constant over the number of polarisations, then, records are expected as to be gridded at a double or quadruple rate respectively when compared to quad-polarisation gridding. Instead, as reported in Figure \ref{fig:res-pol-comp} the record gridding rate increased only by a factor of 1.1$\times$ for single-polarisation gridding of large convolution functions. The culprit is believed to be the retrieval of convolution function data that is discussed in section \ref{sec:lim}.

\subsection{Main limiting factor of the gridder}
\label{sec:lim}
It is claimed that the main limiting factor of the gridder is the retrieval of the convolution function numerical data from texture. This happens for each grid point update. Romein \cite{Romein2012}, argues that the main limiting factor of his scenario is atomic operations. This behaviour is not observer in the implementation  presented in this thesis, since there is little change in performance when the oversampling factor is varied. A reduction in oversampling rate results in a reduction in the number of records that are gridded without causing any atomic commits, since more records get compressed. Thus for each grid point update the likelihood of an atomic commit is increased. If atomic operations were the main limiting factor, the real gridding rate should drastically decrease with decreasing oversampling factor. Instead, the rate increases for large convolution functions (refer to Figures \ref{fig:res-sam-set0}, \ref{fig:res-sam-set1}, \ref{fig:res-sam-set2}). For smaller convolution functions,  the rate does decrease but not in the order that would be expected. A similar argument can be made for the polarisation results (batch 4). Decreasing polarisations reduces atomic operations per gridded record in proportion to the decrease in the number of polarisations. No substantial increase in record gridding rate resulted when the number of polarisations is decreased. 

The polarisation results also rule out the floating point operations required for grid point update as a main limiting factor. The payload is constant to eight flops per grid point and independent on the number of polarisations. In view of the heavy reduction in the gridding rate expressed in grid point updates per second, it cannot be the main limiting factor. 

The remaining possible culprits are the retrieval of the convolution function from texture, the gridding logic and access to shared memory. When \textit{compression} is disabled (see Figure \ref{fig:res-batch1}), the real rate increases drastically (around 30G). Payload of shared memory access and gridding logic is fixed for enabled or disabled \textit{compression}. The only variant is the texture performance since it is a cache. When \textit{compression} is disabled the likelihood of a cache hit is increased. This is because the probability that the same numerical data is used by subsequent records is larger than zero. Hence, the claim that the main limiting factor is the retrieval of convolution function data from the texture is proved.

\subsection{\textit{WImager} preparation phase performance}

Figure \ref{fig:res-overall-rate} shows that the preparation phase prepares records at a fast rate. Its impact on the overall performance of the \textit{WImager} algorithm is dependent on the computational intensity required by the gridder. The higher the computational intensity of the gridder phase, the lower is the overall impact of the preparation phase. Figure \ref{fig:res-kernel-time} visualises the point. For the heavy-weight Configuration 2, the preparation phase execution time is negligible when compared to the execution time of the gridder. On the other hand, imaging over Configuration 1 resulted in a preparation phase execution time larger than the gridding time. One must bear in mind that the preparation phase's main job is to reduce logic from  the gridding phase. If the logic is integrated in the gridding phase, the total execution time of the \textit{WImager} algorithm would increase.   

\subsection{Overall performance} 

Overall performance of the \textit{mt-imager} is reported in Figure \ref{fig:timeline}. Results for the multi-GPU scenario are ignored in this section but discussed in section \ref{sec:multiGPU}. 

Figure \ref{fig:timeline} reveals that \textit{mt-imager} synthesised Data Set 2 over Configuration 2 (the hardest of all runs) 94$\times$ faster than \textit{lwimager}. \textbf{This result shows that the main thesis objective of developing a high-performance imaging tool has been achieved.}

This gain is not sustained for all runs. It is a side effect of high performance. Computation is so efficient that the loading of data from disk\footnote{As per section \ref{sec:visibilitymanager}, data is loaded through the casacore ms API which can affect the data loading time.} is a significant limiting factor. During this time, the GPU has to wait for the first chunk of data. Worst case occurred for the simplest run (Data Set 1 over Configuration 1), where only a 6.2$\times$ gain was obtained. Loading of data in this run took most of the time\footnote{Exit time is ignored in this discussion.}. 

It should be stated that for the simplest run (Data Set 1 over Configuration 1), the generation of the 200 convolution functions over the GPU finished nearly at the same time when loading of data was ready. This implies that generation of convolution functions might sometimes limit the performance further. 

One notes the remarkably small time interval, shown in red, for all runs. During this time, the \textit{Visibility Manager} makes the last preparations for the first chunk of data after that all data has been loaded from disk. This time interval is short because the \textit{Visibility Manager} did most of the preparation work while the system is still loading data. It does not make miracles, and for the large Data Set 1 it requires a substantial amount of extra time to prepare the other data chunks. More in-depth analyses reveal that this extra time is needed to sort and convert visibility data (that is the sequence defined as $\{\vis_i\}_p$ is Table \ref{tab:legend}). Visibility data has the highest memory consumption. This extra time can also impose limits on the performance of the imaging tool. The \textit{Visibility Manager} might not supply data chunks at rates faster than the processing of the chunks over the GPU. A case in point is Data Set 2 imaged using Configuration 1, where most of the GPU work is done while the \textit{Visibility Manager} is preparing data.

Results show that the processing time of the \textit{Visibility Manager} is independent of configuration but mostly dependent on the data set being imaged. This is an expected result.

A final observation is the excessive time the imaging tool takes to exit. It is marked in Figure \ref{fig:timeline} as \textit{exit time}. This time interval is a waste of time since by then, all images are finalised and saved to disk. For the multi-GPU scenario,  it amounted to 10 seconds! Detailed analysis revealed that most of this time is consumed by the CUDA Runtime API to reset GPU devices. It is yet unclear who is the culprit, whether a limitation of the CUDA Runtime API, GPUs, or something else.

\subsection{Multi-GPU scalability}
\label{sec:multiGPU}

When imaging over two GPUs (refer to Figure \ref{fig:timeline}), only a performance gain of $1.2\times$ against the GTX670\footnote{Comparison is made against the GTX670 GPU because it is less powerful than the GTX680.} run was obtained. The rather low value is the result of the exceptionally strong performance already obtained by imaging over 1 GPU. Performance  gains obtained from imaging over more than 1 GPU are limited by the time consumed to load data from disk. If the time to load data and the \textit{exit time} are ignored, a speed-up of $1.8\times$ results. Thus, \textit{mt-imager} is scalable over GPUs. Nevertheless, the loading of data from disk, limits the gains severely.

\subsection{Summary}
Table \ref{tab:maintopics} summarises the main performance topics reported in this chapter.

\begin{table}[H]
  \centering
    \begin{tabularx}{\textwidth}{LL}
    \hline
    \textit{Topic} & \textit{Comment} \\
    \hline
	Overall performance & Nearly $100\times$ faster then \textit{lwimager}. \\
	Main limiting factor & Loading of data from disk. \\
	Other limiting factor & \textit{mt-imager} takes substantial time to exit. \\
	Scalability over GPUs & \textit{mt-imager} is scalable over GPU, but most gains are hindered by the main limiting factor. \\
	\textit{WImager} gridder performance & Real gridding rate of ~50 Giga grid point updates/sec for most computationally intensive scenario. \\
	\textit{Compression} performance & Performance obtained from \textit{compression} varies depending on imaging configuration and data set. 3-fold increase in gridding performance were obtained in particular runs.\\
	Gridder main limiting factor & Retrieval of convolution function numeric data from GPU memory.\\  
    \hline
	\end{tabularx}
	\caption{\textit{mt-imager} performance summary} 
	\label{tab:maintopics}
\end{table}
\chapter{Conclusion}
\label{chap:conclusion}

In this thesis,  a new high-performance imaging synthesis tool for radio interferometry was developed. The tool which is called \textit{malta-imager} or \textit{mt-imager} exploits the computational power delivered by GPUs, to achieve unprecedented high performance. The backbone handling numerical calculations was generalised and a new framework was developed called the \textit{General Array Framework} (GAFW).

Test cases presented in this thesis show that the imaging tool is able to synthesis images nearly $100\times$ faster than a common CPU based imaging tool. This clearly shows that the thesis main objective, which is the development of a high-performance imaging tool, has been achieved in full.

\section{Future work}

The achievement and detailed results reported in this thesis open the door for more research and development for the imaging tool. 

The imaging tool still lacks necessary functionality to make it a popular tool of choice. This is especially true in wide-field astronomical observations. Future work has to address this problem. In particular,  the imaging tool lacks a deconvolution process and primary beam correction handling the A-term described in equation \ref{equ:measurment}.

The A-term is normally handled by A-projection \cite{Bhatnagar2013}. Convolution gridding can be applied in a similar way as the \textit{w-projection} algorithm. The main difference is that the A-term is a function of polarisation forcing a different convolution function for each polarisation. In view that the \textit{gridder} is limited by the retrieval of the convolution function data from memory, this method will most probably ill-perform, using the thesis implementation. A better method would be \textit{A-stacking} whereby the \textit{A-term} is corrected directly on the intensity plane.

Another limitation is that synthesised images are limited by GPU memory. It is indispensable to overcome this limitation since current and next generation telescopes support images that do not fit in the GPU memory (SKA images may be as large as $10^9$ pixels \cite{Cornwell2012}). The solution is to scale over a GPU cluster. It will be advantageous if the system scales down to one GPU and still be able to synthesise large images. Such down-scaling will let \textit{mt-imager} work on PCs and non-clustered independent servers. 

As pointed out in chapter \ref{chap:gafw}, the GAFW has been designed in such a way as to support different hardware. In this thesis, only  GPUs are supported, and the natural way forward is to support CPUs and clusters of GPUs and/or CPUs. It is quite desirable that the imaging tool grids over GPU clusters or CPU clusters by virtue of the GAFW.

In the field of high performance, it is extremely beneficial to retain an open mind set. Though this thesis shows that GPUs do provide a suitable solution, it does not mean that they provide the best solution. The possibility of synthesising over CPUs should be analysed. The use of other hardware such as FPGAs or Intel\textregistered Xeon Phi\texttrademark\ \cite{INTELPHI} should also be considered. This argument is particularly significant for the future. With the continuous enhancement of current technology and new ideas that come out on the market, it will not be a surprise if solutions better than GPUs will be available.

\bibliomatter
\bibliographystyle{packages/IEEEtran}
\bibliography{mainbin}
\end{document}